%% file: blend.tex
\input macros.tex

\def\CI{{\chi_I}}
\def\Gr{{Gr}}
\def\Ls{{L\!^*}}
\def\Lb{{L\!^\bullet}}

\def\NuNol{{\overline{N\!u_N}}}
\def\NuIol{{\overline{N\!u_{I}}}}

\def\Nuiol{{\overline{N\!u_\iota}}}

\def\NuolR{{\overline{N\!u_R}}}
\def\Nuolq{\overline{N\!u'}}
\def\Nuolq{\overline{N\!u'}}
\def\Nuols{\overline{N\!u^*}}
\def\Nuol{{\overline{N\!u}}}
\def\Nurol{{\overline{N\!u_\rho}}}

\def\Nus{{N\!u_\sigma}}
\def\Nutol{{\overline{N\!u_\tau}}}

\def\Nuzq{{N\!u'_0}}
\def\Nuzs{{N\!u^*_0}}
\def\Nuz{{N\!u_0}}

\def\Pra{{Pr}}
\def\Ra{{Ra}}
\def\ReI{{Re_I}}
\def\ReF{{Re_F}}
\def\ReN{{Re_N}}

\def\Rev{{Re_\varepsilon}}
\def\Rex{{Re_x}}
\def\Rey{{Re}}

\def\Wz{{\rm W\!}_0}
\def\W{{\rm W}}
\def\diff{{\rm d}}

\def\fctol{{\overline{f_{\tau}}}}

\def\hFol{{\overline{h_F}}}
\def\hNol{{\overline{h_N}}}
\def\hRol{{\overline{h_R}}}
\def\hol{{\overline{h}}}
\def\hqol{{\overline{h'}}}
\def\hqtol{{\overline{h'_\theta}}}
\def\hsol{{\overline{h^*}}}
\def\htol{{\overline{h_\theta}}}
\def\uinf{{u_\infty}}
\def\us{{u^*}}
\def\etal{{et~al.~}}

\centerline{\bf{Mixed Convection From an Isothermal Rough Plate}}
\medskip
\centerline{Aubrey G. Jaffer and Martin S. Jaffer}
\centerline{e-mail: agj@alum.mit.edu}

\beginsection{Abstract}

{\narrower

 This investigation derives formulas to predict the mixed
 convective surface conductance of a flat isotropic surface roughness
 having a convex perimeter in a Newtonian fluid with a steady forced
 flow in the plane of that roughness.

 Heat transfer measurements of a 30.5 cm square rough plate with
 forced air velocities between 0.1 m/s and 2.5 m/s were made by the
 present apparatus in two inclined and all five orthogonal
 orientations.  The present work's formulas are compared with 104
 measurements in twelve data-sets.  The twelve data-sets have
 root-mean-square relative error (RMSRE) values between 1.3\% and 4\%
 relative to the present theory.

 The present work's formulas are also compared with 78 measurements
 in 28 data-sets on five vertical rough surfaces in horizontal flow
 from Rowley, Algren, and Blackshaw (1930).  The five stucco data-sets
 have RMSRE values between 2.5\% and 6.5\%; the other data-sets have RMSRE
 values between 0.2\% and 5\%.

\par}

\bigskip
This research did not receive any specific grant from funding agencies
in the public, commercial, or not-for-profit sectors.

\beginsection{Table of Contents}

\readtocfile

\section{Introduction}

  Natural convection is the flow caused by nonuniform density in a
  fluid under the influence of gravity.  Forced convection is the heat
  or solute transfer to or from a surface induced by forced fluid flow
  parallel to that surface.  Mixed convection is the heat or solute
  transfer when both processes are operating simultaneously.

  Modeling mixed convection from the exterior faces of walls and roofs
  is essential to predicting the thermal performance of buildings and
  determining their heating and cooling requirements.  This
  investigation derives and tests mixed convection formulas for a
  rough, flat exterior face at any inclination, subjected to forced
  flow in the plane of the surface.

  Three modes of forced flow of a Newtonian fluid along a (flat)
  surface are laminar flow, turbulent flow, and rough flow.  Flow
  along flat, smooth plates gradually transitions from laminar to
  turbulent in a continuous
  boundary-layer\numberedfootnote{Schlichting~\cite{schlichting2014}
  describes a boundary-layer: ``In that thin layer the velocity of the
  fluid increases from zero at the wall (no slip) to its full value
  which corresponds to external frictionless flow.''}
  (Lienhard~\cite{10.1115/1.4046795}).

  Surface roughness repeatedly disrupts the boundary-layer in rough
  flow, which occurs along rough surfaces
  (Jaffer~\cite{thermo3040040}).

  Forced convection fluid flow is parallel to the surface.  In natural
  convection the temperature difference between the fluid and surface
  creates an upward or downward fluid flow, which is not necessarily
  parallel to the surface.  Along a vertical plate, ``aiding'' has
  natural and forced flows in the same direction; ``opposing'' flows
  are in opposite directions.

  Natural convection is sensitive to plate inclination, while forced
  convection is not.  Forced convection has different formulas for
  laminar, turbulent, and rough flows, while a single formula governs
  both laminar and turbulent natural convection
  (Fujii and Imura~\cite{fujii1972natural},
   Churchill and Chu~\cite{churchill1975correlating},
   Jaffer~\cite{thermo3010010}).

  There is a symmetry in external natural convection; a cooled plate
  induces downward flow instead of upward flow.  Flow from a cooled
  upper face is the mirror image of flow from a heated lower face.
  Flow from a cooled lower face is the mirror image of flow from a
  heated upper face.

  The rest of this investigation assumes a surface warmer than the
  fluid.

\subsection{Fluid Mechanics}
  In engineering, heat transfer rates for both natural and forced
  convection are expressed using the average surface conductance
  $\hol$ with units ${\rm{W/(m^2\cdot K)}}$.

  In fluid mechanics, the convective heat transfer rate is represented
  by the dimensionless average Nusselt
  number~($\Nuol\equiv{\hol\,L/k}$), where~$k$ is the fluid's thermal
  conductivity with units ${\rm W/(m\cdot K)}$, and $L$ is the
  system's characteristic length (m).

  The Reynolds number~$\Rey$ is dimensionless and proportional to
  fluid velocity.  The Rayleigh number~$\Ra$ is the impetus for
  fluid flow due to temperature difference and gravity.  A fluid's
  Prandtl number~$\Pra$ is its momentum diffusivity per thermal
  diffusivity ratio.
  The system's characteristic length~$L$ scales both~$\Nuol$ and
  $\Rey$; $\Ra$~is scaled by~$L^3$; both $\hol$ and $\Pra$~are
  independent of~$L$.

\subsection{Combining Transfer Processes}
  Formula~\eqref{eq:mixing} is an unnamed form for combining functions
  which appears frequently in heat transfer formulas:
$$F^p=F_1^p+F_2^p\eqdef{eq:mixing}$$

  Churchill and Usagi~\cite{AIC:AIC690180606} stated that such
  formulas are ``remarkably successful in correlating rates of
  transfer for processes which vary uniformly between these limiting
  cases.''
  Convection transfers heat (or solute) between the plate and fluid.

\subsection{The $\ell^p$-norm}
  When $F_1\ge0$ and $F_2\ge0$, taking the~$p$th root of both sides of
  Equation~\eqref{eq:mixing} yields a vector-space functional form
  known as the
  $\ell^p$-norm, which is notated $\|F_1~,~F_2\|_p$~:
$$\left\|F_1~,~F_2\right\|_p\equiv\left[~|F_1|^p+|F_2|^p\right]^{1/p}\eqdef{eq:l^p}$$

  Norms generalize the notion of distance.  Formally, a vector-space
  norm obeys the triangle inequality: $\|F_1,F_2\|_p\le|F_1|+|F_2|$, which
  holds only for $p\ge1$.  However, $p<1$ is also useful.

\medskip

\unorderedlist
 
 \li When $p>1$, the processes modeled by $F_1$ and $F_2$ compete and
 $\|F_1,F_2\|_p\ge\max(|F_1|,|F_2|)$; the most competitive case is
 $\|F_1,F_2\|_{+\infty}\equiv\max(|F_1|,|F_2|)$.

\unorderedlist

 \li Formula~\eqref{eq:h-vertical} uses the $\ell^3$-norm.

 \li Formula~\eqref{eq:h-opp} uses the $\ell^{\sqrt{3}}$-norm.

\endunorderedlist

 \li The $\ell^2$-norm is
 equivalent to root-sum-squared; it models perpendicular competitive
 processes.

\unorderedlist

 \li Formulas~\eqref{eq:h-v} and \eqref{eq:h-up} use the $\ell^2$-norm.

\endunorderedlist

 \li The $\ell^1$-norm models independent processes;
 $\|F_1,F_2\|_1\equiv{|F_1|+|F_2|}$.

 \li When $0<p<1$, the processes cooperate and $\|F_1,F_2\|_p\ge{|F_1|+|F_2|}$.

\unorderedlist

 \li Cooperation between conduction and flow-induced heat transfer
 manifests as the $\ell^{1/2}$-norm in natural convection
 Formula~\eqref{eq:general}.

\endunorderedlist

 \li When $p<0$, $\|F_1,F_2\|_p\le\min(|F_1|,|F_2|)$, with the transition
 sharpness controlled by $p$; the extreme case is
 $\|F_1,F_2\|_{-\infty}\equiv\min(|F_1|,|F_2|)$.
 Negative $p$ can
 model a single flow through serial processes; the most restrictive
 process limits the flow.

\unorderedlist

 \li Formula~\eqref{eq:bi-level-convect} uses the $\ell^{-4}$-norm.

 \li Formula~\eqref{eq:Re-Y} uses the $\ell^{-\sqrt{1/3}}$-norm.

\endunorderedlist

\endunorderedlist

\subsection{Roughness}
  The present theory treats surface roughness as an elevation
  function~$z(x,y)$ defined on an area~$A$ having a convex perimeter.
  Function~$z(x,y)$ has only one value at each $(x,y)$ coordinate;
  thus surfaces with tunnels and overhangs are disqualified, as are
  porous surfaces.  The mean elevation~$\overline{z}$ and
  root-mean-squared (RMS) height-of-roughness~$\varepsilon\ll{L}$ are:
$$\eqalignno{\overline{z}&={\left.\int_A z\,\diff{A}\right/\int_A\diff{A}}&\eqdef{eq:mean-surface-height}\cr
  \varepsilon&=\sqrt{\left.\int_A |z-\overline{z}|^2\,\diff{A}\right/\int_A\diff{A}}&\eqdef{eq:RMS-surface}\cr}$$

\subsection{Prior Work}

 Nearly all of the experimental prior works
 (\cite{HIEBER1973769,wang1982experimental,osti_5873492,Ramachandran1985,lin1990comprehensive,KOBUS19953329,KOBUS20013381})
 concern smooth plates.  The exception is Rowley, Algren, and
 Blackshaw~\cite{rowley1930surface}, the 1930 result of cooperative
 research between the University of Minnesota and the American Society
 of Heating and Ventilation Engineers.  They measured mixed convection
 of 0.305~m square vertical plates in horizontal flow in a wind
 tunnel.  They tested common rough exterior surfaces, specifically
 concrete, brick, stucco, and rough and smooth plaster.

 Mixed convection measurements from the graphs in
 Rowley \etal\cite{rowley1930surface} were captured by measuring the
 distance from each point to its graph's axes, then scaling to the
 graph's units using the ``Engauge'' software (version
 12.1).  \tabref{tab:sources} lists the data-sets to be compared with
 the present theory, where $\theta$ is the angle of the plate from
 vertical and $\psi$ is the angle of the forced flow from the zenith;
 $\psi=90^\circ$ is horizontal.

\subsection{Present Apparatus}
 The present apparatus combines an open intake wind-tunnel, software
 phase-locked loop fan control, and a heated aluminum plate
 centered in the test chamber by a six wire suspension.

 This apparatus measured average mixed convection in air with
 $2300<\ReF<93000$, a 40:1 range.  The wind-tunnel chassis
 (${\rm{1.3~m\times0.61~m\times0.65~m}}$) was small enough to allow
 positioning in horizontal, vertical, and inclined orientations.
 \ref{Appendix C: Apparatus and Measurement Methodology} describes the
 apparatus and measurement methodology.

\subsection{Approaches}
 Rowley \etal\cite{rowley1930surface} provided graphs for engineering
 use which cover both laminar and turbulent flows.  It applies only to
 forced horizontal flow in the plane of a vertical plate.  It lacked a
 roughness metric which would have allowed application to other types
 of rough surfaces.  Unfortunately, forced convection from rough
 plates does not scale simply, being inversely proportional to
 $\log^2(L/\varepsilon)$.

 The present work is primarily theoretical, combining the system-wide
 heat transfer derivations of natural and forced convections from
 Jaffer~\cite{thermo3010010} and Jaffer~\cite{thermo3040040},
 respectively.  It applies to convex flat surfaces at any inclination
 having isotropic roughness with $0<\varepsilon\ll{L}$ and forced flow
 parallel to the surface.

\medskip
\centerline{\bf\tabdef{tab:sources}\quad{Rowley \etal mixed convection data-sets}}
\moveright 0.1\hsize
\vbox{\settabs 10\columns
\toprule
\+ {\bf Surface} &\hfill$\theta~~~$&\hfill$\psi~~~$&  ~~$\Ra\ge$ & ~~$\Ra\le$ & ~~$\Rey\ge$ & ~~$\Rey\le$ &\hfil\bf Count&\cr
\midrule
\input Rowley-et-al-sources.tex
\bottomrule
}

\subsection{Not Empirical}
  Empirical theories derive their coefficients from measurements,
  inheriting the uncertainties from those measurements.  Theories
  developed from first principles derive their coefficients
  mathematically.  For example, Incropera, DeWitt, Bergman, and
  Lavine~\cite{bergman2007fundamentals} (p.~210) gives the thermal
  conductance of one face of a diameter $D$ disk into a stationary,
  uniform medium having thermal conductivity $k$ as exactly
  $8\,k/[\pi\,D]$ (units ${\rm W/(m^2\cdot K)}$).
  The present theory derives from first principles; it is not
  empirical.  Each formula is tied to aspects of the plate geometry
  and orientation, fluid, and flow.

\subsection{RMS Relative Error}
 Root-mean-squared (RMS) relative error (RMSRE) provides an objective,
 quantitative evaluation of experimental data versus theory.  It
 gauges the fit of measurements $g(\Rey_j)$ to function $f(\Rey_j)$,
 giving each of the $n$ samples equal weight in
 Formula~\eqref{eq:RMSRE}.
 Along with presenting RMSRE, charts in the present work split RMSRE
 into the bias and scatter components defined in
 Formula~\eqref{eq:bias}.  The root-sum-squared of bias and
 scatter is RMSRE.

$$\eqalignno{{\rm RMSRE}=&\sqrt{{1\over n}\sum_{j=1}^n\left|{g(\Rey_j)\over f(\Rey_j)}-1\right|^2}&\eqdef{eq:RMSRE}\cr
                            {\rm bias}={1\over n}\sum_{j=1}^n\left\{{g(\Rey_j)\over f(\Rey_j)}-1\right\}&\qquad
  {\rm scatter}=\sqrt{{1\over n}\sum_{j=1}^n\left|{g(\Rey_j)\over f(\Rey_j)}-1-{\rm bias}\right|^2}&
  \eqdef{eq:bias}\cr}$$

\section{Natural Convection}

 Jaffer~\cite{thermo3010010} derived a natural convection formula for
 external flat plates (with convex perimeter) in any orientation from
 its analyses of horizontal and vertical plates.  This investigation
 uses the same approach.

  \figrefs{fig:vertical-flow}, \figrefn{fig:above-flow},
  and~\figrefn{fig:below-flow} show the induced fluid flows around
  heated vertical and horizontal surfaces.

  For a horizontal plate with heated upper face, streamlines
  photographs in Fujii and Imura~\cite{fujii1972natural} show natural
  convection pulling fluid horizontally from above the plate's
  perimeter into a rising central plume.  \figref{fig:above-flow}
  is a diagram of this upward-facing convection.
  Horizontal flow is nearly absent at the elevation of the dashed
  line.

  The streamlines photograph of a vertical plate in Fujii and
  Imura~\cite{fujii1972natural} shows fluid being pulled horizontally
  before rising into a plume along the vertical
  plate.  \figref{fig:vertical-flow} is its diagram.

  Modeled on a streamlines photograph in
  Aihara, Yamada, and End\"o~\cite{AIHARA19722535},
  \figref{fig:below-flow} is a flow diagram for a horizontal plate with
  heated lower face.  Unheated fluid below the plate flows
  horizontally inward.  It rises a short distance, flows outward
  closely below the plate, and flows upward upon reaching the
  plate edge.
  The edge flows self-organize so that they are at the opposing edges
  of the plate which are nearest to each other.

  Horizontal flow in \figref{fig:above-flow} is radial, but not radial
  in \figref{fig:below-flow}.

\medskip

\hbox{
 \vbox{\settabs 3\columns
  \+\hfil\figscale{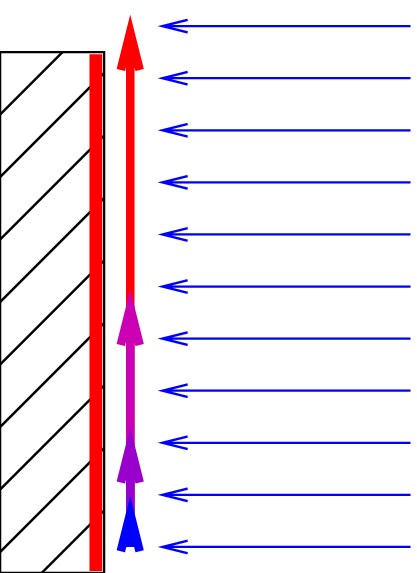}{85pt}&\cr
  \+\hfil{\bf\figdef{fig:vertical-flow}\quad{Vertical plate}}&\cr
  \bigskip
  \medskip
 }
 \vbox{\settabs 2\columns
  \+\hfil\figscale{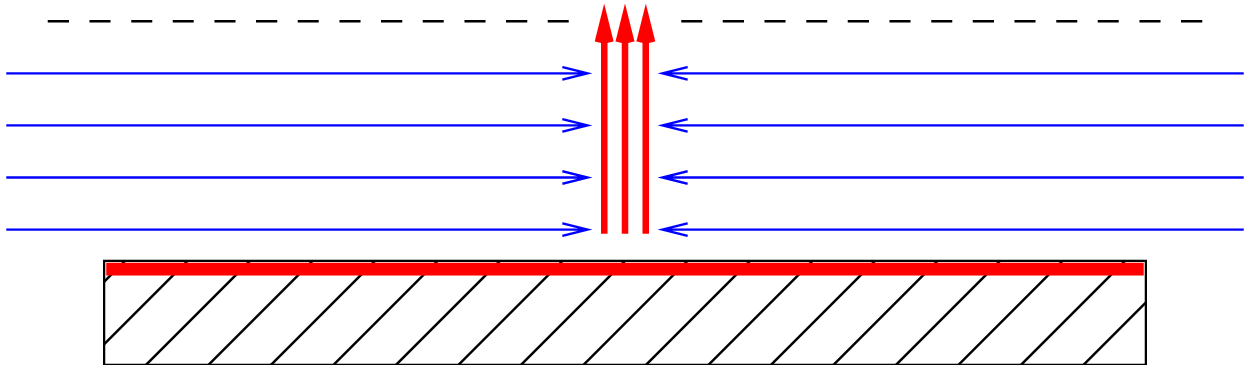}{230pt}&\cr
  \+\hfil{\bf\figdef{fig:above-flow}\quad{Flow above a heated plate}}&\cr
  \medskip
  \+\hfil\figscale{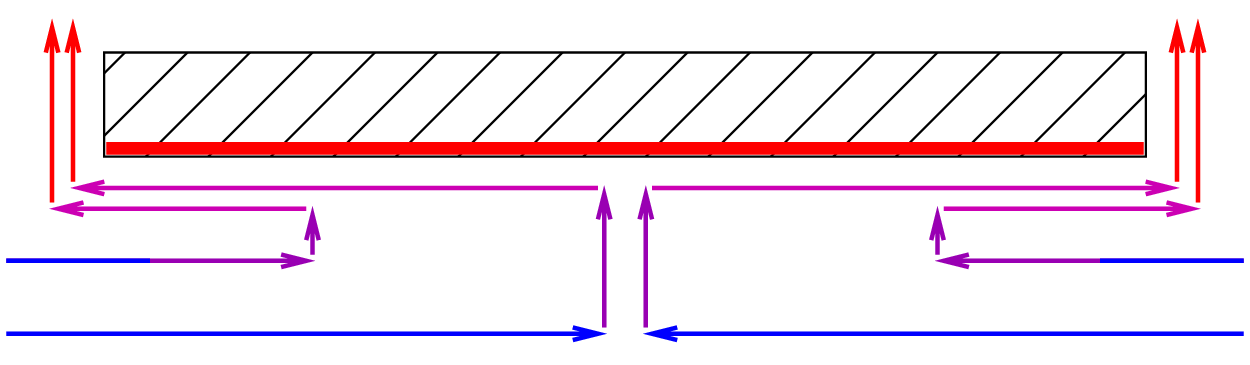}{230pt}&\cr
  \+\hfil{\bf\figdef{fig:below-flow}\quad{Flow below a heated plate}}&\cr
  }
}
\medskip

  An important aspect of all three flow topologies is that fluid is
  pulled horizontally before being heated by the plate.  Pulling
  horizontally expends less energy than pulling vertically because
  the latter does work against the gravitational force.
  Inadequate horizontal (or vertical) clearance around a plate can
  obstruct flow and reduce convection and heat transfer; such a plate
  is not ``external''.

  From thermodynamic constraints, Jaffer~\cite{thermo3010010} derives
  generalized natural convection Formula~\eqref{eq:general} with the
  parameters specified in \tabref{tab:natural convection parameters}:

\unorderedlist

 \li $\theta$ is the angle of the plate from vertical;

 \li $L$ is the characteristic length of a flat plate with convex
 perimeter:

\unorderedlist

 \li face up $\Ls$ is the area-to-perimeter ratio;

 \li vertical $L'$ is the harmonic mean of the perimeter vertical
 spans (the height of a level rectangle);

 \li face down $L_R$ is the harmonic mean of the perimeter
 distances to that bisector which is perpendicular to the shortest
 bisector (1/2 of the shorter side of a rectangle);

\endunorderedlist

 \li $\Nuz$~is the conduction into the fluid when not moving (static);

 \li $\Ra'$ is computed with vertical $L'$; $\Ra^*=\Ra'\,[\Ls/L']^3$;
  $\Ra_R=\Ra'\,[L_R/L']^3$.

  $\Pra$ does not affect upward-facing heat transfer because the
  heated fluid flows directly upward, as does conducted heat.  When
  heated fluid must flow along vertical and downward-facing plates,
  its heat transfer potential is reduced by dividing $\Ra$ by $\Xi$
  from Formula~\eqref{eq:Xi}.

 \li $E$~is the count of $90^\circ$ changes in direction of fluid
  flow;

 \li $B$~is
  the sum of the mean lengths of flows parallel to the plate divided
  by~$L$;

 \li $C$~is the plate area fraction responsible for flow induced
  heat transfer;

 \li $D$~is the effective length of heat transfer contact
  with the plate divided by~$L$;

 \li The $\ell^p$-norm combines the static conduction and induced
  convective heat flows.

\endunorderedlist

\centerline{\bf\tabdef{tab:natural convection parameters}\quad{Natural convection parameters}}
\vbox{\settabs 11\columns
\toprule
  \+\bf Face &$\quad\theta$&$L$  &$\Nuol$   &$\Nuz$   &$\Ra$      &\hfil$E$&\hfil$B$    &\hfil$C$       &\hfil$D$  &\hfil$p$  &\cr
\midrule        
  \+ up      &$-90^\circ$ &$\Ls$&$\Nuols$  &$\Nuzs$  &$\Ra^*$    &\hfil1  &\hfil$2$    &\hfil$1/\sqrt8$&\hfil$1$  &\hfil$1/2$&&\cr
  \+ vertical&$\quad0^\circ$ &$L'$ &$\Nuolq$  &$\Nuzq$  &$\Ra'/\Xi$ &\hfil1  &\hfil$1/2$  &\hfil$1/2$     &\hfil$1/4$&\hfil$1/2$&&\cr
  \+ down    &$+90^\circ$ &$L_R$&$\NuolR$ &$\Nuzq/2$&$\Ra_R/\Xi$&\hfil3  &\hfil$4$    &\hfil$1/2$     &\hfil$2$  &\hfil$1$  &&\cr
\bottomrule
}

$$\eqalignno{
  \Nuol&=\left\|\Nuz\,\bigl[1-C\bigr], \root{2+E}\of{\bigl[{C\,D\,\Nuz}\bigr]^{3+E}{{2\over B}\,\Ra}}~\right\|_p&\eqdef{eq:general}\cr
  \Xi&=\left\|1~,~{0.5\over\Pra}\right\|_{\sqrt{1/3}}\quad\Nuzs={2\over\pi} \quad\Nuzq={2^4\over\root4\of{2}\,\pi^2}&\eqdef{eq:Xi}\cr}$$

\subsection{Effective Vertical Reynolds Number}
 From the derivation in Jaffer~\cite{thermo3010010} with
 $\Ra'/\Xi\gg1$, $\Nuol\approx C\,D\,\Nuz\,\Rey$.  Heat transfer
 $\Nuolq$ is reduced by the self-obstruction factor
 $1/\root3\of{\Xi}$, which grows with~$\Pra$.  However, heat transfer
 is not the same as fluid flow, which increases with decreasing
 $\Pra$.  The $\Xi^3$ factor in Formula~\eqref{eq:Re-from-Nu} makes
 $\ReN$ increase with decreasing $\Pra$.
 Proposed is Formula~\eqref{eq:Re-from-Nu} as the Reynolds number
 associated with the natural convective flow from a vertical plate.
$$\ReN\approx{\Nuol\,\Xi^{2+E}\over\Nuz\,C\,D}={8\,\Nuolq\,\Xi^3\over\Nuzq}
  \qquad\Nuolq\gg\Nuzq\eqdef{eq:Re-from-Nu}$$

\smallskip
\vbox{\settabs 1\columns
\+\hfil\figscale{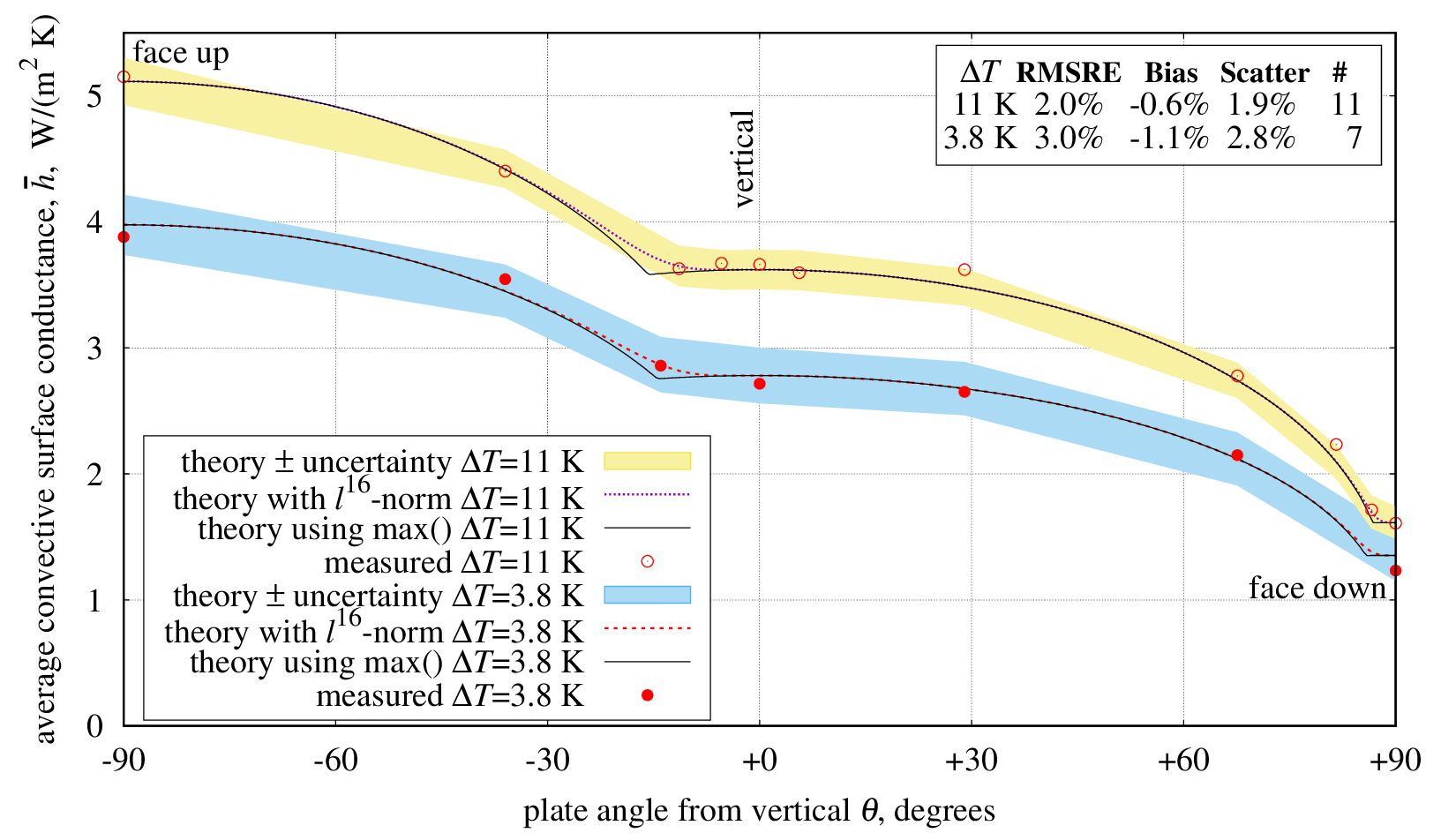}{435pt}&\cr
\+\hfil{\bf\figdef{fig:3mm-angles-correlation}\quad{Natural convection versus angle}}&\cr
}

\subsection{Natural Convection From an Inclined Plate}
 $\Ra$ is proportional to gravitational acceleration.  Following the
 approach of Fujii and Imura~\cite{fujii1972natural}, the $\Ra$
 argument to $\hqol(\Ra)\equiv k\,\Nuolq(\Ra)/L'$ is scaled by
 ${\left|\cos\theta\right|}$, modeling the reduced convection of a
 tilted plate as a reduction in gravitational acceleration.
 Similarly, the $\Ra$ arguments to $\hsol$ and $\hRol$ are scaled by
 ${\left|\sin\theta\right|}$.  An unobstructed plate induces a single
 steady-state mode of natural convection (face up, down, or vertical).
 The instances of $\max()$ in Formula~\eqref{eq:inclined-max} choose
 the largest surface conductance among these modes.
$$\hol=\cases{
 \max\left({\hqol({\left|\cos\theta\right|}\,\Ra'/\Xi)},~{\hsol({\left|\sin\theta\right|}\,\Ra^*)}\right)&$\sin\theta<0$\cr
 \max\left({\hqol({\left|\cos\theta\right|}\,\Ra'/\Xi)},~{\hRol\left({\left|\sin\theta\right|}\,\Ra_R/\Xi\right)}\right)&$\sin\theta\ge0$\cr}
 \eqdef{eq:inclined-max}$$

 In reality, the $\theta$ transition is more gradual using the
 $\ell^{16}$-norm in Formula~\eqref{eq:inclined-natural}:
$$\hol=\cases{
 \left\|{\hqol({\left|\cos\theta\right|}\,\Ra'/\Xi)},~{\hsol({\left|\sin\theta\right|}\,\Ra^*)}\right\|_{16}&$\sin\theta<0$\cr
 \left\|{\hqol({\left|\cos\theta\right|}\,\Ra'/\Xi)},~{\hRol\left({\left|\sin\theta\right|}\,\Ra_R/\Xi\right)}\right\|_{16}&$\sin\theta\ge0$\cr}
 \eqdef{eq:inclined-natural}$$

\subsection{Rough Natural Convection}
The agreement of rough plate measurements with theory over the
$\pm90^\circ$ range in \figref{fig:3mm-angles-correlation} indicates
that Formula~\eqref{eq:inclined-natural} governs plates with RMS
height-of-roughness $0\le\varepsilon\ll{L}$.

\section{Forced Convection}

  Forced convection $\Nuol$ is the heat transfer caused by forced flow
  along (and parallel to) a heated plate.  The surface
  conductance $\hFol\equiv{\Nuol\,k/L}$ grows with $\ReF$, $\Pra$, and
  $k$.  Its characteristic length $L$ is the length of the plate in
  the direction of forced flow.

\subsection{Rough Convection}
  Jaffer~\cite{thermo3040040} derives the forced convection
  $\Nurol$ of rough flow from isotropic, periodic roughness:
$$\Nurol(\ReF)={\ReF\,\Pra_\infty^{1/3}\,w\over6\,[\ln{\left(L/\varepsilon\right)}]^2}
 \qquad\ReF>\left[{0.664\over\varepsilon}\right]^2\,{L_P\,L}
 \eqdef{eq:rough-forced}$$
$$w={\left\| 1, {\varepsilon\over L_W}\right\|_{\sqrt{1/2}}}\eqdef{eq:w}$$

\unorderedlist

  \li $\varepsilon\ll L$ is the root-mean-squared (RMS) height of
  roughness.

  \li $L_P\ll L$ is the isotropic period of the roughness
  (Jaffer~\cite{thermo3040040}).

  \li $L_W$ is the width of the plate (perpendicular to $L$).

  \li $\Pra_\infty$~is the bulk fluid's Prandtl number (far from the plate).

\endunorderedlist

  If the roughness extends to the plate's rim, then it increases the
  effective width of the rough face by more than $2\,\varepsilon$
  because, in addition to the fluid adjacent to plate's face and rim,
  the fluid near the edge between them is affected.
  Thus~$\varepsilon$ and plate width $L_W$ cooperate weakly, leading
  to an effective width of $\|L_W,\varepsilon\|_{\sqrt{1/2}}$.
  Dividing by~$L_W$, Formula~\eqref{eq:w} is the edge roughness
  correction factor~$w$.  \figref{fig:w} graphs~$w$ as a function
  of~$\varepsilon$.

\smallskip
\vbox{\settabs 1\columns
\+\hfil\figscale{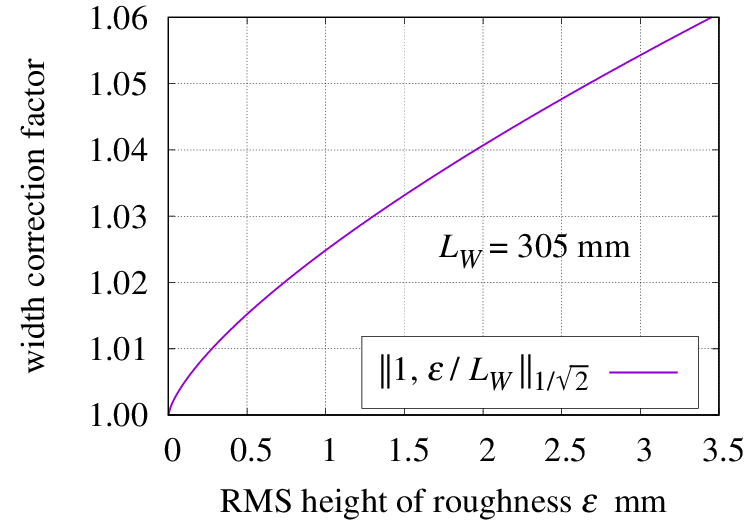}{234pt}&\cr
\+\hfil{\bf\figdef{fig:w}\quad Edge roughness correction factor}&\cr}

\subsection{Plateau Roughness}
 There are isotropic, periodic roughnesses whose convective heat
 transfer differs from Formula~\eqref{eq:rough-forced}.

\unorderedlist
 \li Informally, a ``plateau roughness'' is an isotropic, periodic
 roughness with most of its area at its peak elevation.  A
 quantitative definition is given in Jaffer~\cite{thermo3040040}.

 \li A ``plateau wells roughness'' is an array of co-planar wells
 dropping below a flat surface.

 \li A ``plateau islands roughness'' is an array of co-planar islands.

 The present apparatus plate has plateau islands roughness.
\endunorderedlist

 For a given $\ReF$, a plateau roughness may contain areas
 transferring heat per Formula~\eqref{eq:rough-forced}, and separate
 areas transferring heat as turbulent flow along a smooth plate, but
 with characteristic length $L_P$.

\subsection{Turbulent Forced Convection}
  Jaffer~\cite{thermo3040040} derives the average surface
  conductance, $\hFol\equiv{\Nutol\,k/L}$, of turbulent flow along an
  isothermal plate as Formula~\eqref{eq:turbulent-convect}.
$$\Nutol={\Nuz\,\ReF\,\fctol\over\sqrt{3}}\,\sqrt{\Pra/\sqrt{162}+1\over\sqrt{162}\,\Pra\,\fctol+{1}}\,\root3\of{{\Pra/\Xi\over{\|1,1/\Pra\|_3}}}
  \qquad\sqrt{162}\equiv9\,\sqrt{2}\approx12.7\eqdef{eq:turbulent-convect}$$
$$\fctol={2^{-5/4}\over\left[{\Wz\left(\ReF/\sqrt{3}\right)}-1\right]^2}
 \qquad\Xi={\left\|1,~{0.5\over\Pra}\right\|_{\sqrt{1/3}}}
 \qquad\Nuz={2^4\over\pi^2\,\root4\of2}\eqdef{eq:turbulent-friction}$$
\unorderedlist

  \li The fluid's effective Prandtl number
  $\Pra=\Pra_W^{1/4}\,\Pra_\infty^{3/4}$
  (from \v{Z}ukauskas and \v{S}lan\v{c}iauskas~\cite{Zukauskas1987}).

  $\Pra_W$~is the Prandtl number of fluid at wall (plate) temperature.

  $\Pra_\infty$~is the Prandtl number of fluid at the bulk flow
  temperature.

  \li $\Wz$ is the principal branch of the Lambert $\W$ function,
  defined as $\Wz(\varphi\,\exp{\varphi})=\varphi$ when $\varphi\ge0$.

  \li In Formula~\eqref{eq:turbulent-friction}, $2^{-5/4}$ replaces
  the $\root3\of2/3$ coefficient from Jaffer~\cite{thermo3040040}, a
  $+0.11\%$ correction.

  \li Plateau islands roughness can shed rough and turbulent flow
  simultaneously.
\endunorderedlist

\subsection{Plateau Islands Roughness}

  The plateau islands roughness described in (appendix)
  \ref{Appendix C: Apparatus and Measurement Methodology} has $\Nurol$
  Formula~\eqref{eq:rough-forced} rough convection in the leading
  $\ReI/\ReF$ portion of the plate, and $\NuIol$
  Formula~\eqref{eq:posts-convect} turbulent convection in the rest of
  the plate.  Formula~\eqref{eq:Re_I} $\ReI$ separates the regions,
  where $\Lb$ is the ratio of each (convex) island's area to its
  perimeter.

$$\eqalignno{
  \NuIol=&\left\{{1-\Omega}+\left\|{\Omega\over2},{2\,\varepsilon\,[4\,\Lb]\over L_P^2}\right\|_2\right\}
  {L\over L_P}\,\Nutol\left({\ReF\,L_P\over L}\right)&\eqdef{eq:posts-convect}\cr
  \ReI=&{3^3\,\varepsilon^2\,L^2\over \Lb\,L_P^3}\,\ln{3^3\,\varepsilon^2\,L^2\over\sqrt{3}\,\Lb\,L_P^3}&\eqdef{eq:Re_I}\cr
  \Nuiol=&\NuIol(\ReF)+\Nurol\left(\|\ReF, \ReI\|_{-4}\right)-\NuIol\left(\|\ReF, \ReI\|_{-4}\right)&\eqdef{eq:bi-level-convect}\cr
  }$$

 ``Openness'' $0<\Omega<1$ is the non-plateau area per cell area
 ratio.  Given a $w\times{w}$ matrix of elevations~$S_{s,t}$:
$$\Omega\approx{1\over{w^2}}\sum_{t=0}^{w-1}\,\sum_{s=0}^{w-1}\cases{
        1,& if $S_{s,t}<\max(S)-{\varepsilon^2/L_P}$;\cr
        0,& otherwise.\cr
        }\eqdef{eq:Omega}$$

 In the log-log plots in \figrefs{fig:slopes1mm}
 and~\figrefn{fig:slopes3mm}, the effective $\ReF$ exponent is the
 slope of its line.  For example, the ``$\ReF/200$'' and
 ``$\ReF/333$'' lines have slope 1; thus they are proportional to
 $\ReF^1\equiv\ReF$.  In each plot, the ``bi-level'' trace is $\Nuiol$
 Formula~\eqref{eq:bi-level-convect}.  The ``turbulent~part'' trace is
 Formula~\eqref{turbulent-part}, which is the turbulent component of
 $\Nuiol$ Formula~\eqref{eq:bi-level-convect}:
$$\NuIol(\ReF)-\NuIol\left(\|\ReF, \ReI\|_{-4}\right)\eqdef{turbulent-part}$$

 In both \figrefs{fig:slopes1mm} and~\figrefn{fig:slopes3mm}, the
 slope of the Formula~\eqref{turbulent-part} ``turbulent~part'' trace
 is close to~1 through more than an order of magnitude of $\ReF$.
 The discrepancy at larger $\ReF$ is unimportant because the heat
 transfer is dominated by forced convection in that range.

\medskip
\vbox{\settabs 2\columns
\+\hfil\figscale{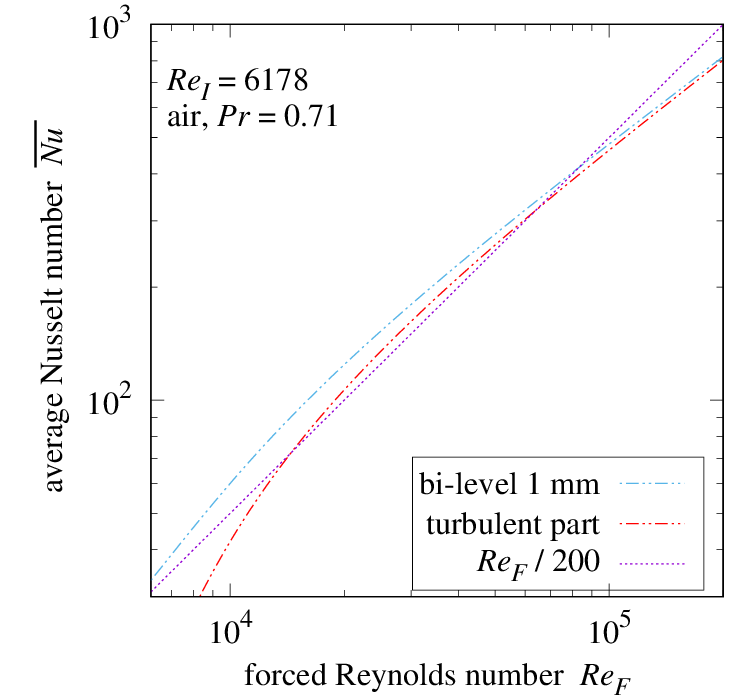}{234pt}&\hfil\figscale{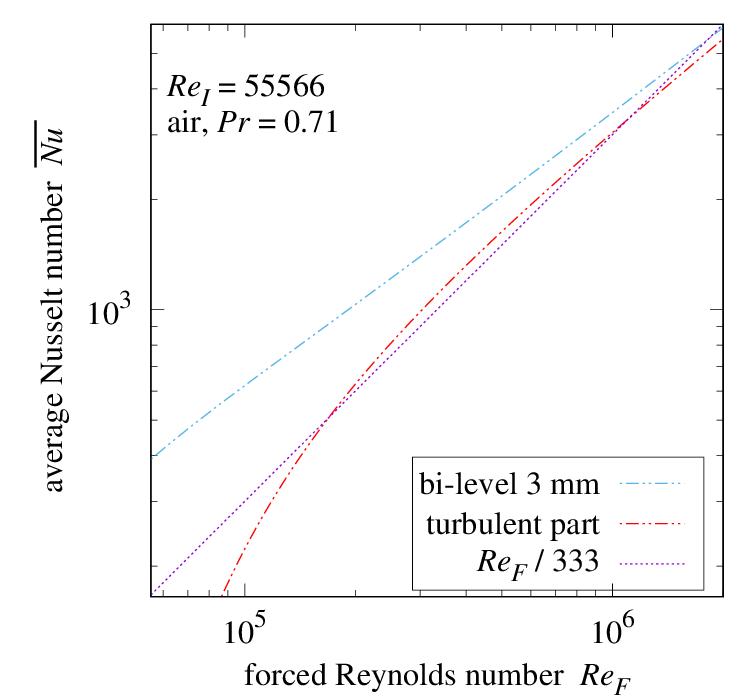}{234pt}&\cr
\+\hfil{\bf\figdef{fig:slopes1mm}\quad Forced convection $\ReI=6178$}&
  \hfil{\bf\figdef{fig:slopes3mm}\quad Forced convection $\ReI=55566$}&
  \cr}

\section{Mixing Natural and Forced Convections}

The previous sections established that:

\unorderedlist
 \li The effective natural Reynolds number $\ReN$ is proportional to
 $\NuNol$ when $\NuNol\gg\Nuz$.

 \li Forced rough convection $\Nurol$ is proportional to $\ReF$ in
 Formula~\eqref{eq:rough-forced}.

 \li And forced turbulent Formula~\eqref{turbulent-part} is nearly
 proportional to~$\ReF$ when $\ReF>\ReI$.
\endunorderedlist

On that basis, this investigation proposes:

\unorderedlist
 \li Surface conductances $\hNol$ and $\hFol$ both being proportional
 to Reynolds numbers indicates that they are commensurate; they can be
 combined using symmetrical functions such as the $\ell^p$-norm.
\endunorderedlist

 One approach to predicting mixed convection would be to compute
 $\Nuol$ from a function of $\ReF$ and $\ReN$.  However, choosing a
 single $\Nuol$ formula is problematic; while $\Rey$ and $\Nuol$ are
 nearly proportional in all three cases, their coefficients are very
 different.  Also, rough convection $\Nurol$ is strongly dependent on
 height-of-roughness $\varepsilon$, but \ref{Natural Convection} found
 that natural convection is insensitive to roughness
 $\varepsilon\ll{L}$.

\unorderedlist
 \li This investigation combines a natural surface conductance $\hNol$
 (specifically $\hsol$, $\hqol$, or $\hRol$) with the forced surface
 conductance $\hFol$ using the $\ell^p$-norm (where $p$ depends on
 plate and flow orientations).  Surface conductance $\hol$ is used
 instead of $\Nuol$ in order to avoid characteristic-length mismatch
 between $\Nuol$ formulas.

\endunorderedlist

This approach departs from prior works (all of which concerned smooth
plates), which compute $\Nuol$ from a ratio of powers of $\ReF$ and
the Grashof number $\Gr=\Ra/\Pra$.

\subsection{Theory and Measurements}
 Figures which follow plot $\Nuol$ at $L=L'$ measurements and
 theoretical curves versus $10^3<\ReF<10^5$ using logarithmic scales
 on both axes.  Logarithmic scales do not include 0; the following
 figures plot the natural convection measurement ($\ReF=0$) at
 $\ReF=10^3$.

 $\theta$ is the angle of the plate from vertical; $-90^\circ$ is face
 up; $+90^\circ$ is face down.

 $\psi$ is the angle of the forced flow from the zenith;
 $\psi=90^\circ$ is horizontal flow; $\psi=0^\circ$ is upward.  In
 this investigation, forced flow is always parallel to the plate;
 hence, horizontal plates have $\psi=90^\circ$.

 RMSRE is calculated from the measurements between the vertical lines,
 $1950<\ReF<5\times10^4$.

 The $\varepsilon=3$~mm plate sheds only rough flow at
 $1950<\ReF<5\times10^4$; its graphs are captioned ``rough''.  The
 $\varepsilon=1.04$~mm plate sheds rough flow at $\ReF<\ReI=6178$ and
 turbulent flow otherwise.  Hence it sheds mostly turbulent flow at
 $1950<\ReF<5\times10^4$; its graphs are captioned ``turbulent''.

\section{Horizontal Forced Flow}

\vbox{\settabs 1\columns
\+\hfil\figscale{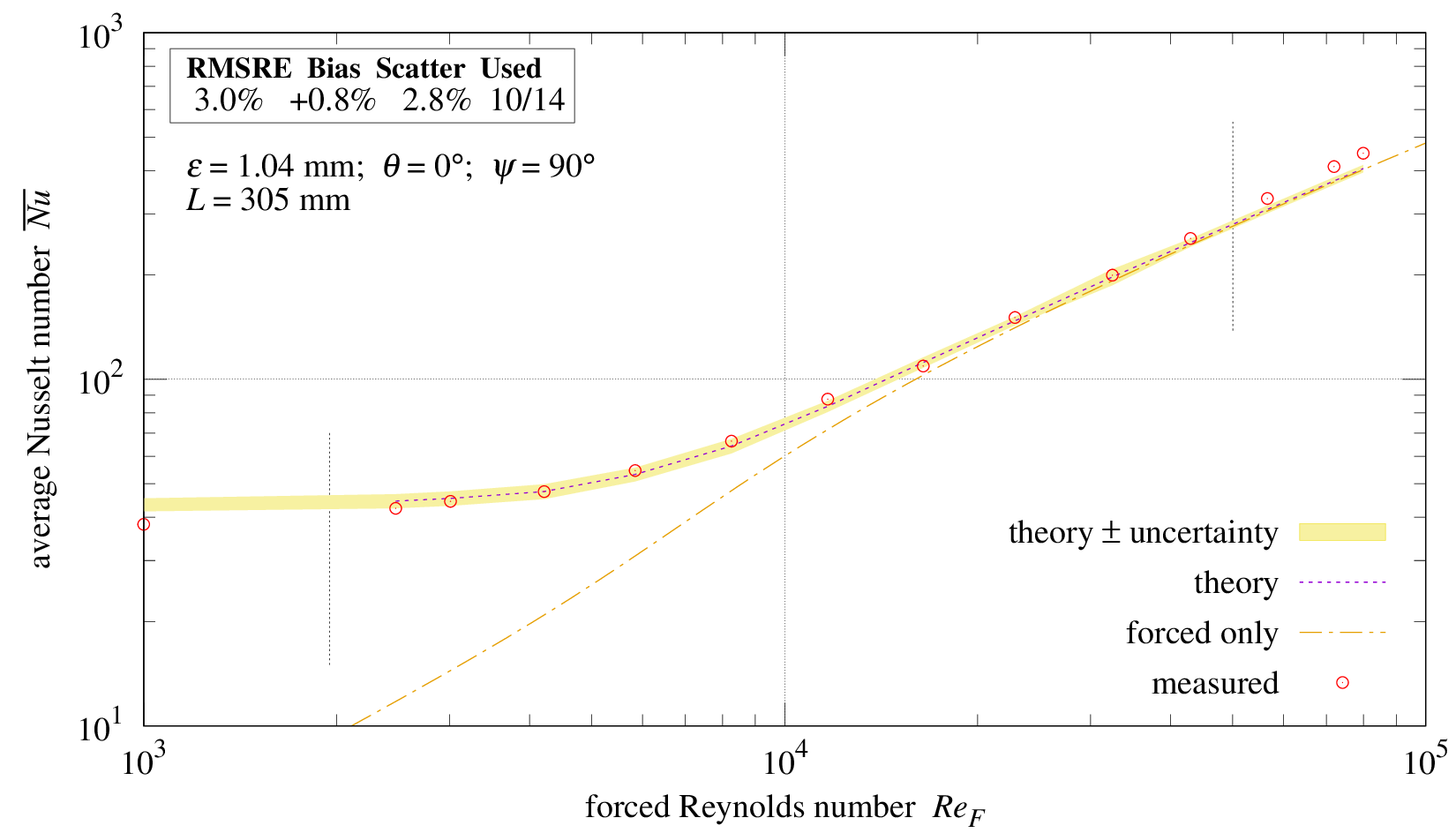}{435pt}&\cr
\+\hfil{\bf\figdef{fig:1mm-mixed-vt-correlation-1}\quad{Vertical plate in horizontal forced turbulent flow}}&\cr
\+\hfil\figscale{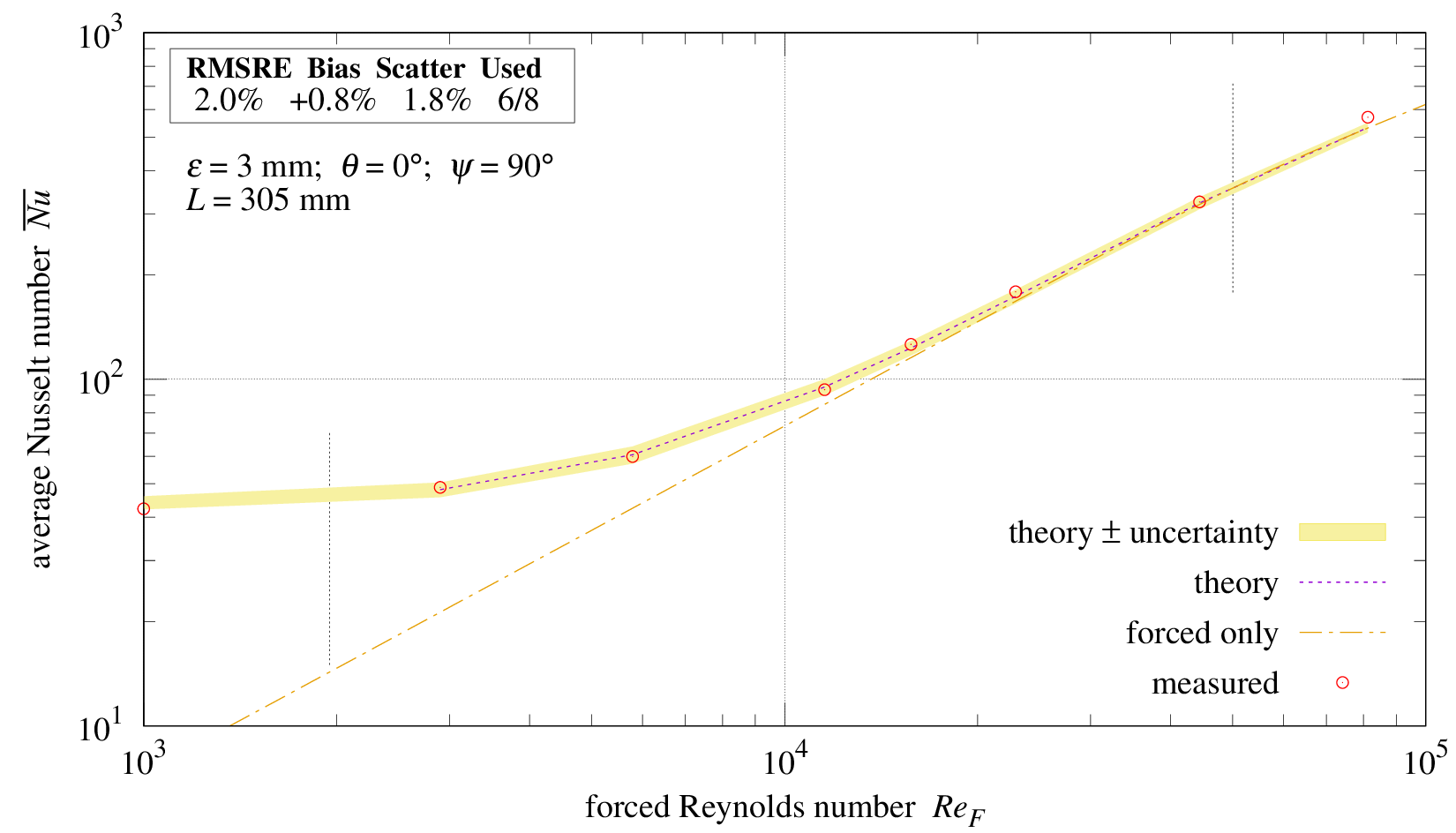}{435pt}&\cr
\+\hfil{\bf\figdef{fig:3mm-mixed-vt-correlation-1}\quad{Vertical plate in horizontal forced rough flow}}&\cr
}

\subsection{Vertical Plate With Horizontal Forced Flow}
 \figref{fig:vertical-flow} shows that fluid is drawn horizontally
 towards the heated surface, then rising.  The forced and natural heat
 flows are thus perpendicular, suggesting the $\ell^2$-norm for
 combining $\hFol$ and vertical $\hqol$:
$$\hol=\left\|{\hFol},{\hqol}\right\|_2\eqdef{eq:h-v}$$

 \figrefs{fig:1mm-mixed-vt-correlation-1}
 and~\figrefn{fig:3mm-mixed-vt-correlation-1} compare
 ($\Nuol\equiv{\hol\,L/k}$) Formula~\eqref{eq:h-v} with measurements
 of vertical plates in horizontal flow shedding turbulent or rough
 flow, respectively.

\subsection{American Society of Heating and Ventilation Engineers}
 Rowley \etal\cite{rowley1930surface} measured mixed convection of
 0.305~m square vertical plates in horizontal flow in a wind tunnel.
 Their graphs report surface conductance of the plate versus~$T_m$,
 the mean of the plate and airflow (Fahrenheit) temperatures.  The
 Rayleigh numbers used by natural convection formulas have the
 temperature difference as a factor.  Through trial and error it was
 found that taking $1.05\,T_m$ as the (Fahrenheit) plate temperature
 and $0.95\,T_m$ as the fluid temperature kept RMSRE values less than
 10\%, which fixed temperature offsets did not.  Using coefficients of
 $1.1$ and $0.9$ or $1.2$ and $0.8$ did not strongly affect RMSRE
 values.

 At a constant rate of airflow, increasing fluid temperature causes
 kinematic viscosity $\nu$ to grow and $h_F$ to shrink because
 $\ReF=V\,L/\nu$.  However, the traces in the graphs from
 Rowley \etal\cite{rowley1930surface} show increasing convective
 conductance with temperature.  Rowley \etal\cite{rowley1930surface}
 reports the airspeed measured at the center of the duct (of which the
 rough plate replaces one side).  However, $V$ is defined as the
 average velocity inside a duct.  Let~$V_\odot$ be the velocity at the
 center of the duct.

 The velocity is 0 at the duct wall, so some velocity near the wall
 must be used as the effective velocity $V$ at the test plate.  The
 velocity profile across the duct develops from flat at the duct
 entrance to the Hagen-Poiseuille parabolic velocity profile for a
 ``fully developed'' flow (Lienhard and Lienhard~\cite{ahtt5e}
 p.~356).  The dimensionless development length is $L_D/D=25.625$,
 where $L_D=5.207~$m is the duct length between the fan and the
 leading edge of the plate and $D=203.2$~mm is the hydraulic-diameter
 of the duct.

 $V=V_\odot$ when $L_D=0$; otherwise $V<V_\odot$.
 Dimensional analysis finds that $V$ must depend on $V_\odot$,
 $L_D/D$, $\nu$, and a viscosity parameter which is independent of
 temperature.  For a gas, let $\nu_0$ be the viscosity at its boiling
 point.

 Dry air is composed of 78.084\%~$\rm{N_2}$, 20.946\%~$\rm{O_2}$, and
 0.934\%~Argon.  $\rm{N_2}$ has kinematic viscosity
 $\nu_0=1.15\times10^{-6}\rm~m^2/s$ at its $77.355$~K boiling point.
 $\rm{O_2}$ has $\nu_0=1.58\times10^{-6}\rm~m^2/s$ at its $90.188$~K
 boiling point; Argon gas has $1.4223\times10^{-6}\rm~m^2/s$ at 100~K.
 Combining these per the air percentages yields
 $\overline{\nu_0}=1.2422\times10^{-6}\rm~m^2/s$.
 Formula~\eqref{eq:effective-V} is the effective $V$ at the plate.
$$V=V_\odot\left/\left[{1+\sqrt2\,{L_D\over D}\,{\overline{\nu_0}\over\nu}}\right]\right.
  \qquad\overline{\nu_0}\approx1.2422\times10^{-6}{\rm~{m^2\over{s}}}\eqdef{eq:effective-V}$$

\centerline{\bf\tabdef{tab:Rowley-parameters}\quad{Assigned parameters}}
\moveright 0.25\hsize\vbox{\settabs 10\columns
\toprule
  \+\bf Figure&\bf Surface&&\hfill$\epsilon$~~~&\hfill$\varepsilon$~~~~&\cr
\midrule        
  \+\figref{fig:Rowley-et-al-stucco}        &stucco         &&\hfill$0.91$ &\hfill$1.47$~mm&\cr
  \+\figref{fig:Rowley-et-al-rough-plaster} &rough-plaster  &&\hfill$0.91$ &\hfill$0.75$~mm&\cr
  \+\figref{fig:Rowley-et-al-brick}         &brick          &&\hfill$0.93$ &\hfill$0.75$~mm&\cr
  \+\figref{fig:Rowley-et-al-concrete}      &concrete       &&\hfill$0.94$ &\hfill$0.55$~mm&\cr
  \+\figref{fig:Rowley-et-al-smooth-plaster}&smooth-plaster &&\hfill$0.91$ &\hfill$0.20$~mm&\cr
\bottomrule
}
\smallskip

 Rowley \etal\cite{rowley1930surface} did not characterize the
 roughnesses other than to note that the forced convection component
 was greatest from stucco, followed by brick and rough-plaster,
 followed by concrete, and the least from smooth-plaster.  This
 investigation has assigned the RMS height-of-roughness
 ($\varepsilon$) parameters shown in \tabref{tab:Rowley-parameters}.

 Rowley \etal\cite{rowley1930surface} did not address thermal
 radiative transfers except to state ``In order to obtain average
 radiation conditions, the inside surface of the test duct was painted
 a dull gray, and all the pipe outside of the refrigerator was covered
 with a one-inch thick blanket of insulating
 material.''  \tabref{tab:Rowley-parameters} shows common values for
 the surface emissivity $\epsilon$ of each rough material tested.
 Experimenting with its value, an effective inside surface emissivity
 of $\epsilon=0.70$ keeps all non-stucco RMSRE values less than 5\%.
 Unexpectedly small for paint, $\epsilon=0.70$ would compensate for
 the wind-tunnel walls being warmer than the forced airflow.

 \figref{fig:Rowley-et-al-stucco} shows the mixed convective
 conductance curves for stucco, the roughest surface
 Rowley \etal\cite{rowley1930surface} tested.  The derivation in
 Jaffer~\cite{thermo3040040} of rough convection
 Formula~\eqref{eq:rough-forced} assumes isotropic roughness.  Stucco
 being non-uniform in its application, it has larger RMSRE than the
 other surfaces.

 \figrefs{fig:Rowley-et-al-rough-plaster},
 \figrefn{fig:Rowley-et-al-brick},
 \figrefn{fig:Rowley-et-al-concrete},
 and \figrefn{fig:Rowley-et-al-smooth-plaster} compare the present
 theory with measurements from rough plaster, brick, concrete, and
 smooth plaster, respectively.\numberedfootnote{One outlying
 measurement for brick at $V=15.65$~m/s was omitted.}  Closer to the
 isotropic ideal, they have RMSRE values smaller than~5\%.

\vfill\eject

 Lacking the actual RMS height-of-roughness and emissivities of the
 original apparatus, while these results lend support to the present
 theory, they are not conclusive.

\smallskip
\vbox{\settabs 1\columns
\+\hfil\figscale{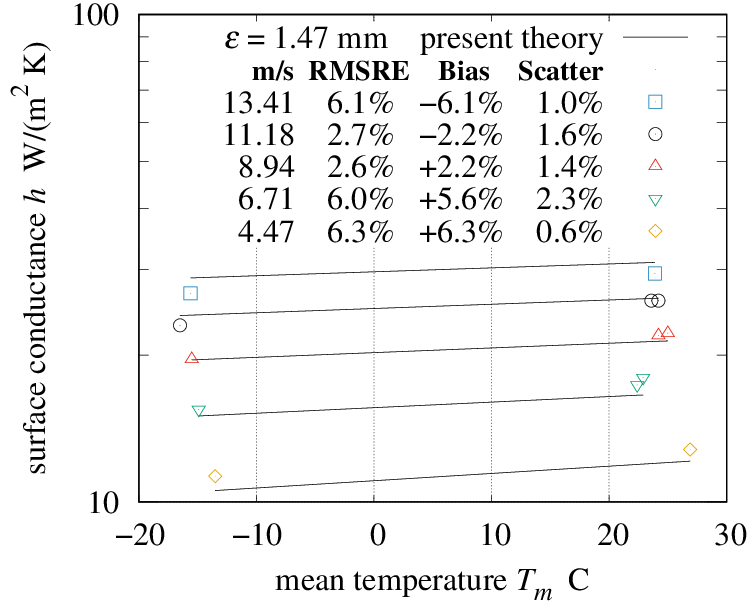}{234pt}&\cr
\+\hfil{\bf\figdef{fig:Rowley-et-al-stucco}\quad{Stucco}}&\cr
}
\vbox{\settabs 2\columns
\+\hfil\figscale{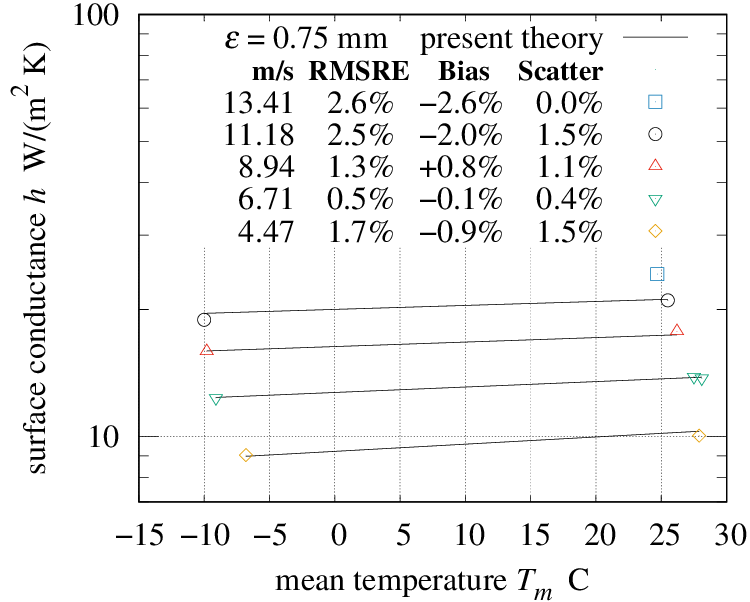}{234pt}&
  \hfil\figscale{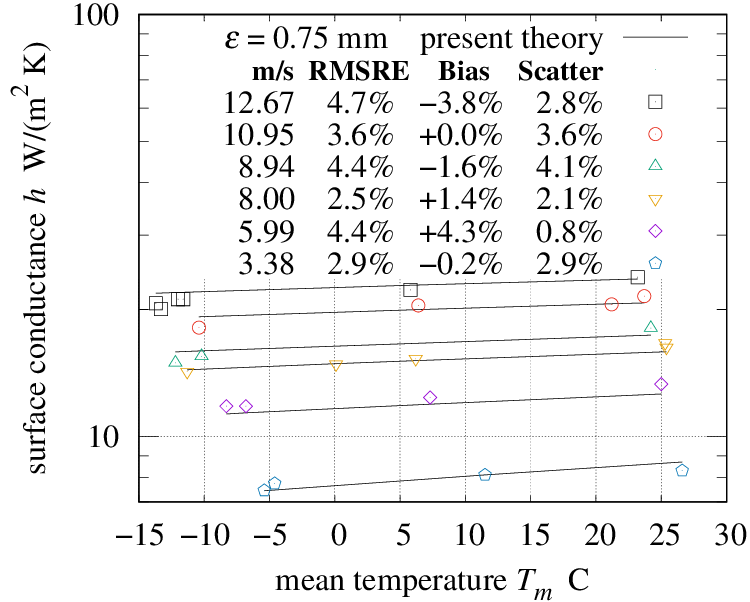}{234pt}&\cr
\+\hfil{\bf\figdef{fig:Rowley-et-al-rough-plaster}\quad{Rough plaster}}&
  \hfil{\bf\figdef{fig:Rowley-et-al-brick}\quad{Brick}}&\cr
\+\hfil\figscale{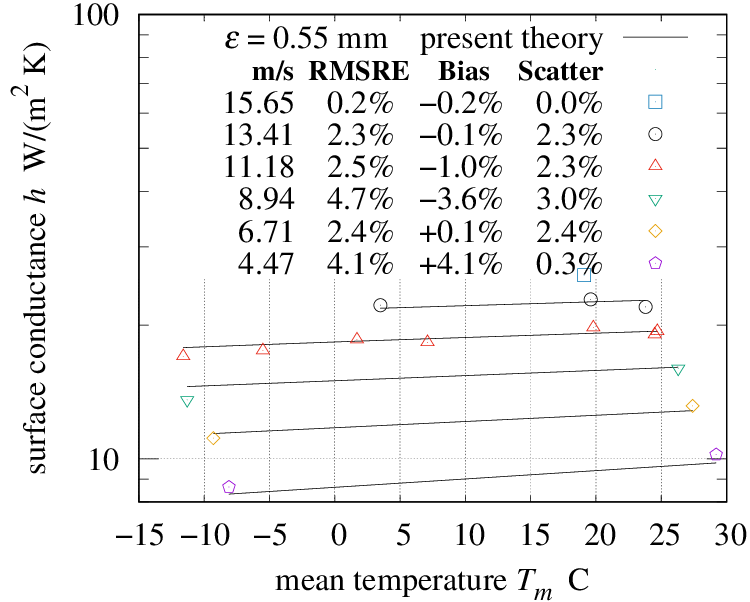}{234pt}&
  \hfil\figscale{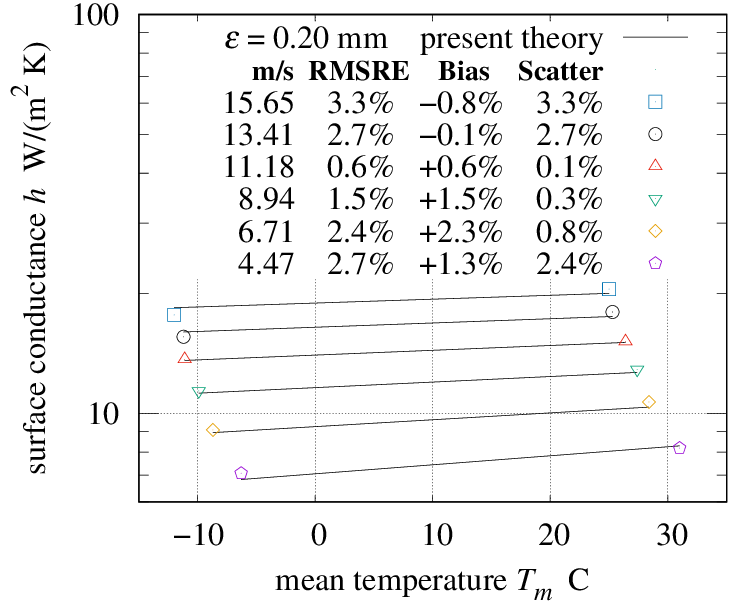}{234pt}&\cr
\+\hfil{\bf\figdef{fig:Rowley-et-al-concrete}\quad{Concrete}}&
  \hfil{\bf\figdef{fig:Rowley-et-al-smooth-plaster}\quad{Smooth plaster}}&\cr
}

\vfill\eject

\subsection{Upward Facing Plate}
 \figref{fig:above-flow} shows that flow is inward above the heated
 surface.  Forced flow parallel to the surface is thus compatible with
 upward natural convection $\hsol$.  Their heat flows are
 perpendicular, suggesting the $\ell^2$-norm for combining $\hsol$ and
 $\hFol$:
$$\hol=\left\|{\hFol},{\hsol}\right\|_2\eqdef{eq:h-up}$$

 \figrefs{fig:1mm-mixed-up-correlation-1}
 and~\figrefn{fig:3mm-mixed-up-correlation-1} compare
 Formula~\eqref{eq:h-up} with measurements of upward-facing plates
 shedding turbulent and rough flow, respectively.

\vbox{\settabs 1\columns
\+\hfil\figscale{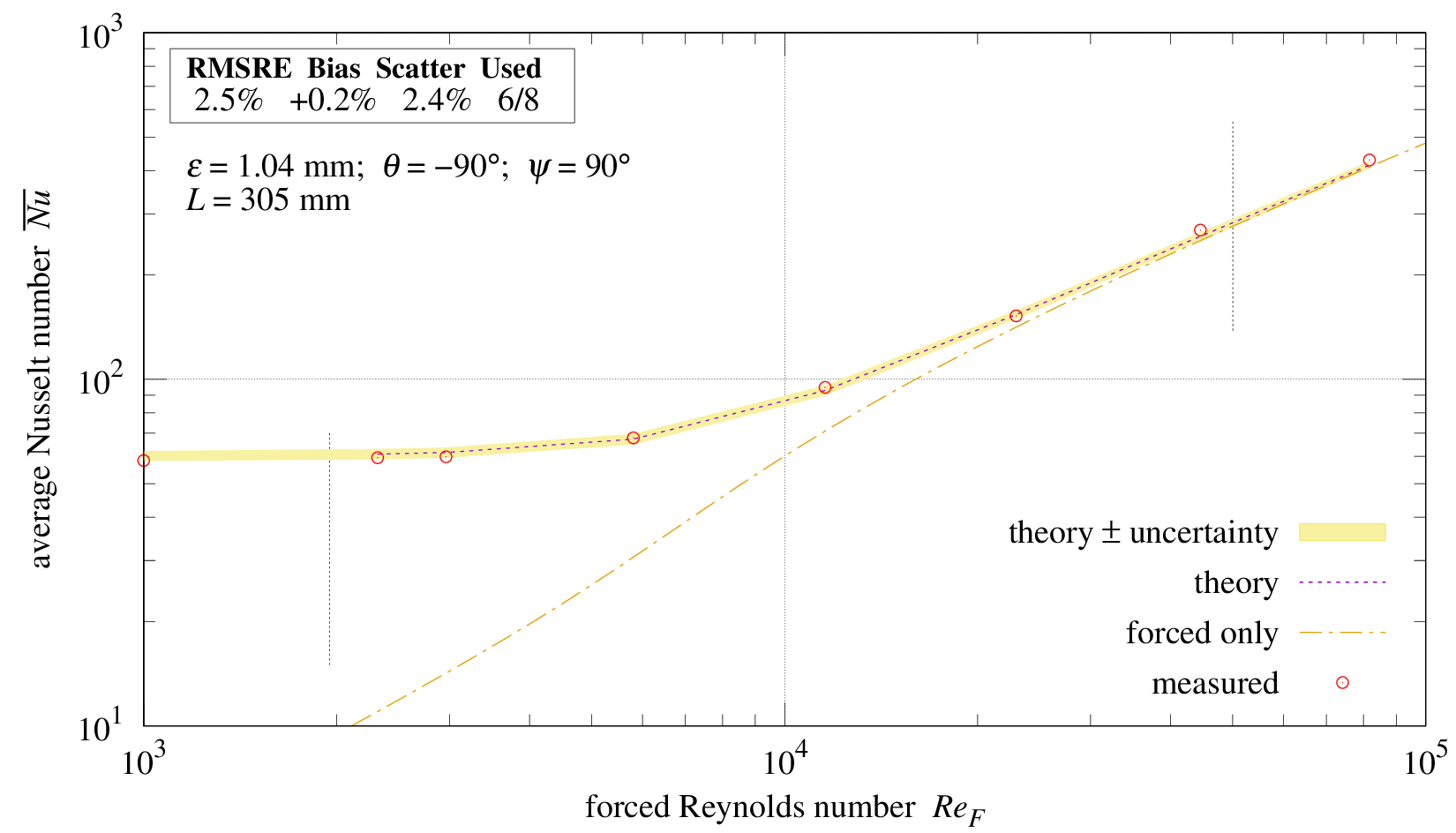}{435pt}&\cr
\+\hfil{\bf\figdef{fig:1mm-mixed-up-correlation-1}\quad{Upward facing plate in horizontal forced turbulent flow}}&\cr
\+\hfil\figscale{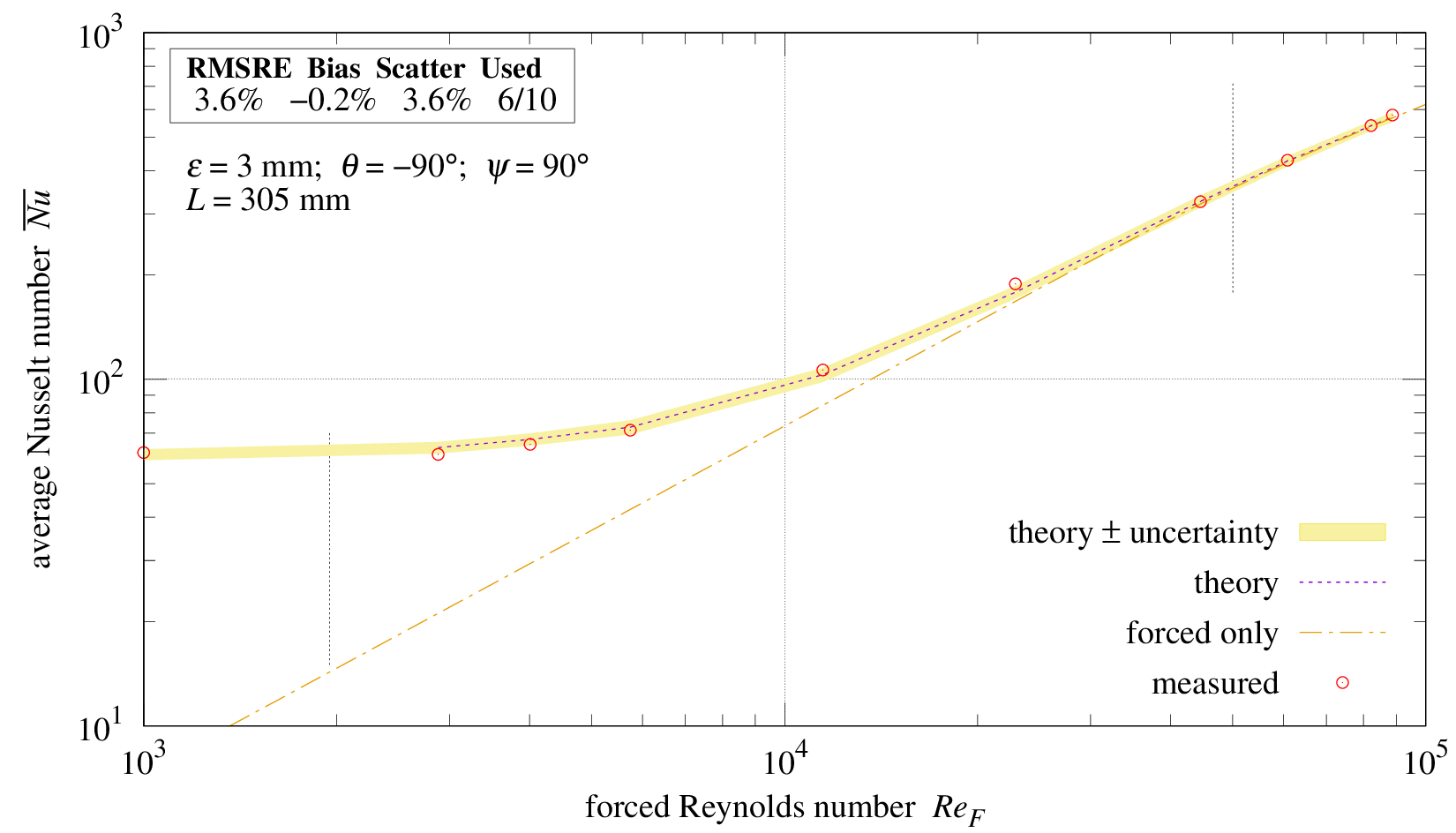}{435pt}&\cr
\+\hfil{\bf\figdef{fig:3mm-mixed-up-correlation-1}\quad{Upward facing plate in horizontal forced rough flow}}&\cr
}

\vfill\eject

\subsection{Downward Facing Plate}
 \figref{fig:below-flow} shows that flow is outward immediately
 beneath the heated surface.  Forced flow parallel to this surface is
 thus incompatible with downward natural convection $\hRol$.  These
 two fluid flows will compete for surface area.  \tabref{tab:natural
 convection parameters} shows that $\NuolR$ is asymptotically
 proportional to $\root5\of{\Ra_R}$.  The $\ell^5$-norm
 combines~$\Ra_R$ with~$\ReF^5$, manifesting the fragility of $\hRol$
 flow because moderate $\ReF$ values can overpower the $\hRol$ term:
$$\hol=\left\|{\hFol},{\hRol}\right\|_5\eqdef{eq:h-down}$$

 \figrefs{fig:1mm-mixed-dn-correlation-1}
 and~\figrefn{fig:3mm-mixed-dn-correlation-1} compare
 Formula~\eqref{eq:h-down} with measurements of downward-facing plates
 shedding turbulent or rough flow, respectively.

\vbox{\settabs 1\columns
\+\hfil\figscale{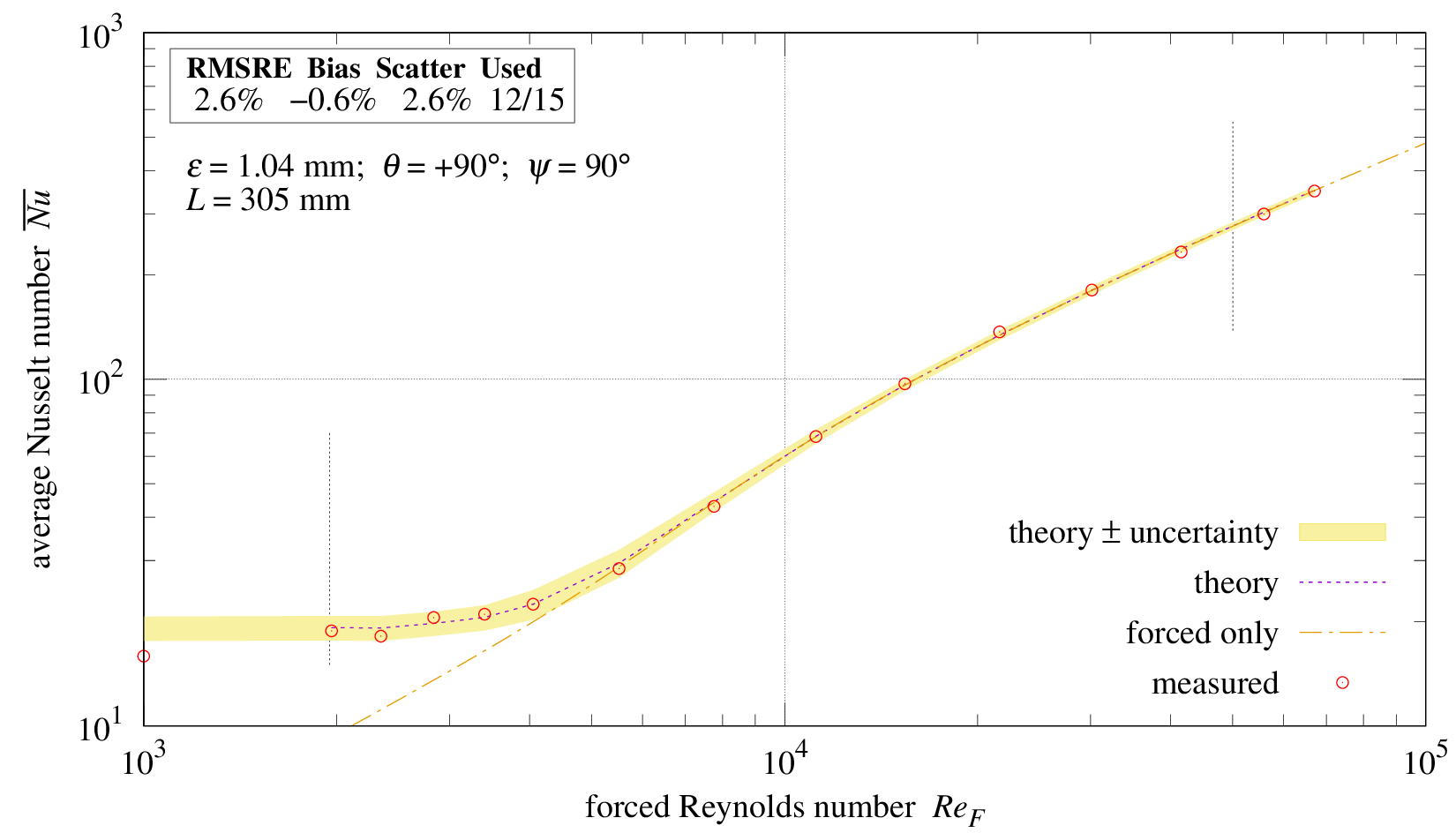}{435pt}&\cr
\+\hfil{\bf\figdef{fig:1mm-mixed-dn-correlation-1}\quad{Downward facing plate in horizontal forced turbulent flow}}&\cr
\+\hfil\figscale{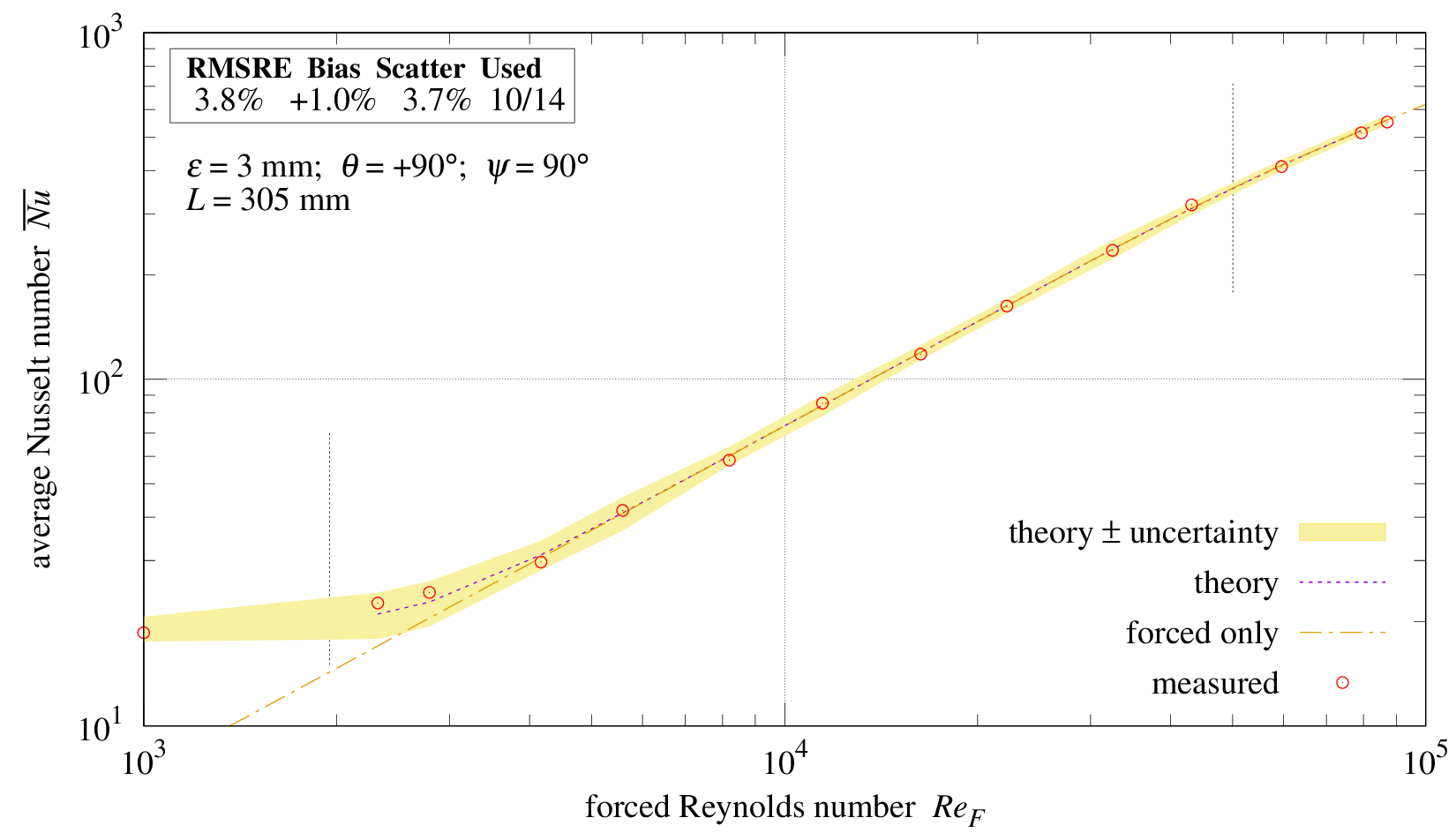}{435pt}&\cr
\+\hfil{\bf\figdef{fig:3mm-mixed-dn-correlation-1}\quad{Downward facing plate in horizontal forced rough flow}}&\cr
}

\section{Vertical Plate With Vertical Forced Flow}

 A vertical plate with vertical forced flow requires a more thorough
 analysis.

\subsection{Velocity Profiles}
 The profile function $u(y)$ is the velocity at $x=L/2$ and
 distance $0<y<\delta$ from the plate, where $\delta$ is the boundary
 layer thickness at $x=L/2$.  Positive $u(y)$ is in the direction of
 forced flow.  The upper plot in \figref{fig:BL-profiles} shows the
 profiles of forced turbulent and natural convection adjacent
 to a vertical 30.5~cm square plate per
 (appendix) \ref{Appendix A: Velocity Profiles}, as well as their sum
 and difference profiles.

 The widest $y$ span of constant $u(y)$ occurs in opposing vertical
 flows when $\ReF=\ReN$.  Because of the laminar sublayer of turbulent
 flows, this cancellation occurs around $u=0.065\rm~m/s$, not~$u=0$.

 The lower plot in \figref{fig:BL-profiles} shows the theoretical
 velocity profiles of forced turbulent flow, and that flow combined
 with laminar natural flow.  The forced $\ReF$ values are double and
 half of natural $\ReN$.

 In the opposing flow cases, the ``forced$\,-\,$natural'' and
 ``forced$\,-\,$natural 21600'' traces both have two inflection points
 near the plate.  These indicate that the boundary layer is split,
 with laminar flow at $0<y<5\rm~mm$ and turbulent flow at
 $5<y<15\rm~mm$.

\vbox{\settabs 1\columns
\+\hfil\figscale{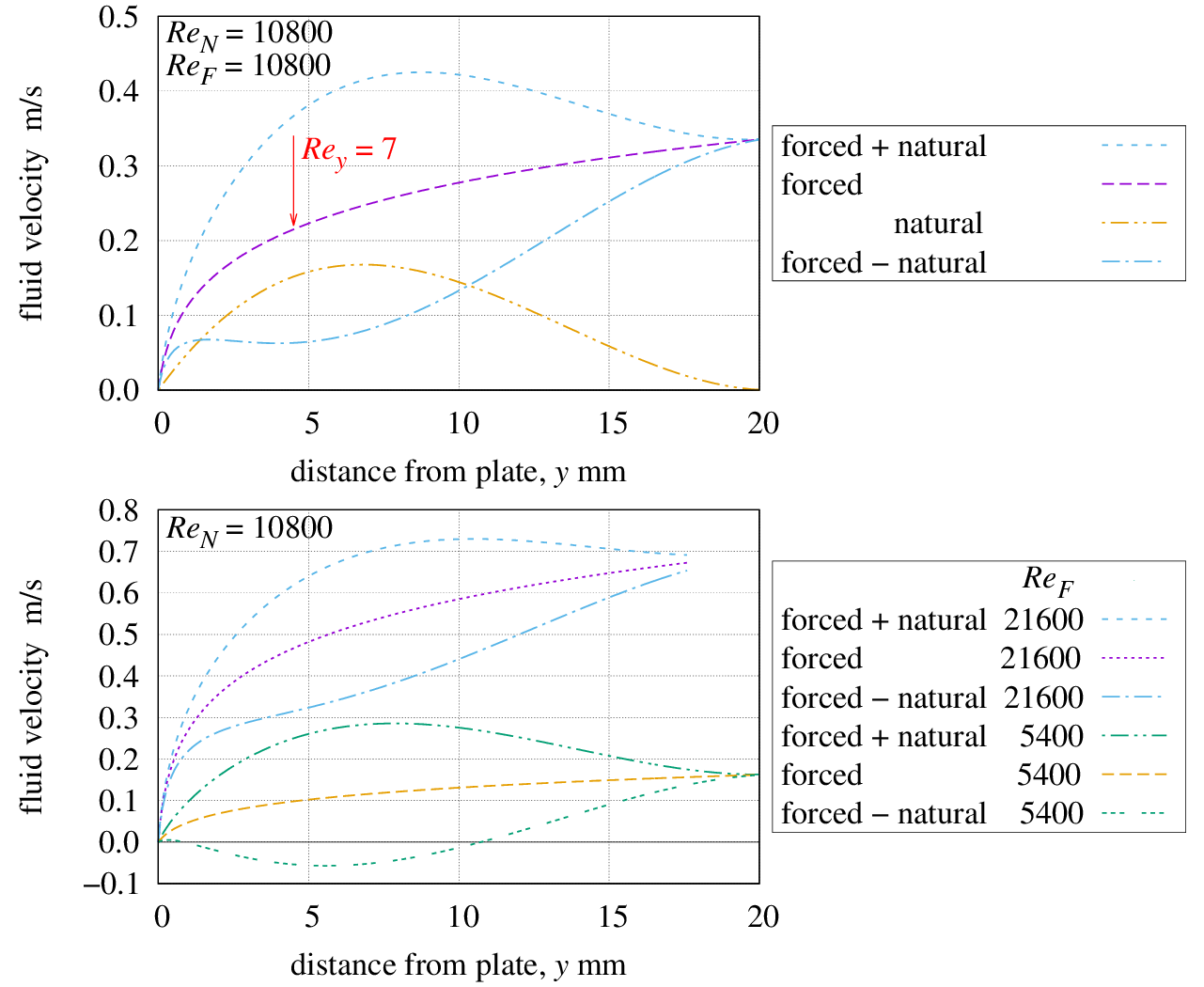}{360pt}&\cr
\+\hfil{\bf\figdef{fig:BL-profiles}\quad Velocity profiles}&\cr}

\subsection{Vertical Plate With Forced Downward Flow}
 Because the net velocity of the ``forced$\,-\,$natural 5400'' curve
 goes negative near the plate, these opposing fluid flows compete for
 plate area.  $\Nuolq$ is asymptotically proportional to
 $\root3\of{\Ra'}$ in \tabref{tab:natural convection parameters}; the
 $\ell^3$-norm combines~$\Ra'$ with $\ReF^3$ (which is more robust
 than the $\ell^5$-norm):
$$\hol=\left\|{\hFol},{\hqol}\right\|_3\eqdef{eq:h-vertical}$$

 The ``forced$\,-\,$natural 21600'' trace indicates that its boundary
 layer is split with laminar natural flow near the plate and forced
 turbulent flow away.  This serves to increase heat transport through
 the boundary layer (compared with pure laminar), exceeding the
 $\ell^2$-norm, but less than the $\ell^{\sqrt2}$-norm.
 The ``mixed $\ell^{\sqrt{3}}$-norm'' curve,
 Formula~\eqref{eq:h-opp}, is close to the upper
 measurements in \figrefs{fig:1mm-mixed-opp-correlation-1}
 and~\figrefn{fig:3mm-mixed-opp-correlation-1}.
$$\hol=\left\|{\hFol},{\hqol}\right\|_{\sqrt3}\eqdef{eq:h-opp}$$

 Both the 1~mm and 3~mm plates had plateau islands roughness, as
 described in \ref{Forced Convection}.  The $\ReI$ arrow marks the
 transition from rough to turbulent flow along the plateau islands
 roughness.

 The $\ReN$ arrow indicates the position of natural convection's
 effective Reynolds number calculated by
 Formula~\eqref{eq:Re-from-Nu}.  ``$\ReN\,\CI$'' is $\ReN$ scaled by the
 roughness correction ($\CI\ge1$) detailed in (appendix)
 \ref{Appendix B: Plateau Islands Roughness Correction}; it
 marks the $\ReF$ lower-bound of the transition between $p=3$ and
 $p=\sqrt3$.

 When the whole plate is shedding turbulent flow,
 $\CI=1$.  When the whole plate is shedding rough flow, $\CI=\chi$,
 which is derived in (appendix) \ref{Appendix A: Velocity Profiles}.
 The close spacing between the arrows
 in \figref{fig:1mm-mixed-opp-correlation-1} indicates that nearly all
 of the plate is shedding turbulent flow.

 \figrefs{fig:1mm-mixed-opp-correlation-1}
 and~\figrefn{fig:3mm-mixed-opp-correlation-1} show the theory curve
 and the measurements switching from the ``mixed $\ell^{3}$-norm'' to
 the ``mixed $\ell^{\sqrt3}$-norm'' at $\ReF>\ReN\,\CI$.

\smallskip
\vbox{\settabs 1\columns
\+\hfil\figscale{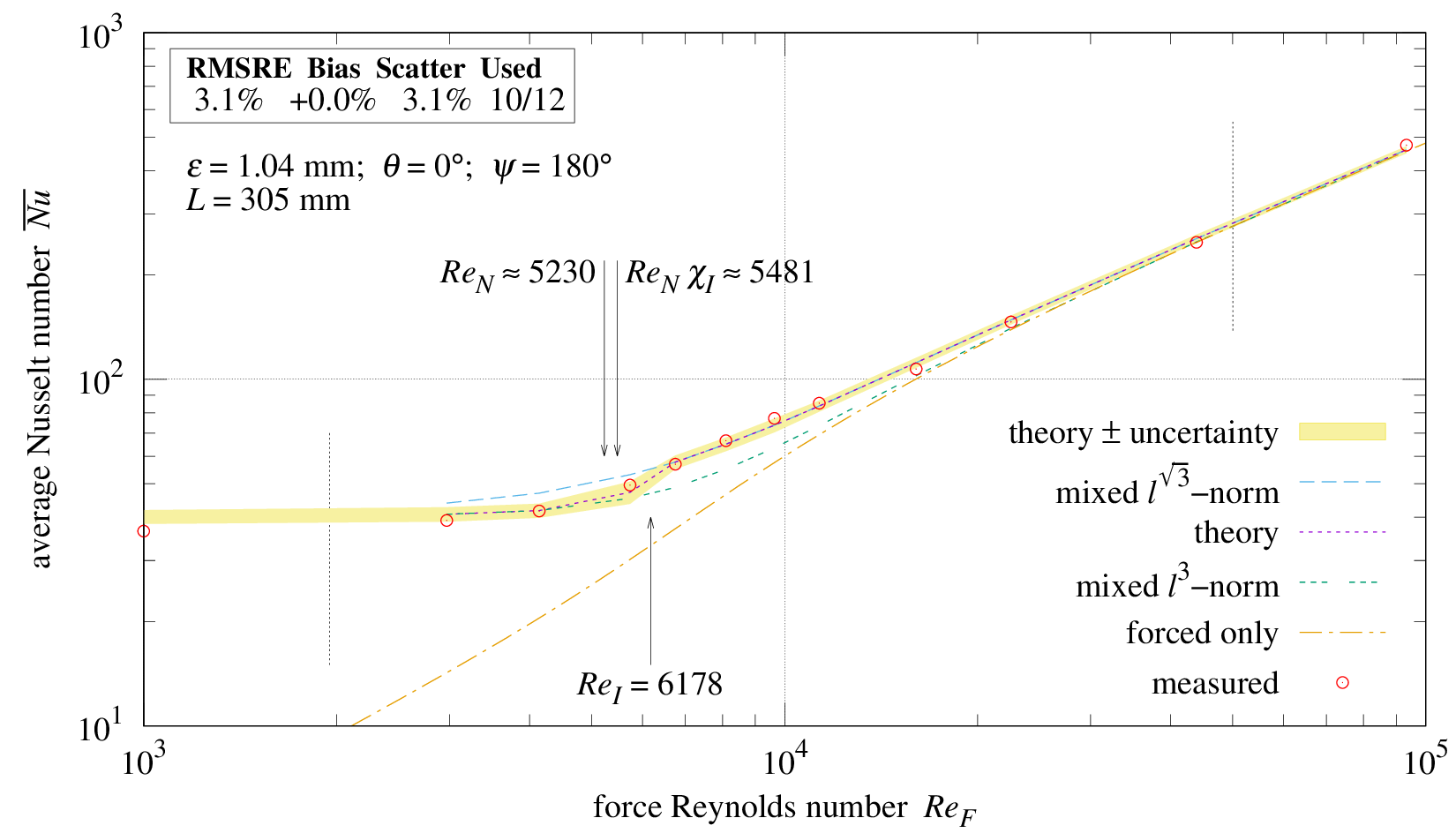}{435pt}&\cr
\+\hfil{\bf\figdef{fig:1mm-mixed-opp-correlation-1}\quad{Vertical plate in opposing forced turbulent flow}}&\cr
\+\hfil\figscale{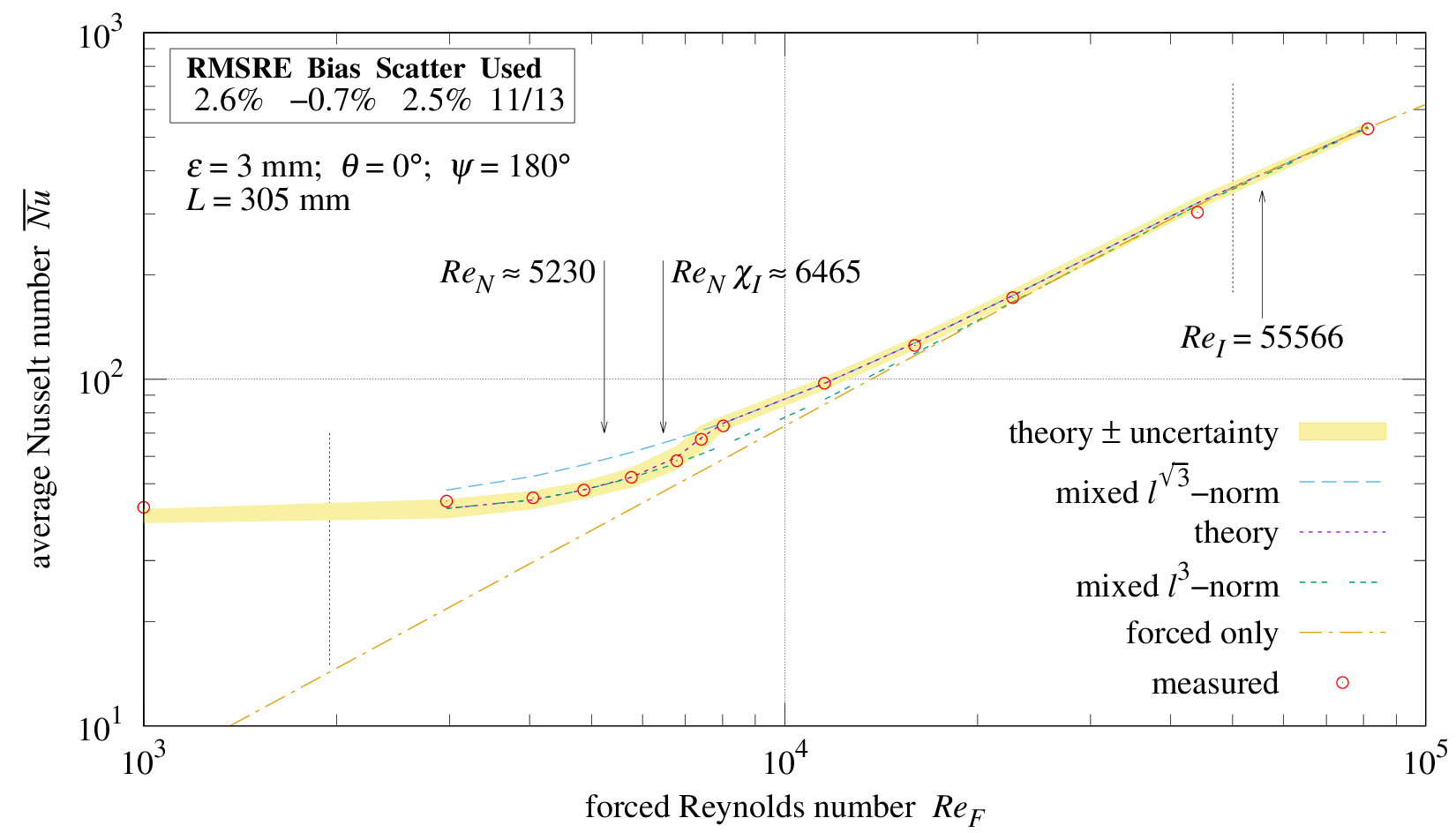}{435pt}&\cr
\+\hfil{\bf\figdef{fig:3mm-mixed-opp-correlation-1}\quad{Vertical plate in opposing forced rough flow}}&\cr
}

\vfill\eject

\subsection{Vertical Plate With Forced Upward Flow}
 At low speeds, the wide separation between the
 ``forced$\,+\,$natural 5400'' and ``forced 5400''
 traces in \figref{fig:BL-profiles} indicates the boundary layer is
 split, with laminar natural flow near the plate and forced turbulent
 flow away, leading to $\hol=\left\|{\hFol},{\hqol}\right\|_{\sqrt3}$.

 The steep slope of the ``forced$\,+\,$natural 21600'' and
 ``forced 21600'' traces near the plate indicates that both
 flows are close to the plate, competing for plate area and leading to
 $\hol=\left\|{\hFol},{\hqol}\right\|_3$.

 \figrefs{fig:1mm-mixed-aid-correlation-1}
 and~\figrefn{fig:3mm-mixed-aid-correlation-1} show the theory curve
 and the measurements switching from the ``mixed $\ell^{3}$-norm'' to
 the ``mixed $\ell^{\sqrt3}$-norm'' at $\ReF$ smaller than $\ReN\,\CI$.

\medskip
\vbox{\settabs 1\columns
\+\hfil\figscale{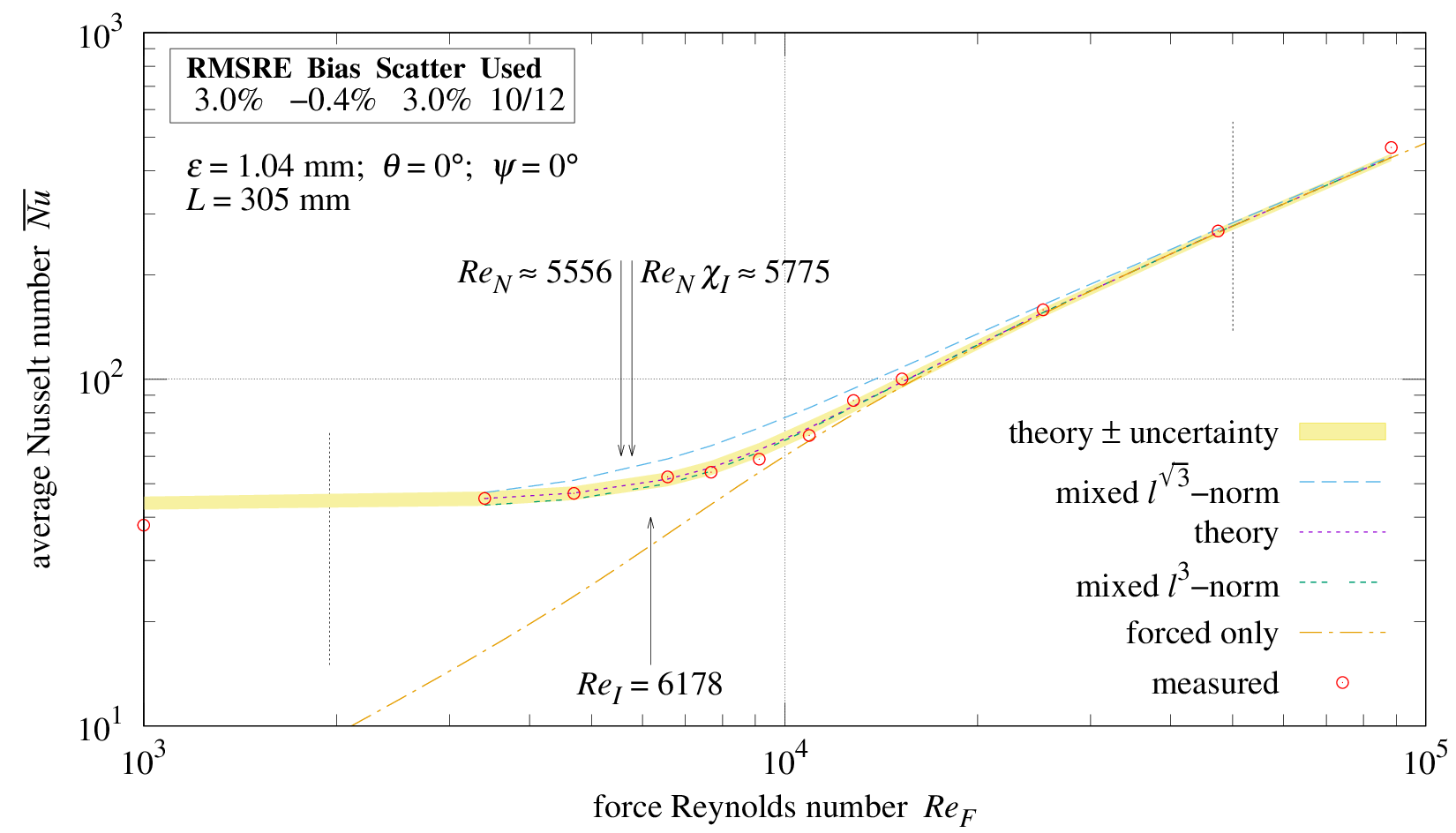}{435pt}&\cr
\+\hfil{\bf\figdef{fig:1mm-mixed-aid-correlation-1}\quad{Vertical plate in aiding forced turbulent flow}}&\cr
\+\hfil\figscale{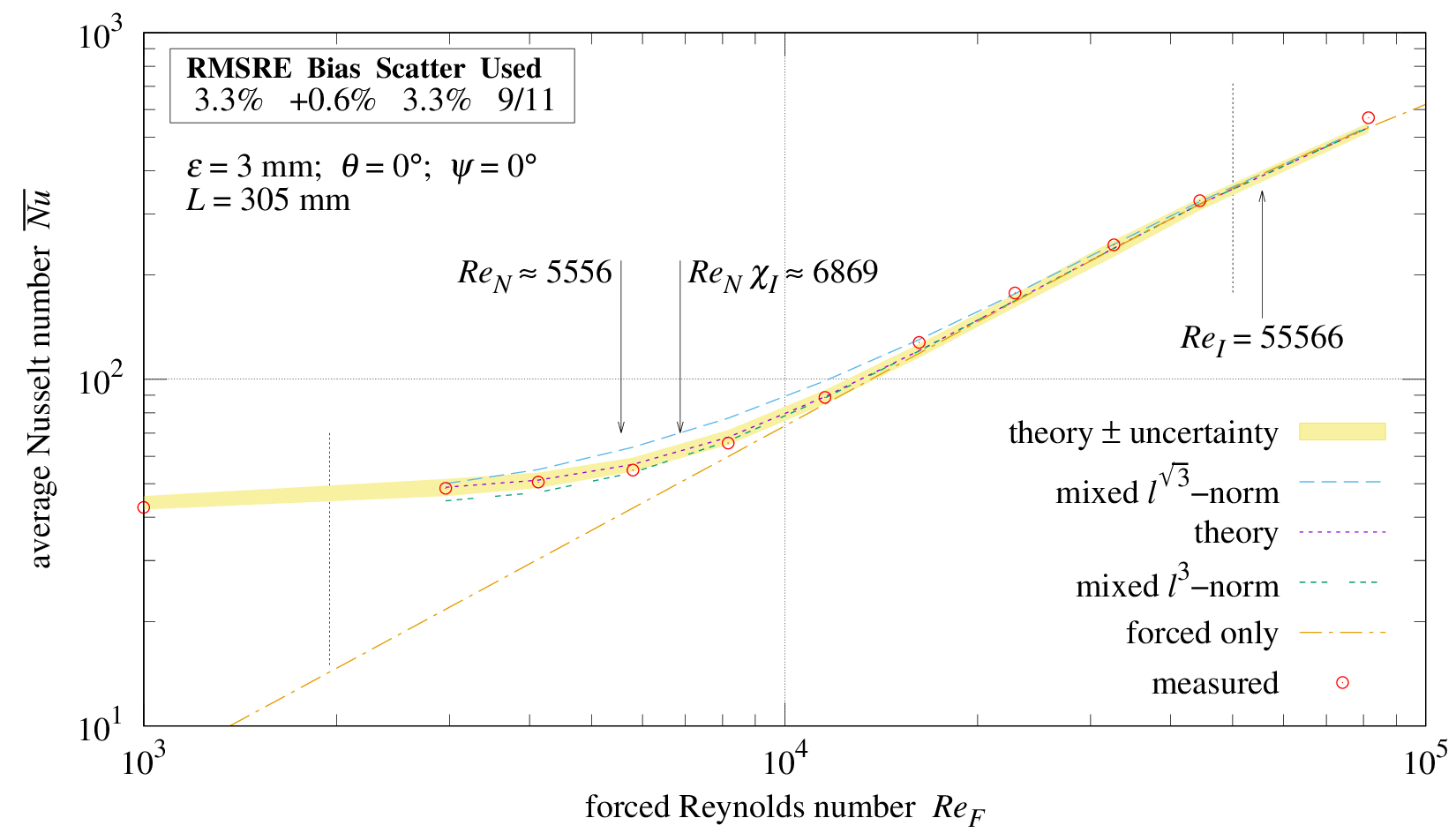}{435pt}&\cr
\+\hfil{\bf\figdef{fig:3mm-mixed-aid-correlation-1}\quad{Vertical plate in aiding forced rough flow}}&\cr
}

\section{Vertical Plate With Forced Flows at any Angle}

 All vertical cases examined so far combine $\hqol$ and $\hFol$ using
 the $\ell^p$-norm with $\sqrt3\le{p}\le3$.  However, $p$ is a
 function of $\ReF$ in the vertical aiding and opposing cases.
 Needed is a function of $\ReF$ which varies smoothly between
 asymptotes $\sqrt3$ and $3$.  This suggests raising 3 to an exponent
 between 1/2 and 1.  With $\zeta>1$ and $\eta>0$, the expression
 ${\eta^\zeta/\zeta}$ varies between 0 and $\infty$, and
 $\exp_\zeta\left({-\eta^\zeta/\zeta}\right)$ varies between 0
 and~1.
 Formula~\eqref{eq:vertical-p} varies between $p=3$ and $p=\sqrt3$,
 with the transition slope controlled by $\zeta$.

 The aiding flow transition is gradual with $\zeta=2$ and
 $\eta=\ReN\,\CI/\ReF$.
 The opposing flow transition is abrupt with $\zeta=16$ and
 $\eta=\ReF/[\ReN\,\CI]$.
 Note that $\eta=\ReN\,\CI/\ReF$ differs from $\eta=\ReF/[\ReN\,\CI]$.
$$p(\zeta,\eta)=
   \exp_3\left(1/2+\exp_\zeta\left({-\eta^\zeta/\zeta}\right)\big/2\right)
   \eqdef{eq:vertical-p}$$

 Introduced in \ref{Mixing Natural and Forced Convections}, $\psi$ is
 the angle between the forced flow and the
 zenith.  \figrefs{fig:vertical-p2} and \figrefn{fig:vertical-p16}
 plot $p$ at $\zeta=2$ and $\zeta=16$, used with $\cos\psi>0$ and
 $\cos\psi<0$, respectively.  \tabref{tab:mode-p} lists $p$ for
 horizontal and vertical plate and flows.

 The theory curve and error statistics
 in \figrefs{fig:1mm-mixed-opp-correlation-1},
 \figrefn{fig:3mm-mixed-opp-correlation-1},
 \figrefn{fig:1mm-mixed-aid-correlation-1},
 and~\figrefn{fig:3mm-mixed-aid-correlation-1}
 employ $p$ Formula~\eqref{eq:vertical-p}.

\smallskip
\vbox{\settabs 2\columns
\+\hfil\figscale{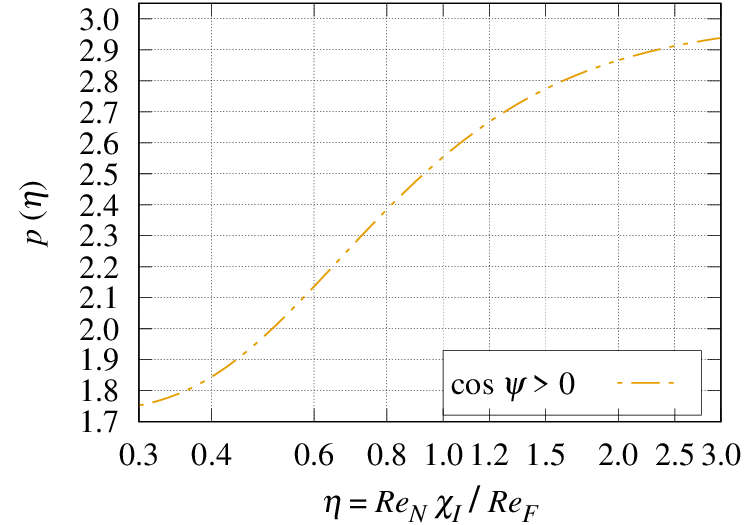}{234pt}&\hfil\figscale{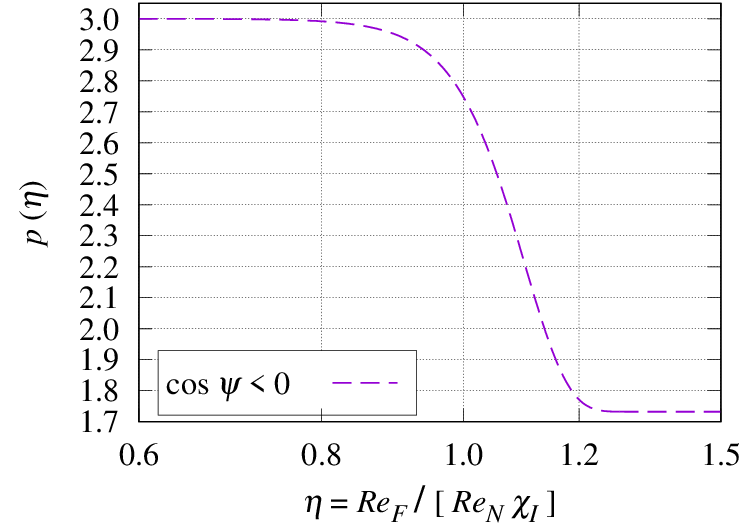}{234pt}&\cr
\+\hfil{\bf\figdef{fig:vertical-p2}\quad{Vertical aiding plate $p$}}&
  \hfil{\bf\figdef{fig:vertical-p16}\quad{Vertical opposing plate $p$}}&\cr
}

\medskip

\centerline{\bf\tabdef{tab:mode-p}\quad{Corner cases $p$}}
\moveright 0.055\hsize
\vbox{\settabs 9\columns
\toprule
\+\bf Description &&\hfill $\theta$~~ &\hfill $\psi$~~ &\qquad $p$&&&&\cr
\midrule
\+ {upward facing plate}       &&\hfill $ -90^\circ$ &\hfill $ 90^\circ$ &\qquad $2$ &\cr
\+ {aiding vertical plate}     &&\hfill $  +0^\circ$ &\hfill $  0^\circ$ &\quad $\exp_3\left(1/2+\exp_2\left({-[{\ReN\,\CI/\ReF}]^{2}/2}\right)\big/2\right)$ &\cr
\+ {vertical plate, level flow}&&\hfill $  +0^\circ$ &\hfill $ 90^\circ$ &\qquad $2$ &\cr
\+ {opposing vertical plate}   &&\hfill $  +0^\circ$ &\hfill $180^\circ$ &\quad $\exp_3\left(1/2+\exp_{16}\left({-[{\ReF/[\ReN\,\CI]}]^{16}/16}\right)\big/2\right)$ &\cr
\+ {downward facing plate}     &&\hfill $ +90^\circ$ &\hfill $ 90^\circ$ &\qquad $5$ &\cr
\bottomrule
}
\smallskip

 At $\psi=0^\circ$, the forced and natural flows align.  As $\psi$
 tilts toward horizontal ($\pm90^\circ$), the forced flow can be split
 into components aligned and perpendicular to the natural upward flow.
 The coefficients of these components are trigonometric functions
 of~$\psi$.  Tilting to the left or right of $\psi=0^\circ$ by equal
 angles must transfer the same amount of heat.  Thus, the
 trigonometric coefficients must be ``even'' functions of $\psi$, that
 is, $F(\psi)=F(-\psi)$.  Formula~\eqref{eq:h3} coefficients
 $[\sin\psi]^2$, $[\cos\psi]^2$, $[\sin\psi]^4$, and $[\cos\psi]^4$
 are even functions of $\psi$.

 Downward tilted flow requires a steeper transition slope around
 $\psi=180^\circ$; this is implemented using $[\sin\psi]^4$ and
 $[\cos\psi]^4$ as the coefficients in the second line of
 Formula~\eqref{eq:h3}.
$$\htol=\cases{[\sin\psi]^2\,\left\|{\hFol},\hqtol\right\|_2+[\cos\psi]^2\,\left\|{\hFol},\hqtol\right\|_{p(2,\ReN\,\CI/\ReF)} & if $0\le\cos\psi$;\cr
               [\sin\psi]^4\,\left\|{\hFol},\hqtol\right\|_2+[\cos\psi]^4\,\left\|{\hFol},\hqtol\right\|_{p(16,\ReF/[\ReN\,\CI])} & if $\cos\psi\le0$.}
 \eqdef{eq:h3}$$

 The vertical natural convection component is independent of $\psi$:
$$\hqtol={\hqol({\left|\cos\theta\right|}\,\Ra'/\Xi)}\eqdef{eq:h4}$$

\section{Mixed Convection From an Inclined Plate}

 To compute mixed convection from an inclined plate,
 Formula~\eqref{eq:inclined-mixed} replaces
 conductance functions $\hqol$, $\hsol$, and $\hRol$ in
 Formula~\eqref{eq:inclined-natural} with
 $\ell^p$-norms mixing each function with $\hFol$.
$$\hol=\cases{\left\|\htol,~
    \left\|{\hFol},
    {\hsol({\left|\sin\theta\right|}\,\Ra^*)}\right\|_2\right\|_{16}
    & $0\le\sin\theta$\cr
    \left\|\htol,~
    \left\|{\hFol},
    {\hRol\left({\left|\sin\theta\right|}\,\Ra_R/\Xi\right)}\right\|_5\right\|_{16}
    & $\sin\theta\le0$\cr}\eqdef{eq:inclined-mixed}$$

 \figref{fig:1mm-mixed-opp+82-correlation-1} shows forced flow opposing
 natural convection at $\theta=+82$ with
 $\psi=98^\circ$.  \figref{fig:1mm-mixed-aid+82-correlation-1} shows
 forced flow aiding natural convection at $\theta=\psi=+82^\circ$.
 The $\theta=\psi=+90^\circ$ curve is shown for comparison.

\vbox{\settabs 1\columns
\+\hfil\figscale{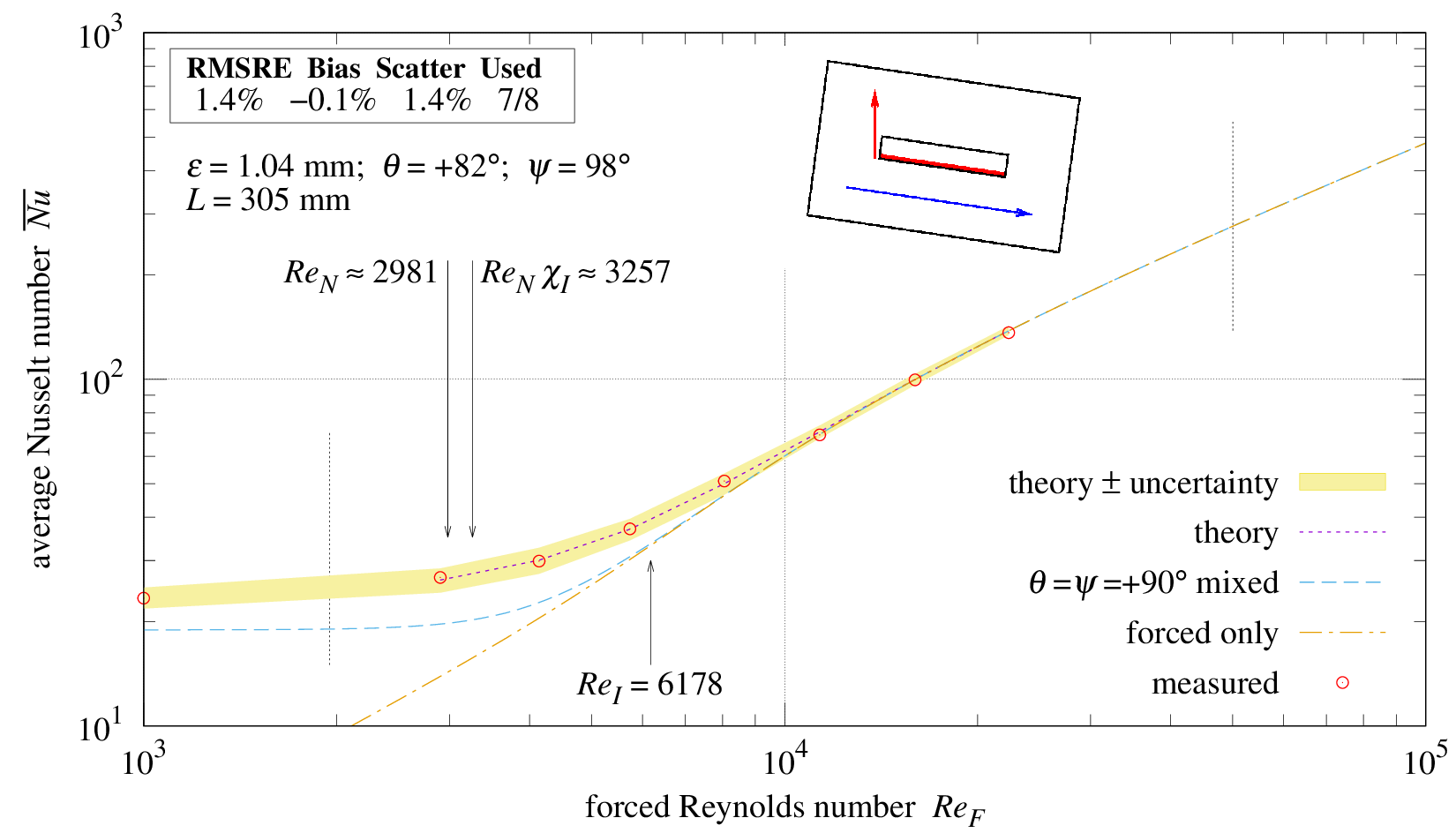}{435pt}&\cr
\+\hfil{\bf\figdef{fig:1mm-mixed-opp+82-correlation-1}\quad{Inclined plate, $\theta=+82^\circ$, opposing forced turbulent flow}}&\cr
\+\hfil\figscale{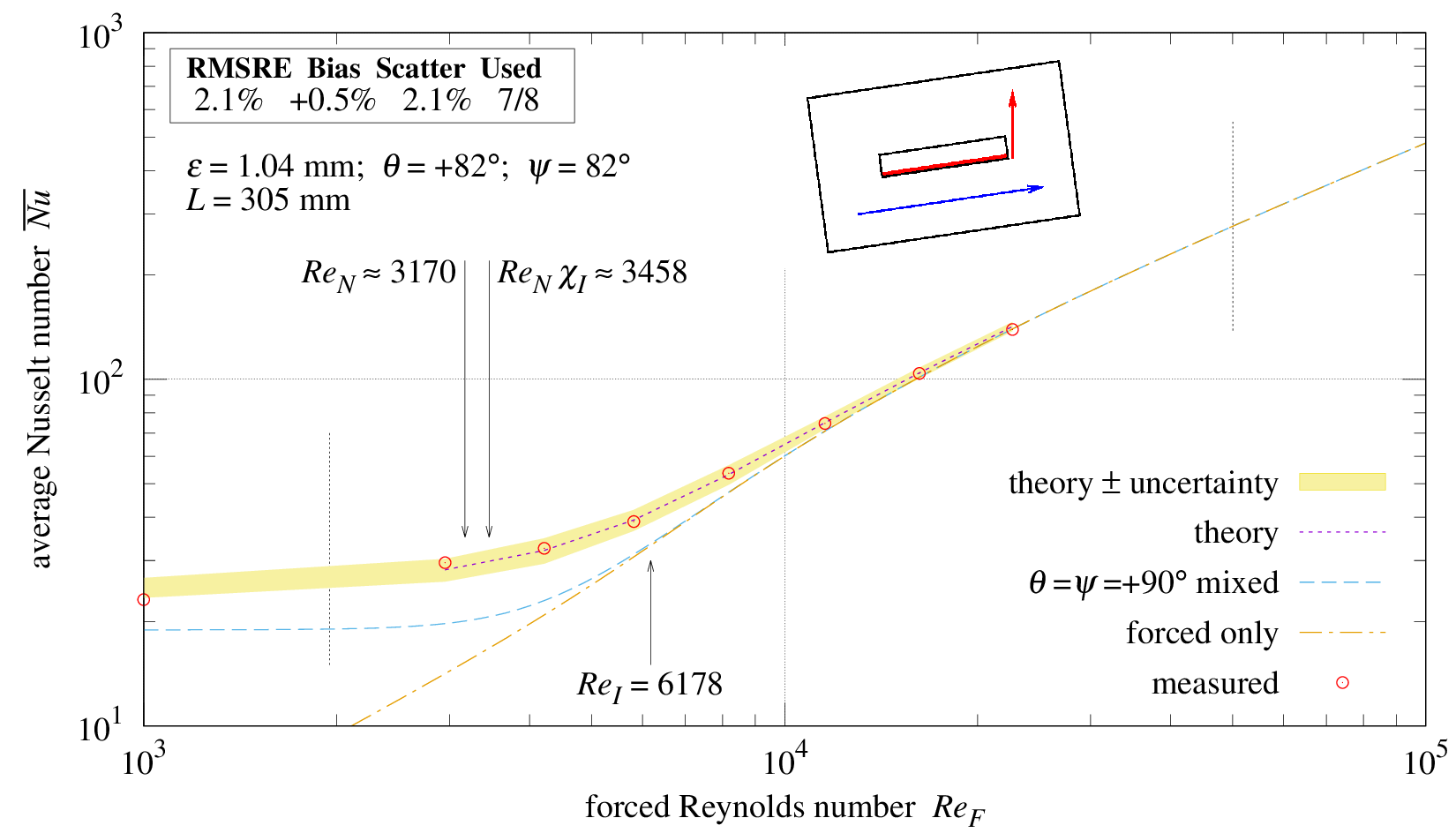}{435pt}&\cr
\+\hfil{\bf\figdef{fig:1mm-mixed-aid+82-correlation-1}\quad{Inclined plate, $\theta=+82^\circ$, aiding forced turbulent flow}}&\cr
}

 With ${\hFol}=0$, mixed Formula~\eqref{eq:inclined-mixed} simplifies
 to natural Formula~\eqref{eq:inclined-natural}.

 When $\theta$ and $\psi$ are multiples of $90^\circ$,
 Formula~\eqref{eq:inclined-mixed} simplifies to
 $\left\|{\hFol},{\hNol}\right\|_p$, with $p$
 from \tabref{tab:mode-p}.

\section{Practice}

 The natural convection heat transfer formulas for $\hsol$, $\hqol$,
 and $\hRol$ were presented in \ref{Natural Convection}.  The formulas
 for forced convection heat transfer $\hFol$ were presented
 in \ref{Forced Convection}.  These are combined using the
 $\ell^p$-norm:
$$\left\|F_1,F_2\right\|_p\equiv\left[~|F_1|^p+|F_2|^p\right]^{1/p}\eqdef{eq:l-p-norm}$$

 $\theta$ is the angle of the plate from vertical; $-90^\circ$ is face
 up; $+90^\circ$ is face down.  Coefficient $|\sin\theta|$ scales
 $\Ra^*$ and $\Ra_R$ to model the effect of the plate's inclination as
 a reduction in the gravitational acceleration.
$$\hol=\cases{\left\|\htol,~
    \left\|{\hFol},
    {\hsol({\left|\sin\theta\right|}\,\Ra^*)}\right\|_2\right\|_{16}
    & $0\le\sin\theta$\cr
    \left\|\htol,~
    \left\|{\hFol},
    {\hRol\left({\left|\sin\theta\right|}\,\Ra_R/\Xi\right)}\right\|_5\right\|_{16}
    & $\sin\theta\le0$\cr}\eqdef{eq:h-bar}$$

 $\psi$ is the angle of the forced flow from the zenith;
 $\psi=0^\circ$ is upward flow; $\psi=90^\circ$ is horizontal flow;
 $\psi=180^\circ$ is downward flow.  The forced flow is always
 parallel to the plate.
$$\htol=\cases{[\sin\psi]^2\,\left\|{\hFol},\hqtol\right\|_2+[\cos\psi]^2\,\left\|{\hFol},\hqtol\right\|_{p(2,\ReN\,\CI/\ReF)} & if $0\le\cos\psi$;\cr
               [\sin\psi]^4\,\left\|{\hFol},\hqtol\right\|_2+[\cos\psi]^4\,\left\|{\hFol},\hqtol\right\|_{p(16,\ReF/[\ReN\,\CI])} & if $\cos\psi\le0$.}
               \eqdef{eq:h-theta}$$

 The vertical natural convection component $\hqtol$ is independent of $\psi$:
$$\hqtol={\hqol({\left|\cos\theta\right|}\,\Ra'/\Xi)}\eqdef{eq:h-prime-theta}$$

 However, its combination with $\hFol$ depends on function
 $p(\zeta,\eta)$, specifically ${p(2,\ReN\,\CI/\ReF)}$ when
 $\cos\psi\ge0$, and $p(16,\ReF/[\ReN\,\CI])$ when $\cos\psi\le0$:
$$p(\zeta,\eta)= \exp_3\left(1/2+\exp_\zeta\left({-\eta^\zeta/\zeta}\right)\big/2\right)\qquad\exp_b(\varphi)\equiv b^\varphi
 \eqdef{eq:p}$$

 The $\CI$ factor models the longer path which forced flow takes along
 plateau islands roughness.  The $\chi$ factor models the longer path
 which forced flow takes along non-plateau roughness.  $\ReI$
 Formula~\eqref{eq:Re_I} is in \ref{Forced Convection}.  Use
 $\ReI=+\infty$ (which implies $\CI=\chi$) for non-plateau roughness.
$$\chi=1-3\,\sqrt3\,{\varepsilon\over L}\,\ln{\varepsilon\over L} \qquad
  \CI=\exp_{\chi}\!\left(\exp_4\left(-\left[{\ReF/\ReI}\right]^4\right)\right)
  \eqdef{eq:chi}$$

 $\ReN$ is the effective Reynolds number of vertical natural
 convection:
$$\ReN\approx{8\,\Nuolq\,\Xi^3\over\Nuzq}
 \qquad\Xi\equiv\left\|1~,~{0.5\over\Pra}\right\|_{\sqrt{1/3}}
 \qquad\Nuolq\equiv{\hqol\,L\over k}
 \qquad\Nuzq\equiv{2^4\over\root4\of{2}\,\pi^2}
 \eqdef{eq:Re-N}
 $$

\section{Results}

 \tabref{tab:coverage} shows the combinations of flow types and
 orthogonal orientations.

 Configurations measured by the present apparatus with its 0.305~m
 square plates are marked with $\bullet$.

 Configurations having turbulent natural convection are marked with
 $\bigcirc$.  These would require either a larger plate and
 wind-tunnel or higher plate temperatures than the present apparatus
 supports.

\medskip
\vbox{\settabs 1\columns
\+\hfil{\bf\tabdef{tab:coverage}\quad{Mixed convective modes}}&\cr
\smallskip
\+\vbox{\settabs 7\columns
\toprule
\+\hfil\bf Natural&\hfil\bf Forced&\hfil\bf Vertical&\hfil\bf Up&\hfil\bf Down&\hfil\bf Opposing&\hfil\bf Aiding&\cr
\midrule
\+\hfil laminar&\hfil turbulent&\hfil$\bullet$\figref{fig:1mm-mixed-vt-correlation-1}&\hfil$\bullet$\figref{fig:1mm-mixed-up-correlation-1}&\hfil$\bullet$\figref{fig:1mm-mixed-dn-correlation-1}&\hfil$\bullet$\figref{fig:1mm-mixed-opp-correlation-1}&\hfil$\bullet$\figref{fig:1mm-mixed-aid-correlation-1}&\cr
\+\hfil laminar&\hfil rough&\hfil$\bullet$\figref{fig:3mm-mixed-vt-correlation-1}&\hfil$\bullet$\figref{fig:3mm-mixed-up-correlation-1}&\hfil$\bullet$\figref{fig:3mm-mixed-dn-correlation-1}&\hfil$\bullet$\figref{fig:3mm-mixed-opp-correlation-1}&\hfil$\bullet$\figref{fig:3mm-mixed-aid-correlation-1}&\cr
\+\hfil turbulent&\hfil turbulent&\hfil$\bigcirc$&\hfil$\bigcirc$&\hfil$\bigcirc$&\hfil$\bigcirc$&\hfil$\bigcirc$&\cr
\+\hfil turbulent&\hfil rough&\hfil$\bigcirc$&\hfil$\bigcirc$&\hfil$\bigcirc$&\hfil$\bigcirc$&\hfil$\bigcirc$&\cr
\bottomrule
}&\cr
}

 \tabrefs{tab:convection-conformance-3mm}
 and \tabrefn{tab:convection-conformance-1mm} summarize the present
 theory's conformance with 104 measurements in twelve data-sets from
 the present apparatus's two plates.

 The ``$\varepsilon$'' column identifies the 30.5~cm square plate
 used.  The 3.00~mm plate had rough flow over the
 $1950<\ReF<5\times10^4$ range.  The 1.04~mm plate had turbulent flow
 over nearly all of the same range.  The ``Used'' column is the count
 of measurements having $1950<\ReF<5\times10^4$ out of the count of
 measurements.  Measurements at $\ReF>5\times10^4$ were practically
 unaffected by mixing.

 The $\varepsilon=3.00$~mm data-sets have RMSRE values between $2.0\%$
 and $3.8\%$.

 The $\varepsilon=1.04$~mm data-sets have RMSRE values between $1.3\%$
 and $3.1\%$.

\smallskip
\centerline{\bf\tabdef{tab:convection-conformance-3mm}\quad{Convection measurements versus present theory, forced rough flow}}
\vbox{\settabs 10\columns
\toprule
\+\bf Description&&&\hfil $\varepsilon$ &\hfil $\theta$ &\hfil $\psi$ &\hfill\bf RMSRE&\hfill\bf Bias &\hfill\bf Scatter&\hfil\bf Used&\cr
\midrule
\+ {downward facing plate} &&&\hfil $3.00$ mm &\hfil $ +90.0^\circ$ &\hfil $ 90.0^\circ$ &\hfill $ 3.8\%$ &\hfill $ +1.0\%$ &\hfill $ 3.7\%$ &\hfil 10/14 &\cr
\+ {upward facing plate} &&&\hfil $3.00$ mm &\hfil $ -90.0^\circ$ &\hfil $ 90.0^\circ$ &\hfill $ 3.6\%$ &\hfill $ -0.2\%$ &\hfill $ 3.6\%$ &\hfil  6/10 &\cr
\+ {vertical plate, level flow} &&&\hfil $3.00$ mm &\hfil $  +0.0^\circ$ &\hfil $ 90.0^\circ$ &\hfill $ 2.0\%$ &\hfill $ +0.8\%$ &\hfill $ 1.8\%$ &\hfil  6/8 &\cr
\+ {opposing vertical plate} &&&\hfil $3.00$ mm &\hfil $  +0.0^\circ$ &\hfil $180.0^\circ$ &\hfill $ 2.6\%$ &\hfill $ -0.7\%$ &\hfill $ 2.5\%$ &\hfil 11/13 &\cr
\+ {aiding vertical plate} &&&\hfil $3.00$ mm &\hfil $  +0.0^\circ$ &\hfil $  0.0^\circ$ &\hfill $ 3.3\%$ &\hfill $ +0.6\%$ &\hfill $ 3.3\%$ &\hfil  9/11 &\cr
\bottomrule
}
\smallskip
\centerline{\bf\tabdef{tab:convection-conformance-1mm}\quad{Convection measurements versus present theory, forced turbulent flow}}
\vbox{\settabs 10\columns
\toprule
\+\bf Description&&&\hfil $\varepsilon$ &\hfil $\theta$ &\hfil $\psi$ &\hfill\bf RMSRE&\hfill\bf Bias &\hfill\bf Scatter&\hfil\bf Used&\cr
\midrule
\+ {downward facing plate} &&&\hfil $1.04$ mm &\hfil $ +90.0^\circ$ &\hfil $ 90.0^\circ$ &\hfill $ 2.6\%$ &\hfill $ -0.6\%$ &\hfill $ 2.6\%$ &\hfil 12/15 &\cr
\+ {upward facing plate} &&&\hfil $1.04$ mm &\hfil $ -90.0^\circ$ &\hfil $ 90.0^\circ$ &\hfill $ 2.5\%$ &\hfill $ +0.2\%$ &\hfill $ 2.4\%$ &\hfil  6/8 &\cr
\+ {vertical plate, level flow} &&&\hfil $1.04$ mm &\hfil $  +0.0^\circ$ &\hfil $ 90.0^\circ$ &\hfill $ 3.0\%$ &\hfill $ +0.8\%$ &\hfill $ 2.8\%$ &\hfil 10/14 &\cr
\+ {opposing vertical plate} &&&\hfil $1.04$ mm &\hfil $  +0.0^\circ$ &\hfil $180.0^\circ$ &\hfill $ 3.1\%$ &\hfill $ +0.0\%$ &\hfill $ 3.1\%$ &\hfil 10/12 &\cr
\+ {aiding vertical plate} &&&\hfil $1.04$ mm &\hfil $  +0.0^\circ$ &\hfil $  0.0^\circ$ &\hfill $ 3.0\%$ &\hfill $ -0.4\%$ &\hfill $ 3.0\%$ &\hfil 10/12 &\cr
\midrule
\+ {opposing inclined plate} &&&\hfil $1.04$ mm &\hfil $ +82.0^\circ$ &\hfil $ 98.0^\circ$ &\hfill $ 1.4\%$ &\hfill $ -0.1\%$ &\hfill $ 1.4\%$ &\hfil  7/8 &\cr
\+ {aiding inclined plate} &&&\hfil $1.04$ mm &\hfil $ +82.0^\circ$ &\hfil $ 82.0^\circ$ &\hfill $ 2.1\%$ &\hfill $ +0.5\%$ &\hfill $ 2.1\%$ &\hfil  7/8 &\cr
\bottomrule
}

 \tabref{tab:Rowley conformance} summarizes the present theory's
 conformance with 78 measurements in 28 data-sets on five vertical
 rough surfaces in horizontal flow from
 Rowley \etal\cite{rowley1930surface}.  The five stucco data-sets have
 RMSRE values between 2.5\% and 6.5\%; the other data-sets have RMSRE
 values between 0.2\% and 5\%.

\vfill\eject
\centerline{\bf\tabdef{tab:Rowley conformance}\quad{Rowley \etal mixed convection measurements}}
\moveright 0.1\hsize
\vbox{\settabs 10\columns
\toprule
\+ {\bf Surface} &\hfill$\epsilon~~$&\hfil$\varepsilon$&\hfil$V$&\hfil\bf RMSRE&\hfill\bf Bias &\hfill\bf Scatter&\hfil\bf Count&\cr
\midrule
\input Rowley-et-al-results.tex
\bottomrule
}

\section{Discussion}

 Developing this theory was difficult due to the lack of photographs
 of mixed convection streamlines along rough surfaces.  Analysis of
 the measurements made clear that mixed convection from rough plates
 was different from that of smooth plates.  The flow patterns had
 to be inferred from these measurements and knowledge of natural and
 forced convections.

 Many theories were tried and discarded concerning the forced vertical
 flow cases.  Examination of hypothetical velocity profiles sparked
 the present theory, which explains the aiding and opposing flow cases
 both having $\ell^{\sqrt3}$-norm and $\ell^{3}$-norm asymptotes.

\subsection{Heat Transfer Bounds}
 All of the $\ell^p$-norms combining natural and forced heat transfer
 have $\sqrt3\le{p}\le5$.  The mixed heat transfer is thus bounded
 between $\|{\hFol},{\hNol}\|_5$ and $\|{\hFol},{\hNol}\|_{\sqrt3}$.

\subsection{Horizontal Flow Obstruction}
 The fan pulling air through the chamber is sufficient to counter the
 effect of the wind-tunnel's obstructions to horizontal flow, except
 in the case of the vertical plate with opposing flow.  In order to
 draw some air upward at slow (downward) fan speeds, the air's
 momentum must be reversed.  This is modeled by increasing parameter
 $B$ of \tabref{tab:natural convection parameters} by twice the
 vertical distance from the plate to the test chamber upper edge,
 normalized by $L$ and the ratio of the upper edge perimeter to the
 plate width.
 This same correction applies to still air in the vertical tunnel.

\subsection{Effective Vertical Reynolds Number}
 Aiding and opposing vertical plate measurements in a fluid other than
 air are needed to further test the effective vertical Reynolds number,
 $\ReN$ Formula~\eqref{eq:Re-from-Nu}.

\subsection{Rough Velocity Profiles}
 The hypothetical forced flows in the
 \ref{Vertical Plate With Vertical Forced Flow} velocity profiles were
 turbulent flows.  Measurements of both vertical plates in vertical
 flow conforming to the present theory suggests that the rough and
 turbulent velocity profiles are similar.

\subsection{Duct Velocity Profile}
 The aggregate boiling point kinematic viscosity $\overline{\nu_0}$ in
 Formula~\eqref{eq:effective-V} may be useful in developing formulas
 for pipe and duct velocity profiles as a function of duct length.

\section{Conclusions}

 Formulas were presented for predicting the mixed convective surface
 conductance of a flat isotropic surface roughness having a convex
 perimeter in a Newtonian fluid with a steady forced flow in the plane
 of that roughness.

 The prerequisites are the RMS height-of-roughness $0<\varepsilon$,
 angle $\theta$ of the surface from vertical, angle $\psi$ of the
 forced flow from the zenith, $\Ra/L^3$ and $\Pra$ of the fluid, and
 the characteristic-length $L>0$ and $\Rey>0$ of the forced flow.

\unorderedlist

 \li RMS height-of-roughness $\varepsilon$ is the correct metric
 for predicting forced convective surface conductance.

 \li Roughness $\varepsilon\ll{L}$ does not affect the natural
 component of mixed convection.

 \li Plate inclination does not affect the forced component of
 mixed convection.

 \li When $\Rey=0$, the mixed convection is the same as its natural
 component.

\endunorderedlist

 The present work's formulas were compared with 104 measurements in
 twelve data-sets from the present apparatus in two inclined and all
 five corner case orientations.  The twelve data-sets had RMSRE
 values between 1.3\% and 4\% relative to the present theory.

 The present work's formulas were compared with 78 measurements in 28
 data-sets on five vertical rough surfaces in horizontal airflow from
 Rowley \etal\cite{rowley1930surface}.  The five stucco data-sets had
 RMSRE values between 2.5\% and 6.5\%; the other data-sets had RMSRE
 values between 0.2\% and 5\%.

\section{Nomenclature}

\nomenclature[A]{$A$}{surface area ($\rm m^2$)}
\nomenclature[A]{$\Gr$}{Grashof number}
\nomenclature[A]{$\hol$}{average convective surface conductance (${\rm W/(m^2\cdot K)}$)}
\nomenclature[A]{$\hFol$}{forced convective surface conductance (${\rm W/(m^2\cdot K)}$)}
\nomenclature[A]{$\hNol$}{natural convective surface conductance (${\rm W/(m^2\cdot K)}$)}
\nomenclature[A]{$\hsol$}{upward natural convective surface conductance (${\rm W/(m^2\cdot K)}$)}
\nomenclature[A]{$\hqol$}{vertical plate natural convective surface conductance (${\rm W/(m^2\cdot K)}$)}
\nomenclature[A]{$\htol$}{vertical mode of inclined natural convective surface conductance (${\rm W/(m^2\cdot K)}$)}
\nomenclature[A]{$\hqtol$}{vertical component of $\htol$ (${\rm W/(m^2\cdot K)}$)}
\nomenclature[A]{$\hRol$}{downward natural convective surface conductance (${\rm W/(m^2\cdot K)}$)}
\nomenclature[A]{$k$}{fluid thermal conductivity (${\rm W/(m\cdot K)}$)}
\nomenclature[A]{$L$}{characteristic length (m)}
\nomenclature[A]{$L_P$}{roughness spatial period (m)}
\nomenclature[A]{$\Ls$}{ratio of plate area to its perimeter (m)}
\nomenclature[A]{$\Lb$}{ratio of island area to its perimeter (m)}
\nomenclature[A]{$L_W$}{width of plate (m)}
\nomenclature[A]{$\Nuol$}{average Nusselt number}
\nomenclature[A]{$\NuNol$}{average Nusselt number of natural convection}
\nomenclature[A]{$\Nuolq$}{average Nusselt number of vertical plate natural convection}
\nomenclature[A]{$\Nuzq$}{Nusselt number of vertical plate conduction}
\nomenclature[A]{$p$}{exponent in $\ell^{p}$-norm: $\left\{|F_1|^p+|F_2|^p\right\}^{1/p}$}
\nomenclature[A]{$\Pra$}{Prandtl number of the fluid}
\nomenclature[A]{$\Ra$}{Rayleigh number}
\nomenclature[A]{$\Ra'$}{vertical plate Rayleigh number}
\nomenclature[A]{$\Ra^*$}{upward Rayleigh number}
\nomenclature[A]{$\Ra_R$}{downward Rayleigh number}
\nomenclature[A]{$\ReF$}{Reynolds number of the forced flow parallel to the plate}
\nomenclature[A]{$\ReI$}{Reynolds number of rough turbulent transition to forced turbulent flow}
\nomenclature[A]{$\ReN$}{effective Reynolds number of vertical natural convection}
\nomenclature[A]{$\Rey_y$}{friction Reynolds number}
\nomenclature[A]{$u(y)$}{velocity at $x=L/2$ and distance $y$ from the plate (m/s)}
\nomenclature[A]{$u_N$}{effective natural flow speed $={\nu\,\ReN/L}$ (m/s)}
\nomenclature[A]{$\us$}{friction velocity (m/s)}
\nomenclature[A]{$\Wz$}{principal branch of the Lambert $\W$ function}
\nomenclature[A]{$y$}{distance from plate (m)}

\subsection{Greek Symbols}
\unskip

\nomenclature[G]{$\delta$}{boundary layer thickness (m)}
\nomenclature[G]{$\delta_\lambda$}{laminar boundary layer thickness (m)}
\nomenclature[G]{$\delta_\tau$}{turbulent boundary layer thickness (m)}
\nomenclature[G]{$\epsilon$}{surface emissivity}
\nomenclature[G]{$\varepsilon$}{surface RMS height-of-roughness (m)}
\nomenclature[G]{$\eta$}{ratio of $\ReN$ and $\ReF$ (either order)}
\nomenclature[G]{$\kappa$}{von K\'{a}rm\'{a}n constant $\approx0.41$}
\nomenclature[G]{$\Omega$}{ratio of non-plateau area to cell area (${\rm m^2/m^2}$)}
\nomenclature[G]{$\nu$}{fluid kinematic viscosity (${\rm m^2/s}$)}
\nomenclature[G]{$\nu_0$}{gas kinematic viscosity (${\rm m^2/s}$) at boiling point}
\nomenclature[G]{$\psi$}{angle between the forced flow and the zenith; $0^\circ$ is aiding flow; $180^\circ$~is opposing flow}
\nomenclature[G]{$\theta$}{angle of the plate surface from vertical; face up is $-90^\circ$; face down is $+90^\circ$}
\nomenclature[G]{$\Xi$}{natural convection self-obstruction factor}
\nomenclature[A]{$\chi$}{roughness velocity correction factor for forced flow}
\nomenclature[A]{$\CI$}{plateau islands roughness correction factor}

\subsection{Abbreviations}
 The following abbreviations are used in this manuscript:

\abbreviation{ARM}{computer processor architecture}
\abbreviation{LM35C}{temperature sensor integrated circuit}
\abbreviation{MIC-6 Al}{a nearly pure aluminum alloy}
\abbreviation{MPXH6115A6U}{air pressure sensor integrated circuit}
\abbreviation{PIR}{polyisocyanurate foam}
\abbreviation{RAM}{random access memory}
\abbreviation{RMS}{root-mean-squared\quad$\sqrt{\sum_{i=1}^n x_i^2/n}$}
\abbreviation{RMSRE}{root-mean-squared relative error (\%)}
\abbreviation{RSS}{root-sum-squared\quad$\sqrt{\sum_{i=1}^n x_i^2}$}
\abbreviation{STM}{STMicroelectronics, an integrated circuit manufacturer}
\abbreviation{USB}{Universal Serial Bus}
\abbreviation{XPS}{extruded polystyrene foam}

\beginsection{Supplementary Materials}

 A zip archive of PDF files of graphs and estimated measurement
 uncertainties of each 102-minute time-series producing a convection
 measurement can be downloaded from:

\vbox{\settabs 1\columns
\+\hfil{\tt http://people.csail.mit.edu/jaffer/convect}&\cr}

 A zip archive of the aggregate measurements are also available from
 that site.

\beginsection{Acknowledgments}

 Thanks to John Cox (1957-2022) and Doug Ruuska for machining the
 bi-level plate.  Thanks to Roberta Jaffer for assistance and
 problem-solving suggestions.  Thanks to anonymous reviewers for their
 useful suggestions.

\section{Appendix A: Velocity Profiles}

 This investigation assumes that the laminar natural boundary layer
 thickness $\delta_\lambda$ is the same as the forced laminar
 thickness calculated using the effective vertical $\ReN$.
 Formulas from Lienhard and Lienhard~\cite{ahtt5e} lead to natural
 convection velocity profile $u(y)$
 Formula~\eqref{eq:natural-u(y)}, where $y$ is the horizontal distance
 from the mid-line of the vertical plate, $\nu$ is the fluid's
 kinematic viscosity, and $u_N={\nu\,\ReN/L}$ is the effective natural
 flow speed.
$${u(y)}\approx4\,u_N\,{y\over\delta_\lambda}\left[1-{y\over\delta_\lambda}\right]^2
 \qquad 0<y<\delta_\lambda\approx{4.92\,L\over\sqrt{\ReN}}\eqdef{eq:natural-u(y)}$$

 In forced turbulent flow along a smooth plate, let friction velocity
 ${\us}\approx\uinf\,\sqrt{\fctol/2}$, and $\Rey_y={\us\,y/\nu}$,
 with $\fctol$ from Formula~\eqref{eq:turbulent-friction}.
 Lienhard and Lienhard~\cite{ahtt5e} gives the viscous sublayer
 velocity profile as Formula~\eqref{eq:viscous-law}, and the log-layer
 velocity profile as Formula~\eqref{eq:log-law}.  The von
 K\'{a}rm\'{a}n constant $\kappa\approx0.41$.
$$\eqalignno{{u(y)\over\us}&\approx\Rey_y\qquad\qquad\qquad\qquad~~\Rey_y<7&\eqdef{eq:viscous-law}\cr
  {u(y)\over\us}&\approx\left[{1\over\kappa}\,\ln\left(\Rey_y\right)+5.5\right]
  \qquad\Rey_y>30&\eqdef{eq:log-law}}$$

 Lienhard and Lienhard~\cite{ahtt5e} does not tell how to interpolate
 these two formulas.  The $7<\Rey_y<30$ range to be interpolated is
 large, and the transition must be gradual.  Adapting the
 staged-transition formula from Jaffer~\cite{thermo3040040} by using
 the $\ell^{-\sqrt{1/3}}$-norm instead of the $\ell^{-4}$-norm in
 $\Rey_Y$ Formula~\eqref{eq:Re-Y} yields Formula~\eqref{eq:forced1}.
$$\eqalignno{&\Rey_Y=\|7, \Rey_y\|_{-\sqrt{1/3}}
  \qquad y<\delta_\tau\approx{0.37\,L\over\root5\of{\Rey}}
  &\eqdef{eq:Re-Y}\cr
  {u(y)\over\us}=&\left[5.5+{\ln\left(\Rey_y\right)\over\kappa}\right]+{\Rey_Y\over\Rey_y}\left[\Rey_Y-5.5-{\ln\left(\Rey_Y\right)\over\kappa}\right]
  &\eqdef{eq:forced1}\cr}$$

 Formula~\eqref{eq:forced-W0} is a proposed alternative to
 Formula~\eqref{eq:forced1} based on the Lambert $\Wz$ function.
$${u(y)\over\us}={8\over\sqrt{3}}\,\Wz(\sqrt{3}\,\Rey_y)
  \eqdef{eq:forced-W0}$$

  The ``interpolated'' Formula~\eqref{eq:forced1} and
  Formula~\eqref{eq:forced-W0} curves are nearly identical
  in \figref{fig:BL-forced}.  It is not surprising that a formula for
  turbulent flow involves the Lambert $\Wz$ function.
  Formula~\eqref{eq:forced-W0} uses $\Wz(\sqrt{3}\,\Rey_y)$ while
  $\fctol$ Formula~\eqref{eq:turbulent-friction} uses
  $\Wz(\ReF/\sqrt{3})$.

  While interesting, these curves are employed only for estimating the
  net vertical flow near the plate.  None of the convective surface
  conductance formulas quantitatively depend on them.

\smallskip
\vbox{\settabs 1\columns
\+\hfil\figscale{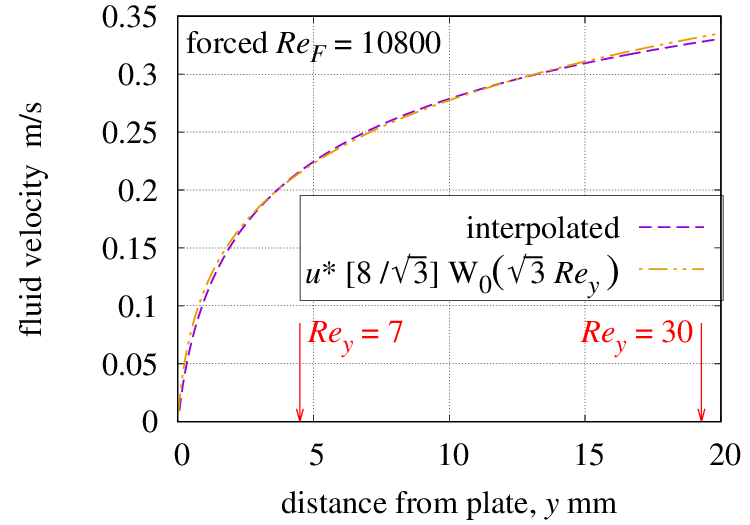}{234pt}&\cr
\+\hfil{\bf\figdef{fig:BL-forced}\quad{Forced convection velocity profile}}&\cr
}

\subsection{Rough Plates}
  Because roughness $\varepsilon\ll{L}$ has negligible effect on
  natural convection, $u_N$ should be the same from smooth and rough
  plates.  Hence, their $\ReN\equiv{{u_N}\,\varepsilon/\nu}$ should
  also be equal.

  Forced flow along a rough surface traverses a path longer than $L$.
  The effective $\ReN/\ReF$ ratio of a rough surface should be
  increased by a function of the ``roughness Reynolds number''~$\Rev$
  Formula~\eqref{eq:roughness-Re}.
$$\Rev={\us\,\varepsilon\over\nu}
={\Rey\over\sqrt{3}\,[L/\varepsilon]\,\ln(L/\varepsilon)} \eqdef{eq:roughness-Re}$$

  Proposed is $\ReN/\Rey$ scale factor $\chi$
  Formula~\eqref{eq:rough-correct}, where $\Rey$ is the solution of
  Formula~\eqref{eq:roughness-Re} combined with
  $\Rev=3\,[\varepsilon/L]^2$.  \figref{fig:chi} graphs $\chi$ as a
  function of $\varepsilon$.
$$\chi={\ReF+\Rey\over\ReF}=1-3\,\sqrt3\,{\varepsilon\over L}\,\ln{\varepsilon\over L}\eqdef{eq:rough-correct}$$

\smallskip
\vbox{\settabs 2\columns
\+\hfil\figscale{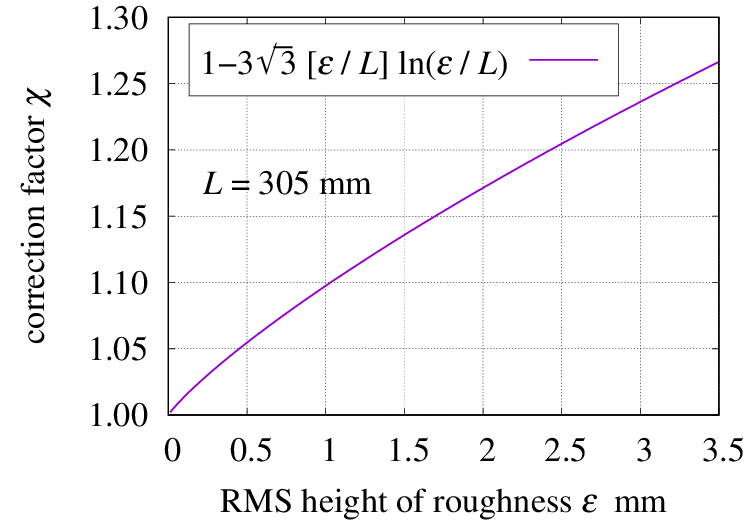}{234pt}&\hfil\figscale{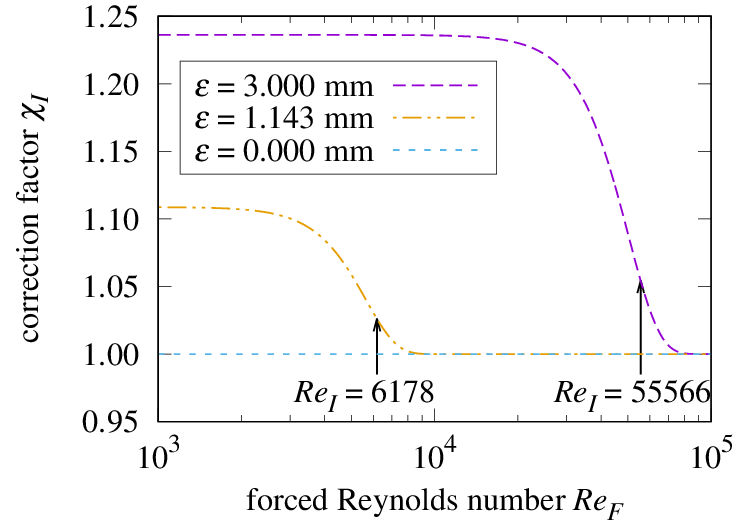}{234pt}&\cr
\+\hfil{\bf\figdef{fig:chi}\quad $\chi$ versus $\varepsilon$}&\hfil{\bf\figdef{fig:Re-correction}\quad $\ReN$ correction factor $\CI$}&\cr}

\section{Appendix B: Plateau Islands Roughness Correction}

  Plates shedding only turbulent flow have $\CI=1$.
  Plates shedding only rough flow have $\CI=\chi$ from
  Formula~\eqref{eq:rough-correct}.

  Plateau roughness (forced) convection transitions from rough flow to
  turbulent flow as the $\ell^{-4}$-norm in
  Formula~\eqref{eq:bi-level-convect}.  The
  scale factor $\CI$ should vary between 1 and $\chi$ as a function
  of~$\ReF$.  Expression $\exp_4\left(-[{\ReF/\ReI}]^4\right)$ varies
  between 0 and 1.
  Proposed is $\CI$, the $\ReN/\ReF$ scale factor:
$$\CI=\exp_{\chi}\!\left(\exp_4\left(-[{\ReF/\ReI}]^4\right)\right)\eqdef{eq:C-w}$$

  Note the similarity of Formula~\eqref{eq:C-w} and
  Formula~\eqref{eq:vertical-p} with $\zeta=4$.

 \figref{fig:Re-correction} plots $\CI$ with $\varepsilon=3$~mm,
 $\varepsilon=1.143$~mm, and $\varepsilon=0$.

\section{Appendix C: Apparatus and Measurement Methodology}

 The original goal of the present apparatus was to measure forced
 convection heat transfer from a precisely rough plate over the widest
 practical span of airflow velocities.  To minimize natural
 convection, it measured downward natural convection mixed with
 horizontal forced flow.  Its measurements are presented in
 Jaffer~\cite{thermo3040040}.

 Although more complicated to analyze, the plate was suspended, not
 embedded, in the wind-tunnel.  The measurements from prior
 investigations which embedded the plate in a wind-tunnel wall were
 largely incompatible with the present theory because their flows were
 not isobaric.

 The small size of the wind-tunnel chassis (${\rm
 1.3~m\times0.61~m\times0.65~m}$) afforded an opportunity to
 characterize mixed convection at other orientations of the plate and
 flow.

\subsection{The Plate}
 \figref{fig:front} shows the rough surface of the test plate; it was
 milled from a slab of MIC-6 aluminum (Al) to have (676 of) square
 $8.33{\rm\,mm}\times8.33{\rm\,mm}\times6{\rm\,mm}$ posts spaced on
 11.7~mm centers over the $30.5~{\rm cm} \times 30.5~{\rm cm}$ plate.
 The area of the top of each post was $0.694{\rm\,cm}^2$, which was
 50.4\% of its $1.38{\rm\,cm}^2$ cell.  The RMS height-of-roughness
 $\varepsilon=3.00{\rm~mm}$.  Openness $\Omega\approx49.6\%$.
 Embedded in the plate are 9 electronic resistors as heating elements
 and a Texas Instruments LM35 Precision Centigrade Temperature Sensor.
 2.54~cm of thermal insulating foam separates the back of the plate
 from a 0.32~mm thick sheet of aluminum with an LM35 at its center.
 \figref{fig:crosect} is a cross-section illustration of the plate assembly.

\medskip
\vbox{\settabs 2\columns
\+\hfil\figscale{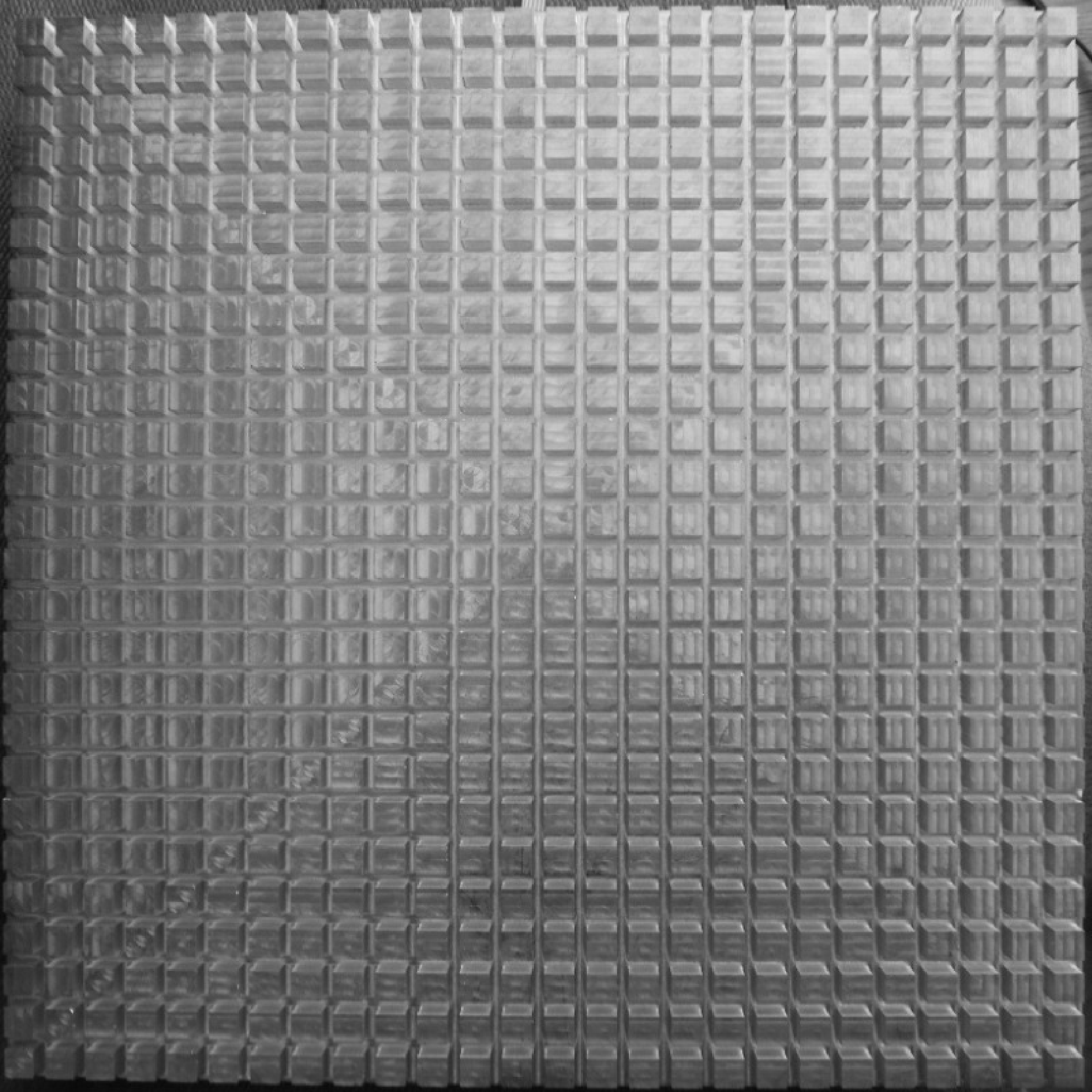}{200pt}&
  \hfil\figscale{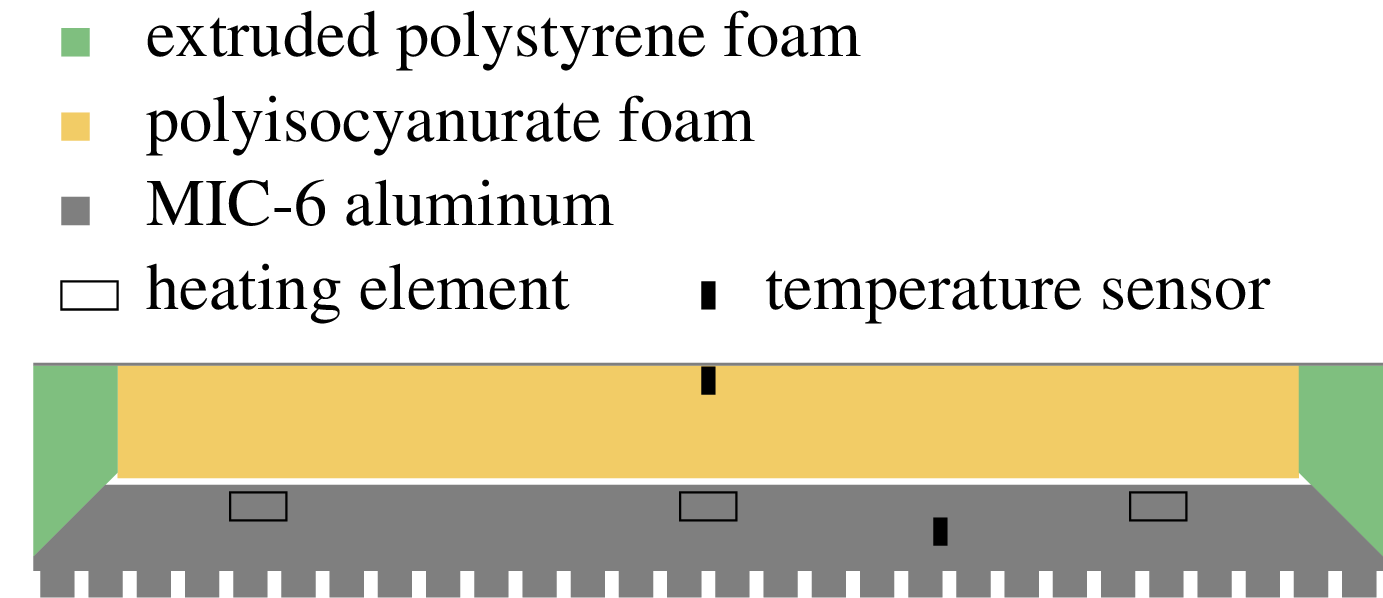}{200pt}&\cr
\+\hfil{\bf\figdef{fig:front}\quad Rough surface of plate}&
  \hfil{\bf\figdef{fig:crosect}\quad Plate assembly cross-section}&\cr
}

\subsection{Wind Tunnel}
 The fan pulls air from the test chamber's open intake through the
 test chamber.  The fan blows directly into a diffuser made of folded
 plastic mesh to disrupt vortexes generated by the fan.  In a
 sufficiently large room, the disrupted vortexes dissipate before
 being drawn into the open intake.

 To guarantee isobaric (no pressure drop) flow, the wind-tunnel must
 be sufficiently large that its test chamber and plate assembly
 boundary-layers do not interact at fan-capable airspeeds.

The wind-tunnel test chamber in \figref{fig:suspend} has a
$61{\rm~cm}\times35.6{\rm~cm}$ cross-section and a 61~cm depth.  This
allows the plate assembly to be centered in the wind-tunnel with 15~cm
of space on all sides.
The fan pulling air through the test chamber produces a maximum
airspeed of 4.65~m/s ($\Rey\approx9.2\times10^4$ along the 30.5~cm
square plate).  Its minimum nonzero airspeed is 0.12~m/s
($\Rey\approx2300$).

 Test chamber laminar and turbulent 99\% boundary-layer
 thicknesses (Schlichting \cite{schlichting2014}) are:
$$\delta_\lambda=4.92\,\sqrt{x\nu\over u}\qquad
  \delta_\tau=0.37\,x^{4/5}\left[\nu\over u\right]^{1/5}\eqdef{eq:tunnel-WTBL}$$

\figref{fig:WTBL} shows that the 15~cm clearance between the plate and
the test chamber walls is sufficient to prevent their boundary-layers
from interacting at airspeeds within the fan's capabilities.

 The plate assembly (face down in \figrefs{fig:crosect}
 and \figrefn{fig:suspend}) is suspended by six lengths of 0.38
 mm-diameter steel piano wire terminated at twelve zither tuning pins
 in wooden blocks fastened to the exterior of the test chamber.
With the plate assembly in the test chamber, the airspeed increases in
proportion to the reduction of test chamber aperture $A_e$ by the
plate's cross-sectional area $A_\times$:
$${u_\times\over u}={A_e\over A_e-A_\times}\approx107.6\%\eqdef{eq:u-aperture}$$

\medskip
\vbox{\settabs 2\columns
\+\hfil\figscale{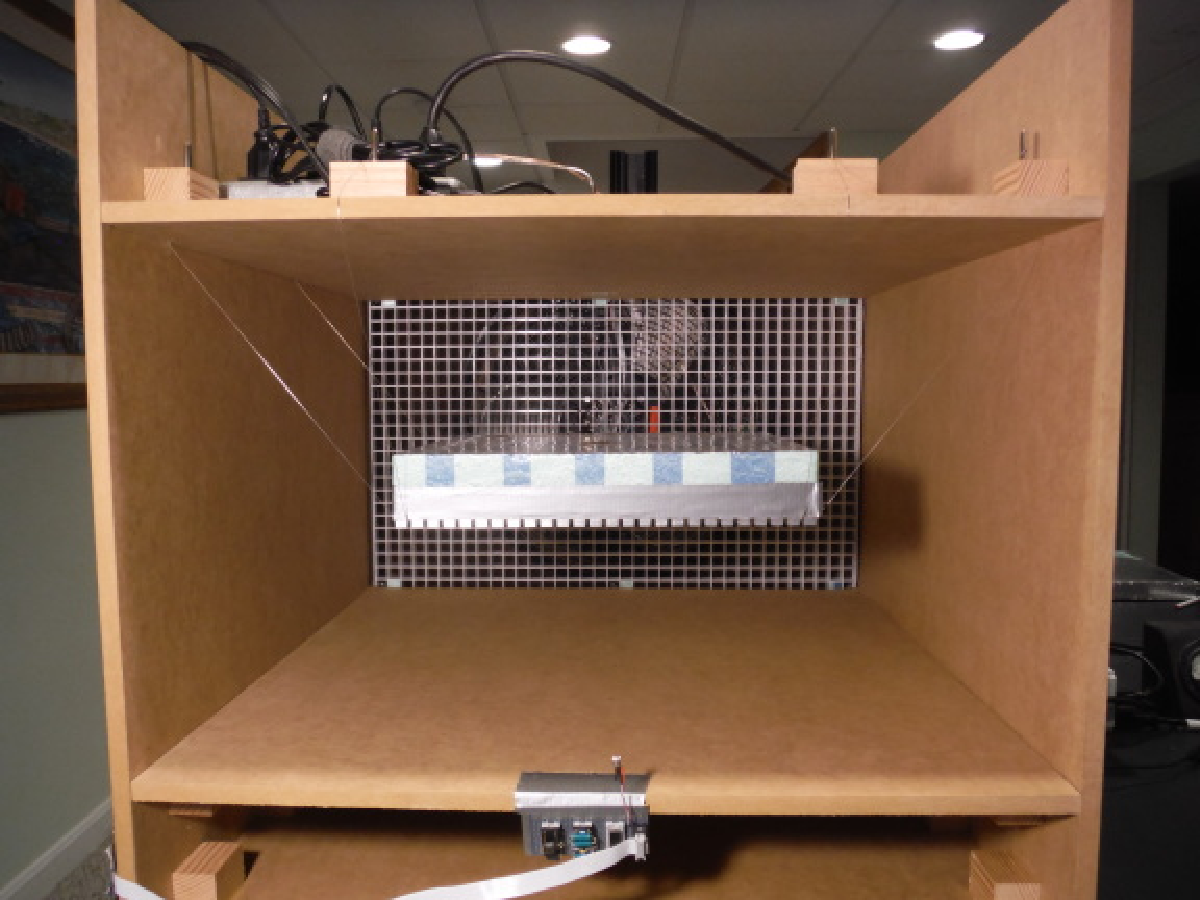}{217pt}&
  \hfil\figscale{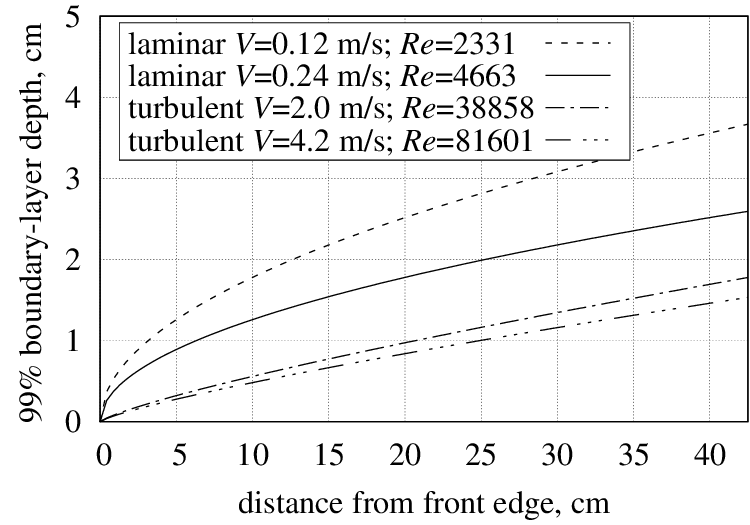}{234pt}\cr
\+\hfil{\bf\figdef{fig:suspend}\quad $\varepsilon=3\rm~mm$ plate in wind-tunnel}&
  \hfil{\bf\figdef{fig:WTBL}\quad Wind-tunnel boundary-layers}\cr
}

\subsection{Automation}
Data capture and control of convection experiments are performed by an
``STM32F3 Discovery 32-Bit ARM M4 72MHz'' development board.
The program written for the STM32F3 captures readings and writes them
to the microprocessor's non-volatile RAM, controls the plate heating,
servos the fan speed, and later uploads its data to a computer through
a USB cable.

Once per second during an experiment, the program calibrates and reads
each on-chip 12~bit analog-to-digital converter 16 times, summing the
sixteen 12~bit readings to create a 16~bit reading per converter.

Rotations of the fan are sensed when a fan blade interrupts
an infrared beam.  The microprocessor controls a solid-state relay
(supplying power to the fan) to maintain a fan rotation rate,
$\omega$, which is dialed into switches.  At~$\omega\le210$~r/min, the
microprocessor pulses power to the fan to phase-lock the beam
interruption signal to an internal clock.  At~$\omega>210$~r/min, the
microprocessor servos the duty cycle of a 7.5~Hz square-wave gating
power to the fan.  This system operates at
$32{\rm~r/min}<\omega<1400{\rm~r/min}$.

\subsection{Calibration}
 The correspondence between fan rotation rate~$\omega$ and test
 chamber airspeed~$u$ was determined using an ``Ambient
 Weather~WM-2'', which specifies an accuracy of $\pm3\%$ of reading.
 After 2017 an ``ABM-200 Airflow \& Environmental Meter'' specifying
 an accuracy of $\pm0.5\%$ of reading between 2.2~m/s and 62.5~m/s,
 was used.

 The ``UtiliTech 20 inch 3-Speed High Velocity Floor Fan'' has three
 blades with maximum radius $r=0.254$~m.  Its characteristic length is
 its hydraulic-diameter, $D_H=0.550$~m.  The velocity of the blade tips
 is $2\,\pi\,r\,\omega/60$, where $\omega$ is the number of rotations
 per minute.  The Reynolds number of the fan is:
$$\Rey_f={2\,\pi\,r\,D_H\,\omega/60\over3\,\nu}\eqdef{eq:fan-Re}$$

 The 3 blade tips trace the whole circumference in only 1/3 of a
 rotation, hence the 3 in the denominator.

 Faster fan rotation $\omega$ yields diminishing increases of
 test-chamber airspeed $u_t$, suggesting Formula~\eqref{eq:Re-tunnel},
 where $u_u$ is the limiting velocity for arbitrarily fast rotation,
 and coefficient $\eta$ converts fan $\Rey_f$ to test-chamber $\Rey_t$.
 \figref{fig:tunnel-fit} gives the parameters and
 measurements at $300{\rm~r/min}\le\omega\le1500{\rm~r/min}$.  The
 ``3mm'' points are the WM-2 measurements of the 3~mm plate in the
 original wind-tunnel; The ``1mm'' points are the ABM-200 measurements
 of the 1~mm plate in the tunnel with a new diffuser and fan cowling.
$$\Rey_t=\|\eta\,\Rey_f,~D_H\,u_u/\nu\|_{-2}\qquad
  u_t=\|\pi\,\eta\,r\,\omega/90,~u_u\|_{-2}\eqdef{eq:Re-tunnel}$$

 Airspeeds slower than 2~m/s should be nearly proportional to
 $\omega$.  Both anemometers show evidence of dry (bearing) friction
 in \figref{fig:tunnel-fit}.  The ABM-200 ``meter predictions'' trace
 plots $1.125\,u_t-0.381$; the WM-2 ``meter predictions'' trace plots
 $1.477\,u_t-0.81$ when $u_t<1.725$ and $u_t$ otherwise.
 A mistake in the 2016 measurement software under-counted fan
 rotations at $\omega>1200$~r/min.  It is compensated by replacing
 $\omega$ in Formulas (\eqrefn{eq:fan-Re}, \eqrefn{eq:Re-tunnel}) with
 $[\,\omega^{-6}-1750^{-6}]^{-1/6}$ in the WM-2 ``meter predictions''.
  The RMSRE and Bias are relative to the ``meter predictions''.  The
  second ``1mm'' row includes the point at 400~r/min.

\medskip
\vbox{\settabs 1\columns
\+\hfil\figscale{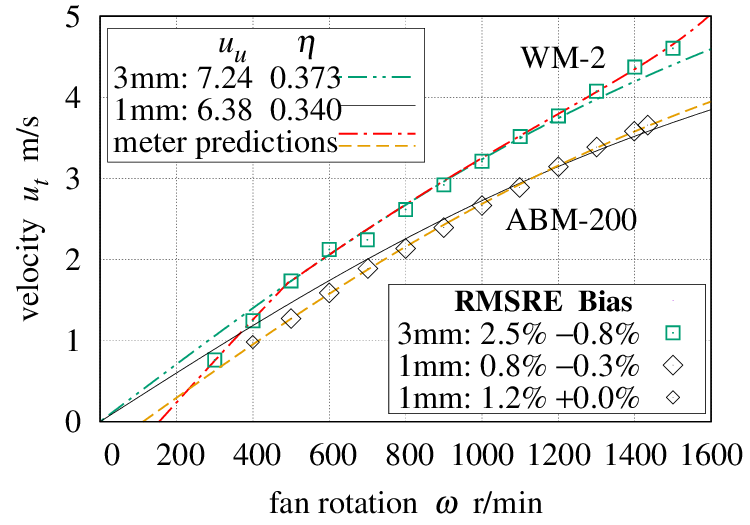}{234pt}&\cr
\+\hfil{\bf\figdef{fig:tunnel-fit}\quad Airspeed versus fan speed}&\cr
}

 \figrefs{fig:rpm-stats-3mm} and \figrefn{fig:rpm-stats-1mm} show the
 fan speed variability in each experiment; these are used in the
 measurement uncertainty calculations.

\vbox{\settabs 2\columns
\+\hfill\figscale{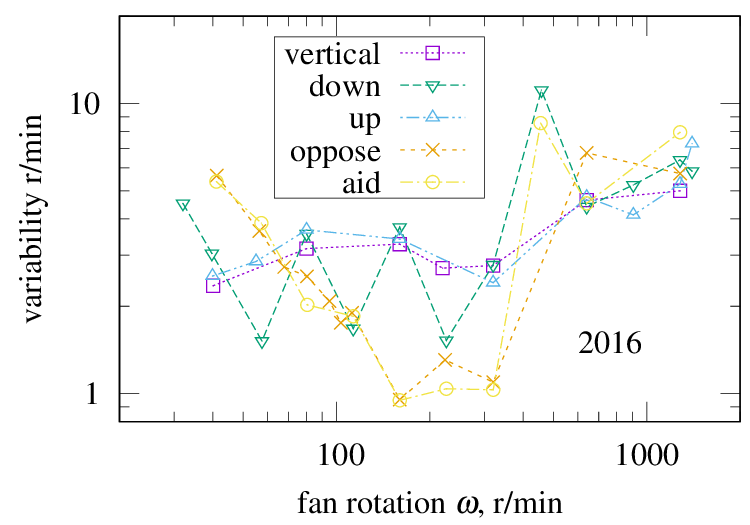}{234pt}\hfill
 &\hfill\figscale{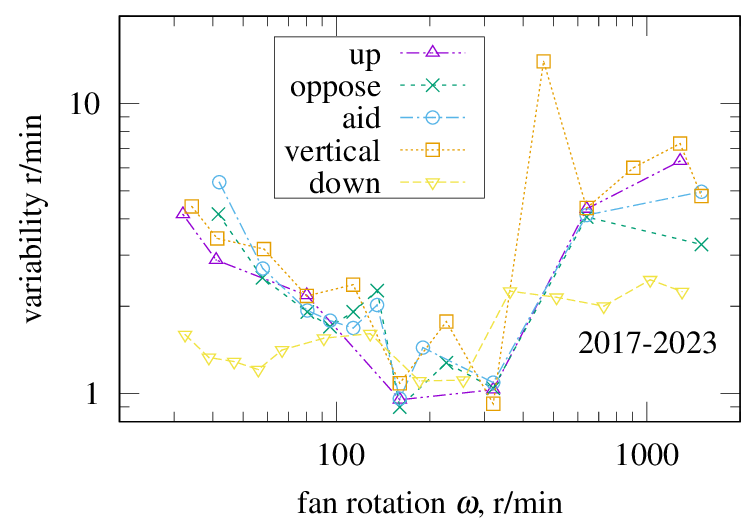}{234pt}\hfill&\cr
\+\hfill{\bf\figdef{fig:rpm-stats-3mm}\quad Fan variability 3~mm plate}\hfill
 &\hfill{\bf\figdef{fig:rpm-stats-1mm}\quad Fan variability 1~mm plate}\hfill&\cr
}

\subsection{Ambient Sensing}
 \figref{fig:ambient} shows the ambient sensor board which was at the
 lower edge of the test chamber in \figref{fig:suspend}.  It measures
 the pressure, relative humidity, and air temperature at the
 wind-tunnel intake.  Wrapped in aluminum tape to minimize radiative
 heat transfer, the LM35 temperature sensor projects into the tunnel.
 To minimize self-heating, the LM35 is powered only while being
 sampled.

\smallskip
\vbox{\settabs 2\columns
\+\hfill\figscale{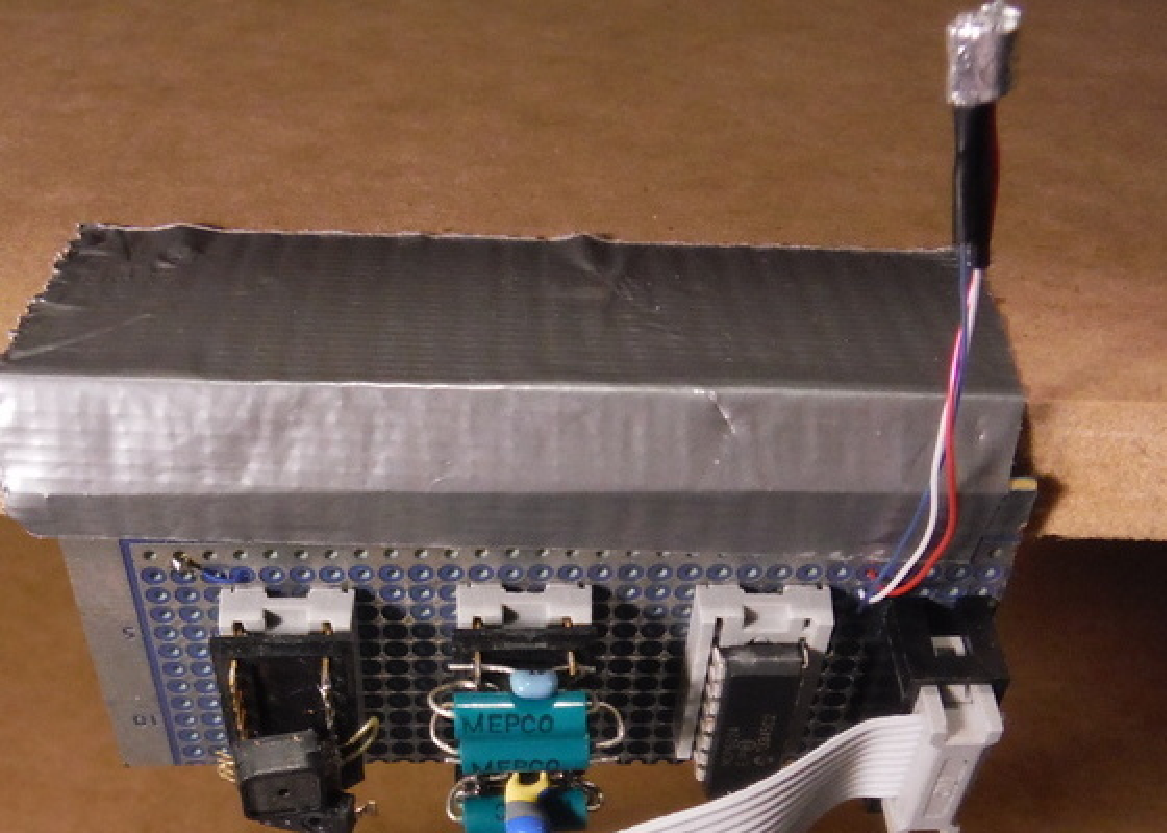}{182pt}\hfill&\hfill\figscale{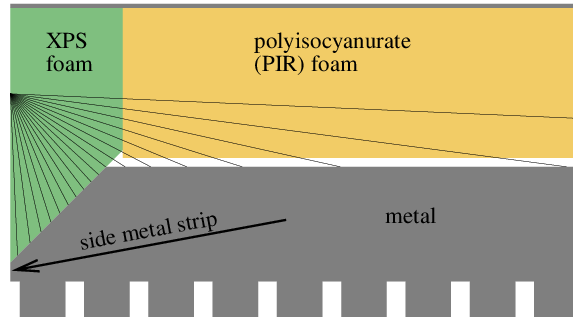}{234pt}&\cr
\+\hfill{\bf\figdef{fig:ambient}\quad Ambient sensors}\hfill&\hfill{\bf\figdef{fig:wedge}\quad XPS wedge conduction}\hfill&\cr
}

\subsection{Physical Parameters}
 \tabref{tab:physical parameters} lists the static parameters
 from measurements and specifications.

 The effective $\epsilon_{wt}$ may differ from the
 medium-density-fiberboard emissivity given by
 Rice~\cite{rice2004emittance} because the temperatures of the test
 chamber surfaces may not be uniform.  Through the open intake, the
 plate also exchanges thermal radiation with objects in the room having
 different temperatures.

\medskip
\centerline{\bf\tabdef{tab:physical parameters}\quad{Physical parameters}}
\moveright 0.0625\hsize\vbox{\settabs 8\columns
\toprule
\+{\bf Symbol}&{\bf Values}&&{\bf Description}&&&&\cr
\midrule
\+$L$&0.305 m&&length of flow along test-surface&&&&\cr
\+$A$&0.093 $\rm m^2$&&area of test-surface&&&&\cr
\+$\varepsilon$&3.00 mm & 1.04 mm & RMS height-of-roughness&&&&\cr
\+$C_{pt}$&4691 J/K & 4274 J/K & plate thermal capacity&&&&\cr
\+$D_{\rm{Al}}$&19.4 $\rm mm$&&metal slab thickness&&&&\cr
\+$D_{\rm{PIR}}$&25.4 $\rm mm$&&polyisocyanurate (PIR) foam thickness&&&&\cr
\+$D_w$&19.05 $\rm mm$&&XPS foam wedge height&&&&\cr
\+$k_{\rm PIR}$&$0.0222~\rm{W/(m\cdot{K})}$&&PIR foam thermal conductivity&&&&\cr
\+$k_{\rm XPS}$&$0.0285~\rm{W/(m\cdot{K})}$&&XPS foam thermal conductivity&&&&\cr
\+$U_I$&0.075 W/K&&front-to-back insulation thermal conductance&&&&\cr
\+$\epsilon_{\rm{Al}}$&0.04&&test-surface (MIC-6 Al) emissivity&&&&\cr
\+$\epsilon_{\rm{XPS}}$&0.515&&XPS foam emissivity (see text)&&&&\cr
\+$\epsilon_{dt}$&0.89&&duck tape emissivity&&&&\cr
\+$\epsilon_{wt}$&0.90&&test chamber interior emissivity&&&&\cr
\bottomrule
}

\subsection{The 1~mm Roughness Plate}
 When the 6~mm posts were milled down to 2~mm height, the four corner
 posts were left at their 6~mm height in order to preserve the wire
 suspension.  This resulted in $\varepsilon=1.04$~mm for the plate as
 a whole.  However, $\ReI$ occurs within the first few rows of posts.
 $\varepsilon=1.143$~mm over the first three rows of posts results in
 $\ReI=6178$.

\subsection{Modeling of Parasitic Heat Flows}
 The plate has six surfaces from which heat can flow.
 At low airflow velocities, the sides of the insulation behind the test
 plate can leak more heat than the test-surface transfers, shrinking
 to 6\% at {1300\rm~r/min}.

 In order to measure natural convection from the (rough) test surface,
 natural convection and thermal radiation from the four sides ($U_S$)
 and back must be deducted from the total heat flow.  Heat from the
 front plate flows through thermal insulating foam to a thin aluminum
 sheet with a temperature sensor at its center.  This heat flow is
 simply $U_I\,[T_P-T_B]$, the product of the foam's thermal
 conductance and the temperature difference across it.

\subsection{Forced Convection Side Model}
 The four sides are not isothermal; a 3.5~mm metal strip (see
 cross-section \figref{fig:wedge}) runs the length of the side; and a
 $D_w$-tall wedge of extruded polystyrene foam (XPS) insulation fills
 the metal slab's $27{\rm~mm}~(=\sqrt{2}\,D_{\rm{Al}})$ $45^\circ$
 chamfer.  The local surface conductance $h_W(z)$ at elevation $z$
 (from the wedge point) is found by averaging the reciprocal distance
 to slab metal with respect to angle~$\theta$:
$$\eqalign{h_W(z)&=\int_0^{\theta_c}{k_{\rm XPS}\over\sqrt{2}\,z\,\theta_c}\,\cos\left(\theta+{\pi\over4}\right)\,\diff{\theta}
 +\int_{\theta_w}^{\theta_W}{k_{\rm PIR}\over z-D_w}\,{\cos\theta\over{\theta_W-\theta_w}}\,\diff{\theta}\cr
 &={k_{\rm XPS}\over\sqrt{2}\,z\,\theta_c}\,\left[\sin\left(\theta_c+{\pi\over4}\right)-\sin{\pi\over4}\right]
  +{k_{\rm PIR}\over z-D_w}\,\left[{\sin\theta_W-\sin\theta_w\over{\theta_W-\theta_w}}\right]\cr
 \theta_c=\arctan&{D_w-z\over D_w}
 \quad\theta_w=\arctan{D_w\over z-D_w}
 \quad\theta_W=\max\left(\theta_w,~{\arctan{L-D_w\over z-D_w}}\right)\cr}\eqdef{eq:h_W}$$

 Forced air flows parallel to the long dimension on two sides, but
 flows into the windward side and away from the leeward side.  Air
 heated by the windward side reduces heat transfer from the
 test-surface; air heated by the test-surface suppresses heat transfer
 from the leeward side.  Hence, the model excludes windward and
 leeward forced convection.
 The average forced convective conductance of the flow-parallel foam
 wedges is calculated by integrating $h_W(z)$ in series (reciprocal of
 the sum of reciprocals, which is also the $\ell^{-1}$-norm) with the
 local surface conductance $k\,\Nus(\Rex)/L$, where $\Nus(\Rex)$ is
 the local pierced-laminar convection from
 Jaffer~\cite{thermo3040040}:
$$\eqalignno{U_W&=\int_0^{D_w}\int_0^L\left\|{h_W(z), {k\,\Nus(\Rex)\over L}}\right\|_{-1}\diff{x}\,\diff{z}&\eqdef{eq:U_W}\cr}$$

\subsection{Other Side Models}
 The heat flow through the four sides $U_S$ will be estimated from the
 plate and ambient temperatures.  While the forced convective surface
 conductance of the sides is modeled by integrating the local forced
 surface conductance, this is not generally possible for natural
 convection.

 Natural convection formulas are known for some convex surfaces.
 The plate's side metal surface is not convex.

 Instead, the effective side width $L_{es}$ and effective
 emissivity $\epsilon_W$ are introduced into the model.  The natural
 convection of each side is calculated for an $L_{es}\times L_C$ area
 instead of its actual $L_S\times L_C$ area.  The black-body radiation
 from each side is calculated for its actual $L_S\times L_C$ area with
 an effective emissivity of~$\epsilon_W$.

 The flow patterns in Fujii and Imura~\cite{fujii1972natural} figures
 14(e) and 14(f) (and schematic drawing~\figref{fig:above-flow}) show
 a plume rising from the center of an upward-facing plate fed by flow
 from the plate's edges.  For the test surface, the upward heat flow
 of 0.467~W/K is more than twice the 0.212~W/K expected from the back
 and sides.  Convective flow from the upward-facing test surface will
 draw in the air heated by the back and sides, reducing heat transfer
 from the test surface.  In order to avoid double counting the
 convected heat from the back and sides, they should not be deducted
 from the plate heat (the thermal radiation is still deducted).  The
 ``reuptake'' of this convected back and side heat should be nearly
 complete; its coefficient was set to~1 to avoid introducing another
 degree-of-freedom into the model.

 Not deducting side convection from upward natural convection has an
 unexpected benefit: the upward convection model is thus insensitive
 to $L_{es}$, allowing $\epsilon_W$ to be determined from only
 upward-facing measurements.

\subsection{Radiative Transfer Side Model}
 The 3~mm roughness plate had its sides wrapped with duck tape, which
 has a different emissivity from the foam wedges forming each side
 surface.  Some of the 1~mm roughness plate runs were with tape and
 some without, requiring different $\epsilon_W$ values.  For taped
 sides $\epsilon_W\approx0.703$; without tape
 $\epsilon_W\approx0.515$.

 \figref{fig:suspend} shows duck tape applied to the lower 54\% of the
 plate's side, which corresponds to 50\% coverage of the XPS foam
 wedge.  For this partial tape coverage, $\epsilon_W$
 Formula~\eqref{eq:emissivity} is the area proportional mean of the
 duck tape emissivity and XPS emissivity.  Barreira, Almeida, and
 Sim\~oes~\cite{s21061961} measured $\epsilon_{dt}$ emissivities of
 0.86 and 0.89 from two brands of ``duck tape''.  The emissivity is
 largely controlled by the exposed polyethylene film, and increases
 with oxidation.  Hence, the larger value is used for the aged duck
 tape on the plate sides.  As of this writing, published emissivity
 measurements of XPS foam have not been located.

$$\epsilon_W=50\%\,\epsilon_{dt}+50\%\,\epsilon_{\rm{XPS}}\eqdef{eq:emissivity}$$

 Natural convection measurements ($u=0$) from the plate assembly over
 the span of inclinations in \figref{fig:3mm-angles-correlation} have
 less than 3.3\% RMSRE when calculated with
 $\epsilon_{\rm{XPS}}=0.515$; the RMSRE increases to either side of
 0.515.  This value is consistent with natural convection measurements
 of the plate assembly without tape.

\subsection{Natural Convection Side Model}
 With $\epsilon_W$ thus determined, $L_{es}$ was the remaining degree
 of freedom.  Trials with vertical and downward plate measurements
 found that $L_{es}$ had a value near the sum of the aluminum slab
 thickness 19.4~mm and the effective height of the side face of the
 roughness $\approx\sqrt{2}\,\varepsilon$.  This makes sense for a
 natural convection dimension; it is used for $L_{es}$.  The 3~mm
 roughness plate has $L_{es}\approx23.6\rm~mm$; 1~mm has
 $L_{es}\approx21.0\rm~mm$.

 In the vertical case, 1/2 of the heated air from the bottom side flows
 along the vertical test surface and would be counted twice.  And 1/2
 of the air drawn by the top side comes from the vertical test surface
 and would be counted twice.  This vertical reuptake coefficient was
 set to~1/2; discrepancy from the actual reuptake coefficient
 will manifest as error in measurements.

Consider the (initially) vertical plate as $\theta$ decreases from
$0^\circ$.  As the bottom side face tilts upward, more (than half of
the) heated air will rise toward the test surface.  That heat will
reduce the convection from the test surface.  When tilted downward,
the heat from the test surface will reduce the convection from the top
side.  To handle these cases, Equation~\eqref{eq:U_S} includes a term
$2\,\cos\theta\,\sin\theta$ whose minimum of $-1$ is reached at
$\theta=-45^\circ$ and a term $-2\,\cos\theta\,\sin\theta$ whose
minimum of $-1$ is reached at $\theta=+45^\circ$.

\medskip
\subsection{Combining Radiative Transfer and Convection}
 A side's radiative emissions, $U_\epsilon$, compete with its
 convective heat transfer.  Both increase with side temperature, but
 both act to lower that side temperature.  Competitive heat transfer
 processes can often be modeled using the $\ell^p$-norm with $p>1$.
 The value of $p$ was adjusted so that the $\Delta{T}=3.8\rm{K}$ and
 $\Delta{T}=11\rm{K}$ data points align with the theory traces
 in \figref{fig:3mm-angles-correlation}.  The optimal range is between
 $p=4/3$ and $p=3/2$; the geometric mean of those values is
 $p=\sqrt{2}$.  The $\ell^{\sqrt{2}}$-norm appears three times in
 $U_S$ Formula~\eqref{eq:U_S}.

Formula~\eqref{eq:U_S} $U_S$ is an amount which will be deducted from the
measured heat flow.  For each side, the $\ell^{\sqrt{2}}$-norm of the
radiative and convective conductances is paired with the product of
the convective conductance and a continuous trigonometric function of
$\theta$ which goes negative when the natural convection would
otherwise be double counted.  Because of the triangle inequality, the
$\ell^{\sqrt{2}}$-norm will be greater than the convective component;
thus, each side's contribution to $U_S$ will be positive.

No more than one reuptake process will be simultaneously active for a
side.  In Formula~\eqref{eq:U_S}, the expressions
$\min(0,\sin\theta,-.5\,\cos\theta,2\,\cos\theta\,\sin\theta)$ and
$\min(0,\sin\theta,-.5\,\cos\theta,-2\,\cos\theta\,\sin\theta)$ return
the negative of the largest magnitude potential
reuptake.  \tabref{tab:natural convection function and parameters}
describes the natural convection parameters and function.

 Note that this analysis applies only to the plate assembly in
 alignment with the wind-tunnel, and oriented to have at least one
 horizontal edge.  Hence, rotation in plane of plate, $\phi$, must be
 an integer multiple of $90^\circ$.
 The only effect of $\phi$ in the equations is to swap arguments $L_F$
 and $L_W$ when $\phi$ is an odd multiple of $90^\circ$.

\medskip
\centerline{\bf\tabdef{tab:natural convection function and parameters}\quad{Natural convection function and parameters}}
\moveright 0.0833\hsize\vbox{\settabs 6\columns
\toprule
\+{\bf Symbol}&&{\bf Description}&&\cr
\midrule
\+$U_N(\theta,L_F,L_W,\phi)$&&natural convective conductance from Jaffer~\cite{thermo3010010}&&\cr
\+$\qquad\theta$&&surface angle from vertical ($-90^\circ$ is face up)&&\cr
\+$\qquad L_F$&&plate length&&\cr
\+$\qquad L_W$&&plate width&&\cr
\+$\qquad\phi$&&rotation in plane of plate; integer multiple of $90^\circ$&&&\cr
\+$L_C=0.305\rm~m$&&plate length = side length&&\cr
\+$L_S=45.8\rm~mm$&&side width&&\cr
\+$L_{es}=19.4{\rm~mm}+\sqrt{2}\,\varepsilon$&&effective side width for natural convection&&\cr
\+$\epsilon_{wt}=0.9$&&wind-tunnel test chamber emissivity&&\cr
\+$\epsilon_W\approx0.703$ taped; 0.515 bare&&effective side emissivity&&\cr
\+$h_R$&&black-body radiative surface conductance&&\cr
\+$U_\epsilon=L_C\,L_S\,\epsilon_W\,\epsilon_{wt}\,h_R$&&radiative emission from a side&&\cr
\bottomrule
}

$$\eqalign{
 U_S&=\left\|U_\epsilon,U_N(\theta-90^\circ,L_C,L_{es},0^\circ)\right\|_{\sqrt2}\cr
 &+U_N(\theta-90^\circ,L_C,L_{es},0^\circ)
 \,\min(0,\sin\theta,-.5\,\cos\theta, -2\,\cos\theta\,\sin\theta)\cr
 &+\left\|U_\epsilon,U_N(90^\circ-\theta,L_C,L_{es},0^\circ)\right\|_{\sqrt2}\cr
 &+U_N(90^\circ-\theta,L_C,L_{es},0^\circ)
 \,\min(0,\sin\theta,-.5\,\cos\theta, 2\,\cos\theta\,\sin\theta)\cr
 &+2\,\left\|U_\epsilon,U_N(0^\circ,L_{es},L_C,\theta)\right\|_{\sqrt2}\cr
 &+2\,U_N(0^\circ,L_{es},L_C,\theta)
 \,\min(0,\sin\theta)\cr
 }\eqdef{eq:U_S}$$

$U_{B0}$ is the test surface reuptake conductance from the back.  Its
$\min(0,\sin\theta)$ term is squared because the heated air from the
back must flow around two right-angle edges to reach the test surface.
$$U_{B0}=-U_N(90^\circ,L_C,L_C,0^\circ)\,\min(0,\sin\theta)^2\eqdef{U_{B0}}$$

\vfill\eject

 \figrefs{fig:upward}, \figrefn{fig:vertical},
 and \figrefn{fig:downward} show upward, vertical, and downward
 convection measurements, respectively.  Taken from 3~mm and 1~mm
 roughness plates over a range of $\Ra$ values, these graphs, in
 combination with \figref{fig:3mm-angles-correlation} test the natural
 convection and radiative transfer side models.

 The traces labeled ``theory'' are Formula~\eqref{eq:general} with
 the appropriate row of \tabref{tab:natural convection parameters}.
 The difference between $\psi=0^\circ$ and $\psi=180^\circ$
 in \figref{fig:vertical} is explained in \ref{Discussion}.

\vbox{\settabs 1\columns
\+\hfill\figscale{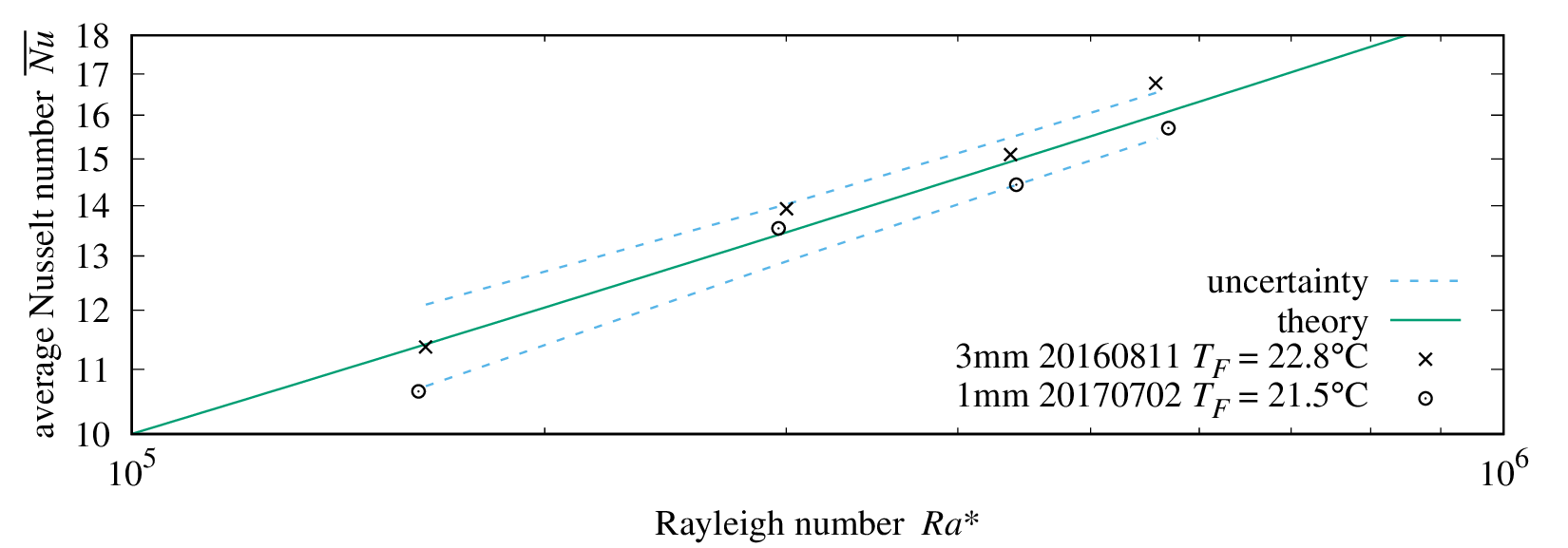}{435pt}\hfill&\cr
\+\hfill{\bf\figdef{fig:upward}\quad Natural convection from upward-facing surface}\hfill&\cr
}

\vbox{\settabs 1\columns
\+\hfill\figscale{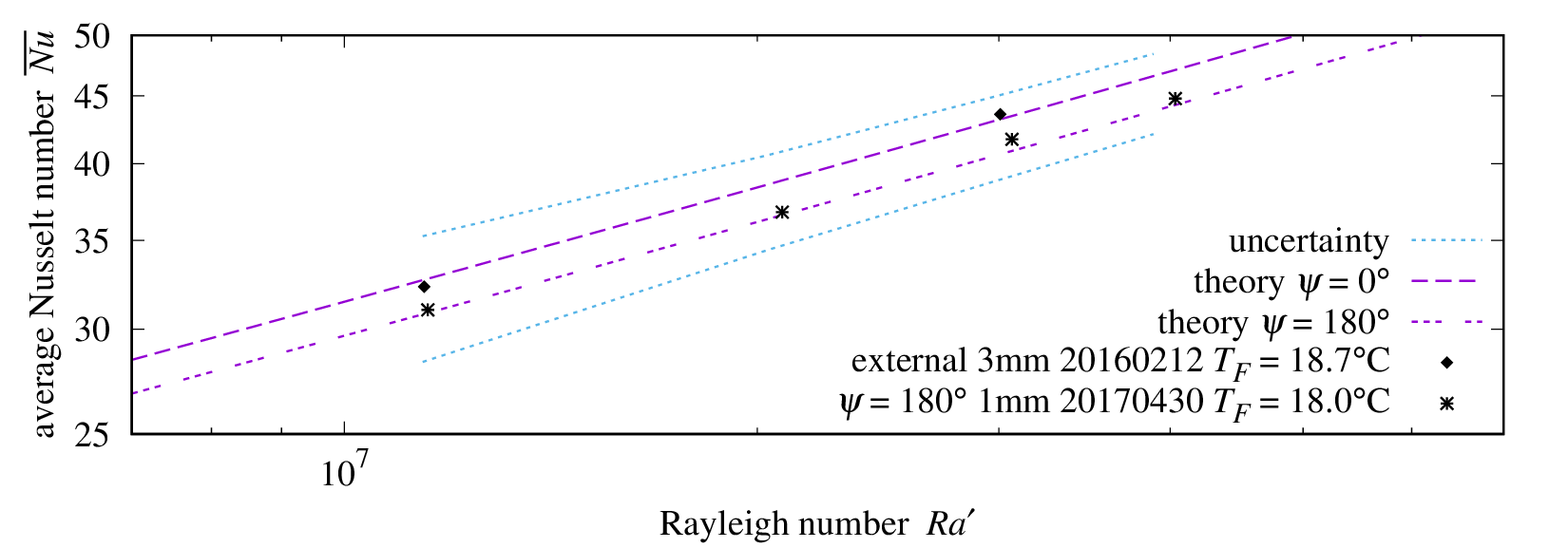}{435pt}\hfill&\cr
\+\hfill{\bf\figdef{fig:vertical}\quad Natural convection from vertical surface}\hfill&\cr
\+\hfill\figscale{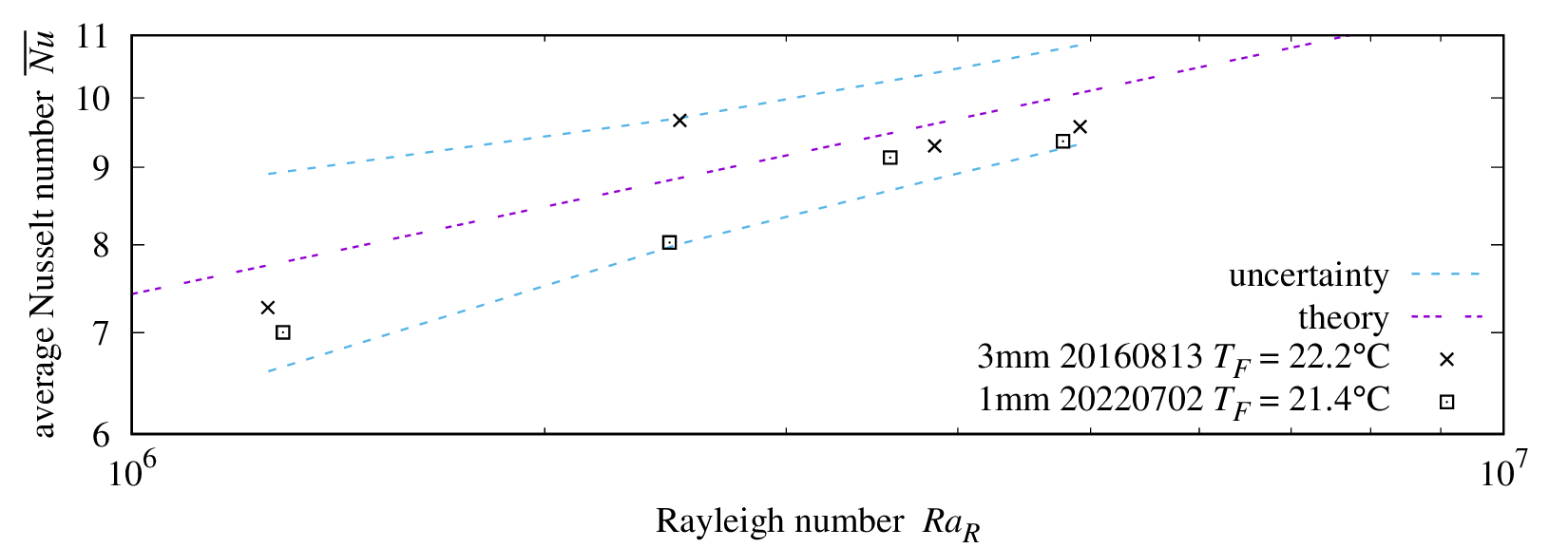}{435pt}\hfill&\cr
\+\hfill{\bf\figdef{fig:downward}\quad Natural convection from downward-facing surface}\hfill&\cr
}

\vfill\eject

\subsection{Mixed Convection Side Model}
Each $\ell^{\sqrt{2}}$-norm instance of a call to $U_N$ is replaced by
a call to $U_M$, with first argument $U_{fl}(u)$ or $U_{ft}(u)$.
In order to ignore forced convection from the leading and trailing
sides, $U_{ft}(u)=0$ when $\psi=90^\circ$ (horizontal flow);
otherwise, $U_{fl}(u)=0$.

The reuptake instances of $U_N(\theta,L_F,L_W,\phi)$ are changed to
the equivalent $U_M(0,\theta,L_F,L_W,\phi,0^\circ)$.

\medskip
\centerline{\bf\tabdef{mixed conductance functions}\quad{Mixed conductance functions and parameters}}
\moveright 0.143\hsize\vbox{\settabs 7\columns
\toprule
\+{\bf Symbol}&&{\bf Description}&&&\cr
\midrule
\+$U_{fl}(u)$&&level flow side forced thermal conductance&&&\cr
\+$U_{ft}(u)$&&tilted flow side forced thermal conductance&&&\cr
\+$\qquad u$&&bulk flow velocity&&&\cr
\+$U_M(U_F,\theta,L_F,L_W,\phi,\psi)$&&mixed convective conductance&&&\cr
\+$\qquad U_F$&&forced thermal conductance&&&\cr
\+$\qquad \theta$&&surface angle from vertical ($-90^\circ$ is face up)&&&\cr
\+$\qquad L_F$&&forced characteristic length&&&\cr
\+$\qquad L_W$&&other plate dimension&&&\cr
\+$\qquad \phi$&&rotation in plane of plate; integer multiple of $90^\circ$&&&\cr
\+$\qquad \psi$&&angle of fluid flow from vertical ($0^\circ$ is upward)&&&\cr
\bottomrule
}
\smallskip

$$\eqalign{
 U_S(u)&=\left\|U_\epsilon,U_M(U_{fl}(u),\theta-90^\circ,L_C,L_{es},0^\circ,\psi)\right\|_{\sqrt2}\cr
 &+U_M(0,\theta-90^\circ,L_C,L_{es},0^\circ,0^\circ)
 \,\min(0,\sin\theta,-.5\,\cos\theta, -2\,\cos\theta\,\sin\theta)\cr
 &+\left\|U_\epsilon,U_M(U_{fl}(u),90^\circ-\theta,L_C,L_{es},0^\circ,\psi)\right\|_{\sqrt2}\cr
 &+U_M(0,90^\circ-\theta,L_C,L_{es},0^\circ,0^\circ)
 \,\min(0,\sin\theta,-.5\,\cos\theta, 2\,\cos\theta\,\sin\theta)\cr
 &+2\,\left\|U_\epsilon,U_M(U_{ft}(u),0^\circ,L_{es},L_C,\theta,\psi)\right\|_{\sqrt2}\cr
 &+2\,U_M(0,0^\circ,L_{es},L_C,\theta,0^\circ)
 \,\min(0,\sin\theta)\cr
 }\eqdef{U_S(u)}$$

When $u$ is large, $U_S(u)$ approaches the sum of the forced
convection conductances.  $U_S(0)\equiv{U_S}$ of Formula~\eqref{eq:U_S}.

\subsection{Measurement Methodology}
 The measurement methodology employed is unusual.  Instead of waiting
 until the plate reaches thermal equilibrium, the plate is heated to
 15~K above ambient, heating stops, the fan runs at the designated
 speed, and convection cools the plate.  All of the sensor readings
 are captured each second during the 102 minute
 process, \tabref{tab:dynamic quantities} lists the dynamic physical
 quantities measured each second.  \tabref{tab:computed quantities}
 lists computed quantities.  Both $U_S(u)$ and
 $\{\epsilon_{\rm{Al}}\,\epsilon_{wt}\,h_R\,A\}$ are subtracted from
 the combined heat flow.  The mean of $\hol(u,t)$ over the time
 interval in which~$\Delta{T}$ drops by half (or exceeds 6142~s total
 time) is the result from that experiment.

\medskip
\centerline{\bf\tabdef{tab:dynamic quantities}\quad{Dynamic quantities}}
\moveright 0.25\hsize\vbox{\settabs 8\columns
\toprule
\+{\bf Symbol} &{\bf Units} & {\bf Description}&&\cr
\midrule
\+$\omega$ &r/min & fan rotation rate&&\cr
\+$T_F$ &K & ambient air temperature&&\cr
\+$T_P$ &K & plate temperature&&\cr
\+$T_B$ &K & back surface temperature&&\cr
\+$P$ &Pa & atmospheric pressure&&\cr
\+$\Phi$ &Pa/Pa & air relative humidity&&\cr
\bottomrule
}
\medskip

\centerline{\bf\tabdef{tab:computed quantities}\quad{Computed quantities}}
\moveright 0.1875\hsize\vbox{\settabs 8\columns
\toprule
\+ {\bf Symbol} &{\bf Units} &{\bf Description} &&&\cr
\midrule
\+ $h_R$ &$\rm W/(m^2K)$ &radiative surface conductance &&&\cr
\+ $U_S(u)$ &W/K &side radiative and convective conductance &&&\cr
\+ $\hol(u,t)$ &$\rm W/(m^2K)$ &convective surface conductance &&&\cr
\bottomrule
}

\vfill\eject

\subsection{Heat Balance}
Collecting into $U_T(u)$ Formula~\eqref{eq:U_T(u)} those terms which have a
factor of temperature difference $\overline{T_P}-\overline{T_F}$,
Formula~\eqref{eq:heat-balance} is the heat balance equation of the
plate during convective cooling:
$$\eqalignno{U_T(u)&=U_S(u)+\{\hol(u)\,A\}+\{\epsilon_{\rm{Al}}\,\epsilon_{wt}\,h_R\,A\}&\eqdef{eq:U_T(u)}\cr
  0&=U_T(u)\,\left[\overline{T_P}-\overline{T_F}\right]+U_I\left[\overline{T_P}-\overline{T_B}\right]+
  C_{pt}\,{\diff{\overline{T_P}}\over \diff{t}}&\eqdef{eq:heat-balance}\cr}$$

The plate and ambient temperatures are functions of time $t$.
Determined experimentally during heating, the temperature group-delay
through the 2.54~cm block of insulation between the slab and back
sheet is~110~s:
$$\overline{T_P}(t)={U_T(u)\,\overline{T_F}(t)+U_I\,\overline{T_B}(t-110{\rm~s})
-C_{pt}\,[\diff{\overline{T_P}(t)}/\diff{t}]
  \over U_T(u)+U_I}\eqdef{eq:delay}$$

To compute Nusselt number $\Nuol=\hol\,L/k$, Equation~\eqref{eq:delay} is
solved for the $\{\hol(u,t)\,A\}$ term from Equation~\eqref{eq:U_T(u)}.
$$\eqalignno{\varsigma(t)&=-U_I\,\left[\overline{T_P}(t)-\overline{T_B}(t-110{\rm~s})\right]&\eqdef{eq:eta(u,t)}\cr
  \{\hol(u,t)\,A\}&={\varsigma(t)-C_{pt}\,\left[\overline{T_P}(t)-\overline{T_P}(t')\right]/[t-t']
         \over \overline{T_P}(t)-\overline{T_F}(t)}
        -\{\epsilon_{\rm{Al}}\,\epsilon_{wt}\,h_R\,A\}-U_S(u)&\eqdef{eq:U_P(t)}}$$
 where $t'$ is the previous value of $t$.  In
 Equations~\eqref{eq:eta(u,t)} and \eqref{eq:U_P(t)}, $\overline{T_P}(t)$,
 $\overline{T_F}(t)$, and $\overline{T_B}(t)$ are the 15-element
 cosine averages of plate and fluid temperatures (centered at
 time~$t$).

In order to simulate $T_P$ from the other dynamic inputs,
\eqref{eq:delay} is solved as a finite-difference equation where
$\diff{t}=t-t'=1$:
$${T_P}(t)={U_T(u)\,\overline{T_F}(t)+U_I\,\overline{T_B}(t-110{\rm~s})
+C_{pt}\,{{T_P}(t')}
  \over U_T(u)+U_I+C_{pt}}\eqdef{eq:T_P}$$

In Equation~\eqref{eq:T_P}, $T_P(t')$ is the previous simulated value,
not a measured value.

\subsection{Measurement Uncertainty}
Following Abernethy, Benedict, and Dowdell~\cite{Abernethy1985}, the
final steps in processing an experiment's data are:
\orderedlist

 \li Using Equation~\eqref{eq:U_P(t)}, calculate the sensitivities of
 convected power $\hol\,A\,\Delta{T}$ per each parameter's
 average over the measurement time-interval;

 \li multiply the absolute value of each sensitivity by its estimated
 parameter bias to yield component uncertainties;

 \li calculate combined bias uncertainty as the root-sum-squared (RSS)
 of the component uncertainties;

 \li calculate the RSS combined measurement uncertainty as the RSS of
 the combined bias uncertainty and twice the product of the rotation
 rate sensitivity and variability.

\endorderedlist

\vfill\eject

 \tabrefs{tab:20160830T024010} and \tabrefn{tab:20230712T011216} list
 the sensitivity, bias, and uncertainty for each component
 contributing more than 0.20\% uncertainty for downward-facing 3~mm
 and 1~mm roughness plates, respectively.
 \figrefs{fig:measure-vs-theory-3} and \figrefn{fig:measure-vs-theory-1} show
 the measurements relative to the present theory for rough flow
 and turbulent flow, respectively.

 The supplementary data contains these graphs and tables for each
 data-set.

\medskip
\input 20160830T024010.tex
\medskip
\input 20230712T011216.tex
\medskip

\vbox{\settabs 2\columns
\+\hfill\figscale{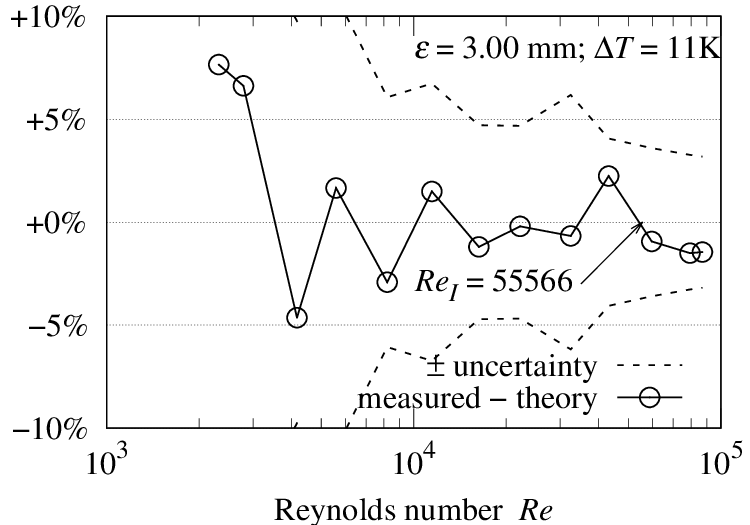}{234pt}\hfill&
 \hfill\figscale{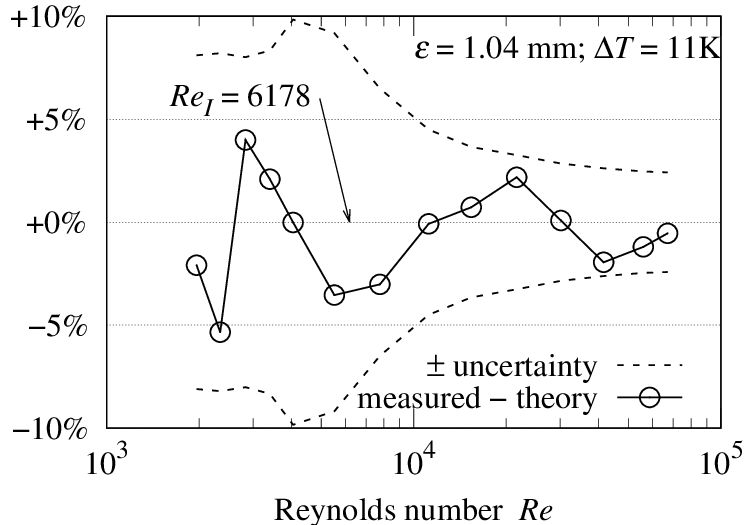}{234pt}\hfill&\cr
\+\hfill{\bf\figdef{fig:measure-vs-theory-3}\quad Measured versus theory $\varepsilon=3$~mm}\hfill&
 \hfill{\bf\figdef{fig:measure-vs-theory-1}\quad Measured versus theory $\varepsilon=1$~mm}\hfill&\cr
}

\subsection{Details}
  Documentation, photographs, electrical schematics, and software
  source-code for the apparatus, as well as calibration and
  measurement data are available from:

  {\tt{http://people.csail.mit.edu/jaffer/convect}}

\section{References}

\bibliographystyle{unsrtDOI}
\bibliography{citations}

\bye

%% file: macros.tex
\input eplain

\def\toprule{\vskip1.5pt\hrule height0.8pt\vskip1.5pt}
\def\midrule{\vskip1.5pt\hrule\vskip1.5pt}
\def\bottomrule{\vskip1.5pt\hrule height0.8pt\vskip1.5pt}

\def\nomenclature[#1]#2#3{\line{{#2\hfill} \hbox to0.85\hsize {#3\hfill}}}

\def\abbreviation#1#2{\line{{#1\hfill} \hbox to0.75\hsize {#2\hfill}}}

\newcount\fignumber
\def\figdef#1{\global\advance\fignumber by 1 \definexref{#1}{\number\fignumber}{figure}\ref{#1}}
\def\figdefn#1{\global\advance\fignumber by 1 \definexref{#1}{\number\fignumber}{figure}}
\let\figref=\ref
\let\figrefn=\refn
\let\figrefs=\refs

\newcount\tabnumber
\def\tabdef#1{\global\advance\tabnumber by 1 \definexref{#1}{\number\tabnumber}{table}\ref{#1}}
\def\tabdefn#1{\global\advance\tabnumber by 1 \definexref{#1}{\number\tabnumber}{table}}
\let\tabref=\ref
\let\tabrefn=\refn
\let\tabrefs=\refs

\ifx\pdfoutput\undefined
\input epsf

\def\figscale#1#2{\epsfxsize=#2\epsfbox{#1.eps}}
\else

\def\figscale#1#2{\pdfximage width#2 {#1.pdf}\pdfrefximage\pdflastximage}
\fi

\newcount\scount \scount=0
\newcount\sscount \sscount=0

\makeatletter
\def\section#1\par{
  \vskip\z@ plus.3\vsize\penalty-250
  \vskip\z@ plus-.3\vsize\bigskip\vskip\parskip
  \global\advance\scount by1
  \sscount=0
  \writenumberedtocentry{section}#1{\the\scount}
  \definexref#1{\the\scount}{section}
  \message{#1}
  \noindent\the\scount.\quad{\bf #1}\nobreak\smallskip\noindent}
\makeatother

\def\subsection#1{
  \global\advance\sscount by1
  \smallskip
  \noindent{~~\the\scount.\the\sscount~~{\bf{#1.~}}}}

%% file: Rowley-et-al-sources.tex
\+smooth-plaster &\hfill $   0.0^\circ$ &\hfill $  90.0^\circ$~~& $2.8\times10^{6}$ & $1.1\times10^{7}$ & $7.9\times10^{4}$ & $8.3\times10^{4}$ & \hfil  2&\cr
\+smooth-plaster &\hfill $   0.0^\circ$ &\hfill $  90.0^\circ$~~& $3.1\times10^{6}$ & $1.1\times10^{7}$ & $6.7\times10^{4}$ & $7.1\times10^{4}$ & \hfil  2&\cr
\+smooth-plaster &\hfill $   0.0^\circ$ &\hfill $  90.0^\circ$~~& $3.2\times10^{6}$ & $1.1\times10^{7}$ & $5.6\times10^{4}$ & $5.9\times10^{4}$ & \hfil  2&\cr
\+smooth-plaster &\hfill $   0.0^\circ$ &\hfill $  90.0^\circ$~~& $3.7\times10^{6}$ & $1.1\times10^{7}$ & $4.5\times10^{4}$ & $4.7\times10^{4}$ & \hfil  2&\cr
\+smooth-plaster &\hfill $   0.0^\circ$ &\hfill $  90.0^\circ$~~& $4.1\times10^{6}$ & $1.2\times10^{7}$ & $3.4\times10^{4}$ & $3.5\times10^{4}$ & \hfil  2&\cr
\+smooth-plaster &\hfill $   0.0^\circ$ &\hfill $  90.0^\circ$~~& $5.0\times10^{6}$ & $1.2\times10^{7}$ & $2.2\times10^{4}$ & $2.3\times10^{4}$ & \hfil  2&\cr
\+concrete &\hfill $   0.0^\circ$ &\hfill $  90.0^\circ$~~& $1.1\times10^{7}$ & $1.1\times10^{7}$ & $7.9\times10^{4}$ & $7.9\times10^{4}$ & \hfil  1&\cr
\+concrete &\hfill $   0.0^\circ$ &\hfill $  90.0^\circ$~~& $7.9\times10^{6}$ & $1.1\times10^{7}$ & $6.7\times10^{4}$ & $6.9\times10^{4}$ & \hfil  3&\cr
\+concrete &\hfill $   0.0^\circ$ &\hfill $  90.0^\circ$~~& $2.9\times10^{6}$ & $1.1\times10^{7}$ & $5.6\times10^{4}$ & $5.9\times10^{4}$ & \hfil  7&\cr
\+concrete &\hfill $   0.0^\circ$ &\hfill $  90.0^\circ$~~& $3.1\times10^{6}$ & $1.1\times10^{7}$ & $4.5\times10^{4}$ & $4.7\times10^{4}$ & \hfil  2&\cr
\+concrete &\hfill $   0.0^\circ$ &\hfill $  90.0^\circ$~~& $3.9\times10^{6}$ & $1.1\times10^{7}$ & $3.4\times10^{4}$ & $3.5\times10^{4}$ & \hfil  2&\cr
\+concrete &\hfill $   0.0^\circ$ &\hfill $  90.0^\circ$~~& $4.3\times10^{6}$ & $1.2\times10^{7}$ & $2.2\times10^{4}$ & $2.4\times10^{4}$ & \hfil  2&\cr
\+brick &\hfill $   0.0^\circ$ &\hfill $  90.0^\circ$~~& $2.0\times10^{6}$ & $1.1\times10^{7}$ & $6.4\times10^{4}$ & $6.7\times10^{4}$ & \hfil  6&\cr
\+brick &\hfill $   0.0^\circ$ &\hfill $  90.0^\circ$~~& $3.4\times10^{6}$ & $1.1\times10^{7}$ & $5.5\times10^{4}$ & $5.8\times10^{4}$ & \hfil  4&\cr
\+brick &\hfill $   0.0^\circ$ &\hfill $  90.0^\circ$~~& $2.7\times10^{6}$ & $1.1\times10^{7}$ & $4.5\times10^{4}$ & $4.7\times10^{4}$ & \hfil  3&\cr
\+brick &\hfill $   0.0^\circ$ &\hfill $  90.0^\circ$~~& $3.1\times10^{6}$ & $1.1\times10^{7}$ & $4.0\times10^{4}$ & $4.2\times10^{4}$ & \hfil  5&\cr
\+brick &\hfill $   0.0^\circ$ &\hfill $  90.0^\circ$~~& $4.3\times10^{6}$ & $1.1\times10^{7}$ & $3.0\times10^{4}$ & $3.2\times10^{4}$ & \hfil  4&\cr
\+brick &\hfill $   0.0^\circ$ &\hfill $  90.0^\circ$~~& $5.3\times10^{6}$ & $1.1\times10^{7}$ & $1.7\times10^{4}$ & $1.8\times10^{4}$ & \hfil  4&\cr
\+rough-plaster &\hfill $   0.0^\circ$ &\hfill $  90.0^\circ$~~& $1.1\times10^{7}$ & $1.1\times10^{7}$ & $6.7\times10^{4}$ & $6.7\times10^{4}$ & \hfil  1&\cr
\+rough-plaster &\hfill $   0.0^\circ$ &\hfill $  90.0^\circ$~~& $3.6\times10^{6}$ & $1.1\times10^{7}$ & $5.6\times10^{4}$ & $5.9\times10^{4}$ & \hfil  2&\cr
\+rough-plaster &\hfill $   0.0^\circ$ &\hfill $  90.0^\circ$~~& $3.7\times10^{6}$ & $1.1\times10^{7}$ & $4.5\times10^{4}$ & $4.7\times10^{4}$ & \hfil  2&\cr
\+rough-plaster &\hfill $   0.0^\circ$ &\hfill $  90.0^\circ$~~& $4.0\times10^{6}$ & $1.2\times10^{7}$ & $3.4\times10^{4}$ & $3.5\times10^{4}$ & \hfil  3&\cr
\+rough-plaster &\hfill $   0.0^\circ$ &\hfill $  90.0^\circ$~~& $4.8\times10^{6}$ & $1.2\times10^{7}$ & $2.2\times10^{4}$ & $2.4\times10^{4}$ & \hfil  2&\cr
\+stucco &\hfill $   0.0^\circ$ &\hfill $  90.0^\circ$~~& $1.1\times10^{6}$ & $1.1\times10^{7}$ & $6.7\times10^{4}$ & $7.1\times10^{4}$ & \hfil  2&\cr
\+stucco &\hfill $   0.0^\circ$ &\hfill $  90.0^\circ$~~& $6.9\times10^{5}$ & $1.1\times10^{7}$ & $5.6\times10^{4}$ & $6.0\times10^{4}$ & \hfil  3&\cr
\+stucco &\hfill $   0.0^\circ$ &\hfill $  90.0^\circ$~~& $1.2\times10^{6}$ & $1.1\times10^{7}$ & $4.5\times10^{4}$ & $4.8\times10^{4}$ & \hfil  3&\cr
\+stucco &\hfill $   0.0^\circ$ &\hfill $  90.0^\circ$~~& $1.5\times10^{6}$ & $1.1\times10^{7}$ & $3.4\times10^{4}$ & $3.6\times10^{4}$ & \hfil  3&\cr
\+stucco &\hfill $   0.0^\circ$ &\hfill $  90.0^\circ$~~& $2.1\times10^{6}$ & $1.1\times10^{7}$ & $2.2\times10^{4}$ & $2.4\times10^{4}$ & \hfil  2&\cr

%% file: Rowley-et-al-results.tex
\+smooth-plaster &\hfill$0.91$ &\hfill $0.20$~mm&\hfill$15.65$~m/s&\hfill $ 3.3\%$ &\hfill $ -0.8\%$ &\hfill $ 3.3\%$ &\hfil  2 &\cr
\+smooth-plaster &\hfill$0.91$ &\hfill $0.20$~mm&\hfill$13.41$~m/s&\hfill $ 2.7\%$ &\hfill $ -0.1\%$ &\hfill $ 2.7\%$ &\hfil  2 &\cr
\+smooth-plaster &\hfill$0.91$ &\hfill $0.20$~mm&\hfill$11.18$~m/s&\hfill $ 0.6\%$ &\hfill $ +0.6\%$ &\hfill $ 0.1\%$ &\hfil  2 &\cr
\+smooth-plaster &\hfill$0.91$ &\hfill $0.20$~mm&\hfill$ 8.94$~m/s&\hfill $ 1.5\%$ &\hfill $ +1.5\%$ &\hfill $ 0.3\%$ &\hfil  2 &\cr
\+smooth-plaster &\hfill$0.91$ &\hfill $0.20$~mm&\hfill$ 6.71$~m/s&\hfill $ 2.4\%$ &\hfill $ +2.3\%$ &\hfill $ 0.8\%$ &\hfil  2 &\cr
\+smooth-plaster &\hfill$0.91$ &\hfill $0.20$~mm&\hfill$ 4.47$~m/s&\hfill $ 2.7\%$ &\hfill $ +1.3\%$ &\hfill $ 2.4\%$ &\hfil  2 &\cr
\+concrete &\hfill$0.94$ &\hfill $0.55$~mm&\hfill$15.65$~m/s&\hfill $ 0.2\%$ &\hfill $ -0.2\%$ &\hfill $ 0.0\%$ &\hfil  1 &\cr
\+concrete &\hfill$0.94$ &\hfill $0.55$~mm&\hfill$13.41$~m/s&\hfill $ 2.3\%$ &\hfill $ -0.1\%$ &\hfill $ 2.3\%$ &\hfil  3 &\cr
\+concrete &\hfill$0.94$ &\hfill $0.55$~mm&\hfill$11.18$~m/s&\hfill $ 2.5\%$ &\hfill $ -1.0\%$ &\hfill $ 2.3\%$ &\hfil  7 &\cr
\+concrete &\hfill$0.94$ &\hfill $0.55$~mm&\hfill$ 8.94$~m/s&\hfill $ 4.7\%$ &\hfill $ -3.6\%$ &\hfill $ 3.0\%$ &\hfil  2 &\cr
\+concrete &\hfill$0.94$ &\hfill $0.55$~mm&\hfill$ 6.71$~m/s&\hfill $ 2.4\%$ &\hfill $ +0.1\%$ &\hfill $ 2.4\%$ &\hfil  2 &\cr
\+concrete &\hfill$0.94$ &\hfill $0.55$~mm&\hfill$ 4.47$~m/s&\hfill $ 4.1\%$ &\hfill $ +4.1\%$ &\hfill $ 0.3\%$ &\hfil  2 &\cr
\+brick &\hfill$0.93$ &\hfill $0.75$~mm&\hfill$12.67$~m/s&\hfill $ 4.7\%$ &\hfill $ -3.8\%$ &\hfill $ 2.8\%$ &\hfil  6 &\cr
\+brick &\hfill$0.93$ &\hfill $0.75$~mm&\hfill$10.95$~m/s&\hfill $ 3.6\%$ &\hfill $ +0.0\%$ &\hfill $ 3.6\%$ &\hfil  4 &\cr
\+brick &\hfill$0.93$ &\hfill $0.75$~mm&\hfill$ 8.94$~m/s&\hfill $ 4.4\%$ &\hfill $ -1.6\%$ &\hfill $ 4.1\%$ &\hfil  3 &\cr
\+brick &\hfill$0.93$ &\hfill $0.75$~mm&\hfill$ 8.00$~m/s&\hfill $ 2.5\%$ &\hfill $ +1.4\%$ &\hfill $ 2.1\%$ &\hfil  5 &\cr
\+brick &\hfill$0.93$ &\hfill $0.75$~mm&\hfill$ 5.99$~m/s&\hfill $ 4.4\%$ &\hfill $ +4.3\%$ &\hfill $ 0.8\%$ &\hfil  4 &\cr
\+brick &\hfill$0.93$ &\hfill $0.75$~mm&\hfill$ 3.38$~m/s&\hfill $ 2.9\%$ &\hfill $ -0.2\%$ &\hfill $ 2.9\%$ &\hfil  4 &\cr
\+rough-plaster &\hfill$0.91$ &\hfill $0.75$~mm&\hfill$13.41$~m/s&\hfill $ 2.6\%$ &\hfill $ -2.6\%$ &\hfill $ 0.0\%$ &\hfil  1 &\cr
\+rough-plaster &\hfill$0.91$ &\hfill $0.75$~mm&\hfill$11.18$~m/s&\hfill $ 2.5\%$ &\hfill $ -2.0\%$ &\hfill $ 1.5\%$ &\hfil  2 &\cr
\+rough-plaster &\hfill$0.91$ &\hfill $0.75$~mm&\hfill$ 8.94$~m/s&\hfill $ 1.3\%$ &\hfill $ +0.8\%$ &\hfill $ 1.1\%$ &\hfil  2 &\cr
\+rough-plaster &\hfill$0.91$ &\hfill $0.75$~mm&\hfill$ 6.71$~m/s&\hfill $ 0.5\%$ &\hfill $ -0.1\%$ &\hfill $ 0.4\%$ &\hfil  3 &\cr
\+rough-plaster &\hfill$0.91$ &\hfill $0.75$~mm&\hfill$ 4.47$~m/s&\hfill $ 1.7\%$ &\hfill $ -0.9\%$ &\hfill $ 1.5\%$ &\hfil  2 &\cr
\+stucco &\hfill$0.91$ &\hfill $1.47$~mm&\hfill$13.41$~m/s&\hfill $ 6.1\%$ &\hfill $ -6.1\%$ &\hfill $ 1.0\%$ &\hfil  2 &\cr
\+stucco &\hfill$0.91$ &\hfill $1.47$~mm&\hfill$11.18$~m/s&\hfill $ 2.7\%$ &\hfill $ -2.2\%$ &\hfill $ 1.6\%$ &\hfil  3 &\cr
\+stucco &\hfill$0.91$ &\hfill $1.47$~mm&\hfill$ 8.94$~m/s&\hfill $ 2.6\%$ &\hfill $ +2.2\%$ &\hfill $ 1.4\%$ &\hfil  3 &\cr
\+stucco &\hfill$0.91$ &\hfill $1.47$~mm&\hfill$ 6.71$~m/s&\hfill $ 6.0\%$ &\hfill $ +5.6\%$ &\hfill $ 2.3\%$ &\hfil  3 &\cr
\+stucco &\hfill$0.91$ &\hfill $1.47$~mm&\hfill$ 4.47$~m/s&\hfill $ 6.3\%$ &\hfill $ +6.3\%$ &\hfill $ 0.6\%$ &\hfil  2 &\cr

%% file: 20160830T024010.tex
\centerline{\hfil{\bf\tabdef{tab:20160830T024010}\quad{Estimated measurement uncertainties, bi-level 3mm roughness at $\Rey = 59593$.}}}
\vbox{\settabs 7\columns
\toprule
\+\hfil{\bf Symbol}&\hfil{\bf Nominal}&\hfil{\bf Sensitivity}&\hfil{\bf Bias}&\hfil{\bf Uncertainty}&{\bf Component}&&\cr
\midrule
\+\hfil      $\Delta T$&\hfil    9.47K  &\hfil $+12.2$\%/K  &\hfil   0.10K  &\hfil   1.22\%&       LM35C differential&\cr
\+\hfil             $P$&\hfil    101kPa &\hfil $+0.0009$\%/Pa &\hfil   1.5kPa &\hfil   1.28\%& MPXH6115A6U air pressure&\cr
\+\hfil        $C_{pt}$&\hfil   4.69kJ/K&\hfil $+0.024$\%/(J/K)&\hfil     47J/K&\hfil   1.13\%&   plate thermal capacity&\cr
\+\hfil          $\eta$&\hfil   0.401   &\hfil $+180$\%   &\hfil   0.014   &\hfil   2.52\%&   anemometer calibration&\cr
\+\hfil     $\varsigma$&\hfil   6.00mm  &\hfil $+11285$\%/m  &\hfil   100um  &\hfil   1.13\%&              post height&\cr
\+&&&&\hfil    3.49\%&combined bias uncertainty&\cr
\+\hfil{\bf Symbol}&\hfil{\bf Nominal}&\hfil{\bf Sensitivity}&\hfil{\bf Variability}&\hfil{\bf Uncertainty}&{\bf Component}&&\cr
\midrule
\+\hfil        $\omega$&\hfil     905r/min&\hfil $+0.081$\%/(r/min)&\hfil    5.2r/min&\hfil   0.43\%&        fan rotation rate&\cr
\+&&&&\hfil    3.60\%&RSS combined uncertainty&\cr
\bottomrule

}

%% file: 20230712T011216.tex
\centerline{\hfil{\bf\tabdef{tab:20230712T011216}\quad{Estimated measurement uncertainties, bi-level 1mm roughness at $\Rey = 55935$.}}}
\vbox{\settabs 7\columns
\toprule
\+\hfil{\bf Symbol}&\hfil{\bf Nominal}&\hfil{\bf Sensitivity}&\hfil{\bf Bias}&\hfil{\bf Uncertainty}&{\bf Component}&&\cr
\midrule
\+\hfil      $\Delta T$&\hfil    10.2K  &\hfil $+11.7$\%/K  &\hfil   0.10K  &\hfil   1.17\%&       LM35C differential&\cr
\+\hfil             $P$&\hfil  100.0kPa &\hfil $+0.0008$\%/Pa &\hfil   1.5kPa &\hfil   1.26\%& MPXH6115A6U air pressure&\cr
\+\hfil        $C_{pt}$&\hfil   4.24kJ/K&\hfil $+0.028$\%/(J/K)&\hfil     42J/K&\hfil   1.17\%&   plate thermal capacity&\cr
\+\hfil          $\eta$&\hfil   0.340   &\hfil $+195$\%   &\hfil   0.003   &\hfil   0.66\%&   anemometer calibration&\cr
\+\hfil           $u_u$&\hfil   6.381   &\hfil $+2.44$\%   &\hfil   0.100   &\hfil   0.24\%&diffuser airflow upper bound&\cr
\+\hfil           $L_T$&\hfil   8.34mm  &\hfil $+9365$\%/m  &\hfil   100um  &\hfil   0.94\%&              post length&\cr
\+\hfil           $L_m$&\hfil   3.57mm  &\hfil $+454$\%/m  &\hfil   500um  &\hfil   0.23\%&   side metal strip width&\cr
\+\hfil $\epsilon_{rs}$&\hfil   0.040   &\hfil $+20.4$\%   &\hfil   0.010   &\hfil   0.20\%&  test-surface emissivity&\cr
\+\hfil $\epsilon_{wt}$&\hfil   0.900   &\hfil $+9.05$\%   &\hfil   0.025   &\hfil   0.23\%&   wind-tunnel emissivity&\cr
\+&&&&\hfil    2.44\%&combined bias uncertainty&\cr
\+\hfil{\bf Symbol}&\hfil{\bf Nominal}&\hfil{\bf Sensitivity}&\hfil{\bf Variability}&\hfil{\bf Uncertainty}&{\bf Component}&&\cr
\midrule
\+\hfil        $\omega$&\hfil   1.03kr/min&\hfil $+0.065$\%/(r/min)&\hfil    2.5r/min&\hfil   0.16\%&        fan rotation rate&\cr
\+&&&&\hfil    2.46\%&RSS combined uncertainty&\cr
\bottomrule

}

%% file: blend.bbl
\begin{thebibliography}{10}

\bibitem{10.1115/1.4046795}
V~Lienhard, John~H.
\newblock {Heat Transfer in Flat-Plate Boundary Layers: A Correlation for
  Laminar, Transitional, and Turbulent Flow}.
\newblock {\em Journal of Heat Transfer}, 142(6), 04 2020,
  doi:10.1115/1.4046795.
\newblock 061805.

\bibitem{thermo3040040}
Aubrey Jaffer.
\newblock Skin-friction and forced convection from rough and smooth plates.
\newblock {\em Thermo}, 3(4):711--775, 2023, doi:10.3390/thermo3040040.

\bibitem{fujii1972natural}
Tetsu Fujii and Hideaki Imura.
\newblock Natural-convection heat transfer from a plate with arbitrary
  inclination.
\newblock {\em International Journal of Heat and Mass Transfer},
  15(4):755--764, 1972, doi:10.1016/0017-9310(72)90118-4.

\bibitem{churchill1975correlating}
Stuart~W Churchill and Humbert~HS Chu.
\newblock Correlating equations for laminar and turbulent free convection from
  a vertical plate.
\newblock {\em International journal of heat and mass transfer},
  18(11):1323--1329, 1975, doi:10.1016/0017-9310(75)90243-4.

\bibitem{thermo3010010}
Aubrey Jaffer.
\newblock Natural convection heat transfer from an isothermal plate.
\newblock {\em Thermo}, 3(1):148--175, 2023, doi:10.3390/thermo3010010.

\bibitem{AIC:AIC690180606}
S.~W. Churchill and R.~Usagi.
\newblock A general expression for the correlation of rates of transfer and
  other phenomena.
\newblock {\em AIChE Journal}, 18(6):1121--1128, 1972,
  doi:10.1002/aic.690180606.

\bibitem{schlichting2014}
Hermann Schlichting.
\newblock {\em Boundary-Layer Theory}.
\newblock McGraw Hill, New Delhi, seventh edition, 2014.
\newblock Translated by Kestin, J.

\bibitem{HIEBER1973769}
C.A. Hieber.
\newblock Mixed convection above a heated horizontal surface.
\newblock {\em International Journal of Heat and Mass Transfer},
  16(4):769--785, 1973, doi:https://doi.org/10.1016/0017-9310(73)90090-2.

\bibitem{wang1982experimental}
X.~A. Wang.
\newblock An experimental study of mixed, forced, and free convection heat
  transfer from a horizontal flat plate to air.
\newblock {\em Journal of Heat Transfer}, 104(1):139--144, 1982,
  doi:10.1115/1.3245040.

\bibitem{osti_5873492}
D~L Siebers, R~G Schwind, and R~J Moffat.
\newblock Experimental mixed-convection heat transfer from a large, vertical
  surface in a horizontal flow.
\newblock Technical report, Sandia National Lab., Livermore, CA (United
  States), 7 1983.

\bibitem{Ramachandran1985}
N.~Ramachandran, B.~F. Armaly, and T.~S. Chen.
\newblock {Measurements and Predictions of Laminar Mixed Convection Flow
  Adjacent to a Vertical Surface}.
\newblock {\em Journal of Heat Transfer}, 107(3):636--641, 08 1985,
  doi:10.1115/1.3247471.

\bibitem{lin1990comprehensive}
HT~Lin, WS~Yu, and CC~Chen.
\newblock Comprehensive correlations for laminar mixed convection on vertical
  and horizontal flat plates.
\newblock {\em W{\"a}rme-und Stoff{\"u}bertragung}, 25(6):353--359, 1990.

\bibitem{KOBUS19953329}
C.J. Kobus and G.L. Wedekind.
\newblock An experimental investigation into forced, natural and combined
  forced and natural convective heat transfer from stationary isothermal
  circular disks.
\newblock {\em International Journal of Heat and Mass Transfer}, 38(18):3329 --
  3339, 1995, doi:10.1016/0017-9310(95)00096-R.

\bibitem{KOBUS20013381}
C.J. Kobus and G.L. Wedekind.
\newblock An experimental investigation into natural convection heat transfer
  from horizontal isothermal circular disks.
\newblock {\em International Journal of Heat and Mass Transfer}, 44(17):3381 --
  3384, 2001, doi:10.1016/S0017-9310(00)00330-6.

\bibitem{rowley1930surface}
F.B. Rowley, A.B. Algren, and J.L. Blackshaw.
\newblock Surface conductances as affected by air velocity, temperature and
  character of surface.
\newblock {\em ASHVE Trans.}, 36:429--446, 1930.

\bibitem{bergman2007fundamentals}
F.P. Incropera, D.P. DeWitt, T.L. Bergman, and A.S. Lavine.
\newblock {\em Fundamentals of Heat and Mass Transfer}.
\newblock Wiley, Hoboken, NJ, USA, 2007.

\bibitem{AIHARA19722535}
T~Aihara, Y~Yamada, and S~End\"o.
\newblock Free convection along the downward-facing surface of a heated
  horizontal plate.
\newblock {\em International Journal of Heat and Mass Transfer}, 15(12):2535 --
  2549, 1972, doi:10.1016/0017-9310(72)90145-7.

\bibitem{Zukauskas1987}
A.~\v{Z}ukauskas and A.~\v{S}lan\v{c}iauskas.
\newblock {\em Heat Transfer in Turbulent Fluid Flows}.
\newblock Hemisphere Publishing Corp, Washington, DC, 1987.

\bibitem{ahtt5e}
J.~H. Lienhard, IV and J.~H. Lienhard, V.
\newblock {\em A Heat Transfer Textbook}.
\newblock Phlogiston Press, Cambridge, MA, 5th edition, August 2020.
\newblock Version 5.10.

\bibitem{rice2004emittance}
R.W. Rice.
\newblock Emittance factors for infrared thermometers used for wood products.
\newblock {\em Wood and Fiber Science}, 36:520--526, 2004.

\bibitem{s21061961}
Eva Barreira, Ricardo M. S.~F. Almeida, and Maria~L. Simões.
\newblock Emissivity of building materials for infrared measurements.
\newblock {\em Sensors}, 21(6), 2021, doi:10.3390/s21061961.

\bibitem{Abernethy1985}
R.B. Abernethy, R.P. Benedict, and R.B. Dowdell.
\newblock Asme measurement uncertainty.
\newblock {\em ASME. J. Fluids Eng.}, 107(2):161--164, 1985,
  doi:10.1115/1.3242450.

\end{thebibliography}
